\documentclass[fleqn,usenatbib]{mnras}
\usepackage{newtxtext,newtxmath}
\usepackage[T1]{fontenc}
\usepackage[normalem]{ulem}
\usepackage{epsfig}
\usepackage{mathtools}
\usepackage{cuted}

\title[Thermal evolution in the Primordial Disk]{Thermophysical evolution of planetesimals in the Primordial Disk}

\author[Bj\"{o}rn J. R. Davidsson]{
Bj\"{o}rn J. R. Davidsson,$^{1}$\thanks{E-mail: bjorn.davidsson@jpl.nasa.gov}
\\
$^{1}$Jet Propulsion Laboratory, California Institute of Technology,  M/S 183--401, 4800 Oak Grove Drive, Pasadena, CA 91109, USA
}

\date{Accepted 2021 May 28. Received 2021 May 23; in original form 2021 February 12}

\pubyear{2021}

\begin{document}
\label{firstpage}
\pagerange{\pageref{firstpage}--\pageref{lastpage}}
\maketitle


\begin{abstract}
The Primordial Disk of small icy planetesimals, once located at $15$--$30\,\mathrm{AU}$ from the Sun, was disrupted by giant planet migration in the early Solar System. 
The Primordial Disk thereby became the source region of objects in the current--day Kuiper Belt, Scattered Disk, and Oort Cloud. I present the thermophysics code 
``Numerical Icy Minor Body evolUtion Simulator'', or \texttt{NIMBUS}, and use it to study the thermophysical evolution of planetesimals in the Primordial Disk prior to its disruption. 
Such modelling is mandatory in order to understand the behaviour of dynamically new comets from the Oort Cloud, as well as the activity of Centaurs and short--period comets 
from the Scattered Disk, that return pre--processed to the vicinity of the Sun. I find that bodies in the midst of the Primordial Disk with diameters ranging $4$--$200\,\mathrm{km}$ lost 
all their CO ice on time-scales of order 0.1--$10\,\mathrm{Myr}$ depending on size, through a combination of protosolar and long--lived radionuclide heating. CO and other 
hypervolatiles therefore require a less volatile host for their storage. I consider two possible hosts: amorphous water ice and $\mathrm{CO_2}$ ice. Because of the high 
luminosity of the protosun, some Primordial Disk bodies may have sustained significant crystallisation, $\mathrm{CO:CO_2}$ segregation, and $\mathrm{CO_2}$ sublimation 
in the uppermost few tens of meters. I discuss how this may affect coma abundance ratios and distant activity in dynamically new comets.
\end{abstract}


\section{Introduction} \label{sec_intro}

The Nice model \citep{tsiganisetal05, morbidellietal05, gomesetal05b} postulates that the giant 
planets once were located in a compact orbital configuration roughly $5$--$15\,\mathrm{AU}$ from the Sun. Such a configuration 
appears to be caused by migratory behaviour of the giant planets within the gas disk from which they formed \citep{massetsnellgrove01, morbidellicrida07, 
morbidellietal07, walshetal11}.  Beyond $15\,\mathrm{AU}$ lay a vast population of icy minor bodies with a sharp truncation near $30\,\mathrm{AU}$ known as the Primordial Disk \citep{gomes04}. 
Beyond $30\,\mathrm{AU}$ lay a substantially smaller population that still is present \citep{parkerkavelaars10, parkeretal11} in 
the form of the dynamically cold Kuiper Belt. The Nice model demonstrates that a gravitational instability among the giant planets, 
followed by planetesimal--driven migration to their current locations, disrupted the Primordial Disk.  A fraction of its members was distributed  
between four large populations that formed because of this event and remain to this day: 1) the Trojan swarms in 1:1 mean motion resonance with Jupiter 
\citep{morbidellietal05}; 2) the dynamically hot Kuiper Belt that is superimposed on the previously mentioned dynamically 
cold population \citep{fernandezandip84, malhotra93, levisonetal08}; 3) the Scattered Disk with objects having 
perihelia near the orbit of Neptune and aphelia far beyond the Kuiper Belt \citep{duncanlevison97}; 4) the Oort Cloud \citep{brassermorbidelli13}.

Numerous authors have investigated the thermophysical evolution of bodies in the Kuiper Belt \citep[e.g.][]{desanctisetal01, desanctisetal07, choietal02, 
merkprialnik03, merkprialnik06, guilbertlepoutreetal11, prialniketal08, shchukoetal14}. 
Substantial efforts have also been made to understand the thermophysical evolution of Centaurs \citep[e.g.][]{desanctisetal00, capriaetal00b, capriaetal09, guilbertlepoutre11, saridandprialnik09}. 
Centaurs evolve dynamically from the Scattered Disk into the $5$--$30\,\mathrm{AU}$ region  and a fraction eventually become Jupiter Family Comets, henceforth abbreviated JFCs \citep{fernandez80, 
duncanlevison97, levisonduncan97}. The majority of current--day Centaurs therefore populate the region that once hosted the Primordial Disk. 
However, the solar--driven thermophysical evolution endured by Primordial Disk objects is expected to have been stronger than that of most Centaurs, and much stronger than that of current Kuiper Belt objects, 
for two reasons. Firstly, the luminosity of the protosun at the birthline\footnote{The birthline marks the transition from a Class~I protostar to a visually detectable T~Tauri star (a Class~II protostar), see e.~g., \citet{larson03}. 
This coincides with the end of strong accretion because the envelope and thick disk have been consumed. The thin disk remaining is referred to as the ``Solar Nebula''  in Solar System science 
terminology. The Solar Nebula ceases to exist when most gas has been removed by being heated to escape velocity (photoevaporation), consumed by the giant 
planets, or subjected to slow protosolar accretion.} was $\sim 6.4$ times the current solar luminosity, and it was still $\sim 1.7$ times higher after $1\,\mathrm{Myr}$, only falling below the 
current value $\sim 3\,\mathrm{Myr}$ into Solar System history \citep{pallastahler93}. Therefore, the protosun was substantially more luminous than the current Sun throughout 
the $\sim 3\,\mathrm{Myr}$ lifetime of the Solar Nebula \citep{zuckermanetal95, haischetal01, siciliaaguilaretal06}, during which 
most planetesimals formed and giant planet growth was completed. Consequently, Primordial Disk objects located at $15\,\mathrm{AU}$ could 
initially have evolved as if being placed at $\sim 6\,\mathrm{AU}$ today, and the conditions at $30\,\mathrm{AU}$ could have resembled those currently found at $\sim 12\,\mathrm{AU}$. 
Opaqueness of the Solar Nebula may have shielded planetesimals for some time period, that remains to be determined. Secondly, the estimated mean dynamical half--life of Centaurs with 
semi--major axes $a<29\,\mathrm{AU}$ and aphelia below $40\,\mathrm{AU}$ ranges from 
$2.7\,\mathrm{Myr}$ \citep{horneretal04} to $6.2\,\mathrm{Myr}$ \citep{disistobrunini07}. The lifetime of the Primordial Disk is a matter of debate. The original version 
of the Nice model suggested that the gravitational instability that disrupted the disk coincided with the Late Heavy Bombardment \citep{gomesetal05b}, which would imply a 
lifetime of at least $350$--$450\,\mathrm{Myr}$ and perhaps as much as $700\,\mathrm{Myr}$ \citep{morbidellietal12, bottkeetal12, marchietal13b}. 
In their ``Nice~II'' model scenario, \citet{morbidellietal07} and \citet{levisonetal11} strengthened the case for a late gravitational instability. However, \citet{nesvornyandmorbidelli12} 
demonstrated that models with five giant planets and an early gravitational instability successfully reproduced a number of observable criteria, 
suggesting that the Primordial Disk lifetime could have been as short as $\sim 15\,\mathrm{Myr}$. The shorter estimate of the Primordial Disk lifetime 
is still longer than the typical Centaur dynamical lifetime by a factor of a few \citep[although the lifetime of some resonance--hopping Centaurs may exceed $100\,\mathrm{Myr}$;][]{baileymalhotra09}. 
In conclusion, Primordial Disk objects were thermally processed at a higher radiative flux and for a longer time than most modern--day Centaurs.

For these reasons it is crucial to study the thermophysical evolution of icy minor bodies in the Primordial Disk. This is particularly 
important because: 1) this thermal processing is not only expected to be stronger, but also pre--dates the ones taking place among 
Kuiper Belt objects and Centaurs, thereby providing appropriate initial conditions for thermal studies of those populations; 2) the Primordial 
Disk thermophysical processing would also have affected objects currently placed in the Oort Cloud, thereby challenging the 
concept of dynamically new comets as being largely unevolved; 3) the predicted level of thermal processing in the current Nice model paradigm, 
if found to be at odds with observed properties, may provide novel constraints on the conditions and durations of early Solar System key processes and events.

This paper is therefore devoted to the study of thermophysical processing of minor icy bodies in the Primordial Disk. It focuses on a number of 
specific questions that are inspired by previous work and recent observations, as motivated in the following. Theoretical studies of Kuiper Belt objects have shown that 
hypervolatiles such as $\mathrm{CO}$, when stored within the bodies as separate and pure condensed ices,  readily sublimate at such 
distances. \citet{desanctisetal01} considered a $D=80\,\mathrm{km}$ diameter body orbiting at $41$--$45\,\mathrm{AU}$, having a dust/ice mass 
ratio of 1 or 5, an $80\%$ porosity, containing $1\%$ condensed $\mathrm{CO}$ ice relative water, where the dust component contained the 
long--lived radionuclides $^{40}\mathrm{K}$, $^{232}\mathrm{Th}$, $^{235}\mathrm{U}$, and $^{238}\mathrm{U}$ with chondritic abundances. 
They solved the coupled heat and gas diffusion problem for a spherical--symmetric body. They modelled the body for $10\,\mathrm{Myr}$ and found 
that the $\mathrm{CO}$ sublimation front had withdrawn $5$--$8\,\mathrm{km}$ below the surface at that time, depending on the assumed dust/ice mass ratio. 
Based on the asymptotically declining $\mathrm{CO}$ loss rate, \citet{desanctisetal01} speculated that a steady--state might be reached where sublimation 
is balanced by recondensation and net loss of $\mathrm{CO}$ eventually ceases. If so, an inner core of $\mathrm{CO}$ ice could remain. Among several different 
model bodies, \citet{choietal02} considered a $D=200\,\mathrm{km}$ case at $30\,\mathrm{AU}$, having a dust/ice mass ratio of 1, a $50\%$ porosity, containing 
$10\%$ condensed $\mathrm{CO}$ ice relative water, with long--lived radionuclides included. Their model considered heat conduction for a spherical--symmetric body, but 
assumed immediate escape of vapour because a proper treatment of gas diffusion was considered computationally prohibitive. They modelled the body for the 
Solar System lifetime and found that all $\mathrm{CO}$ was lost in little over $10\,\mathrm{Myr}$. Their other models additionally considered the 
short--lived radionuclide $^{26}\mathrm{Al}$ that, if present at the time of planetesimal formation, could speed up the $\mathrm{CO}$ loss substantially. 
These works, that apparently are the only ones that have attempted to quantify the $\mathrm{CO}$ loss time--scale, inspire to the following related questions.

\begin{trivlist}
\item Q\#1. Under what conditions is loss of condensed $\mathrm{CO}$ ice complete or partial in the Primordial Disk? If loss of condensed $\mathrm{CO}$ ice 
is complete, what are the time--scales for that loss in bodies of different size in the Primordial Disk? 
\end{trivlist}

Next, consider the problem of the activation distance of most comets and Centaurs, as well as extreme exceptions to the general rule. 
There is mounting evidence that a fundamental change in the level of activity of icy minor bodies takes place as they pass 
a transition region $11\pm1\,\mathrm{AU}$ from the Sun inbound. \citet{jewitt09} demonstrated that all Centaurs 
known to be active at the time were confined to the $5$--$12\,\mathrm{AU}$ region and that the lack of active Centaurs 
at larger distances was not an observational bias. Radio telescope searches for gaseous $\mathrm{CO}$ in a combined sample of 18 
Centaurs and Kuiper Belt objects located beyond $\sim 14\,\mathrm{AU}$ by \citet{bockeleemorvanetal01} and \citet{jewittetal08} yielded no detections but  
upper limits only. A Hubble Space Telescope survey of 53 Centaurs and Kuiper Belt objects with perihelia at $15\leq q\leq 30\,\mathrm{AU}$ did not detect 
dust activity in any of those targets \citep{lietal20}. \citet{cabraletal19} observed 20 Centaurs at $6.2$--$25.3\,\mathrm{AU}$, of which 15 were beyond 
$12\,\mathrm{AU}$, and none had a coma. Statistics on the discovery distances of 2,096 dynamically new and long--period comets assembled 
by \citet{meechetal17} shows that all distances within the orbit of Saturn are well represented. However, there is a sharp transition near $10\,\mathrm{AU}$ 
beyond which merely a dozen discoveries have been made. This suggests a sudden systematic onset of activity and brightening near $10\,\mathrm{AU}$. 
Among the Centaurs that do display activity, $\mathrm{CO}$ gas production has been verified in 29P/Schwassmann--Wachmann~1 \citep{senayandjewitt94}, 
95P/Chiron \citep{womackandstern99}, and 174P/Echeclus \citep{wierzchosetal17}. Carbon monoxide is also one of the most common and abundant 
gases observed in JFCs \citep[e.g.][]{bockeleemorvanetal04, ahearnetal12}. As was pointed out by \citet{jewitt09}, the ubiquitous presence 
of $\mathrm{CO}$ outgassing in icy minor bodies near the Sun, but its absence in the majority of such objects beyond the orbit of Saturn, is inconsistent with the 
concept of large deposits of condensed $\mathrm{CO}$ ice. If condensed $\mathrm{CO}$ ice was abundant, the icy minor bodies would display measurable 
activity beyond the orbit of Neptune. The conclusion must be that most currently available $\mathrm{CO}$ is trapped (occluded) within a host medium that does 
not release the $\mathrm{CO}$ unless a boundary near $11\pm1\,\mathrm{AU}$ is crossed. \citet{jewitt09} proposed that amorphous water ice is 
the most likely host, particularly because the sunward emission of $\mathrm{CO}$ inferred from the blue--shifted $\mathrm{CO}$ lines in 29P/Schwassmann--Wachmann~1 
requires a host located sufficiently close to the surface to be sensitive to diurnal temperature variations. The same sunward emission is seen in 
174P/Echeclus \citep{wierzchosetal17}. This excludes a deep source of condensed $\mathrm{CO}$ ice in combination with thermal lag effects as 
a viable explanation for the delayed onset of $\mathrm{CO}$ outgassing. The capability of amorphous water ice to trap hypervolatiles like $\mathrm{CO}$, and 
the detailed conditions governing their release, have been well documented in laboratory experiments \citep[e.g.][]{barnunetal85, barnunetal87}. 
The inclusion of amorphous water ice, crystallisation, and release of occluded species are standard features of advanced comet nucleus thermophysical 
codes since the seminal papers by \citet{prialnikandbarnun87, prialnikandbarnun90} and 
used, e.~g., to explain cometary outbursts. A systematic investigation of the level of amorphous water ice crystallisation at different distances and 
body latitudes, for a range of spin axis obliquities, was made by \citet{guilbertlepoutre12}. She finds that crystallisation is an efficient source of 
comet activity at $10$--$12\,\mathrm{AU}$, and that crystallisation--driven activity beyond that distance is very unlikely. This reinforces the 
suggestion by \citet{jewitt09} that most activity in Centaurs and distant comets may be governed by the crystallisation of amorphous water ice. 
This, in fact, has been the working assumption in comet science for decades \citep[e.~g., reviews by][]{prialniketal04, prialniketal08}, 
though empirical evidence has been lacking.

The surveys of activity in distant dynamically new or long--period comets by \citet{sarneczkyetal16} and \citet{kulyketal18} confirm that 
development of comae and even tails are common at $5$--$12\,\mathrm{AU}$ and that observationally confirmed activity beyond such distances is 
very rare. There are, however, notable exceptions. Comet C/2010~U3~(Boattini) was active at a record distance of $24.6$--$25.8\,\mathrm{AU}$ inbound 
with a dust production of $\sim 0.5\,\mathrm{kg\,s^{-1}}$, apparently keeping a similar level of activity all the way down to $\sim 8.5\,\mathrm{AU}$ from the Sun 
\citep{huietal19}. Comet C/2017~K2~(PANSTARRS) was active at $23.7\,\mathrm{AU}$ inbound, with an estimated dust production rate of $60\,\mathrm{kg\,s^{-1}}$ 
at $15.9\,\mathrm{AU}$ \citep{jewittetal17}. \citet{huietal18} estimate the dust production rate for the $16.2$--$23.7\,\mathrm{AU}$ segment of the 
orbit to $240\pm 110\,\mathrm{kg\,s^{-1}}$. Evidently, Comet C/2017~K2 is producing at least two orders of magnitude more dust than Comet Boattini 
at comparable heliocentric distances. Thermophysical modelling by \citet{meechetal17}, including an estimate of the dust production rate and the 
associated coma brightness, shows that the observed magnitude of C/2017~K2 from $28.7\,\mathrm{AU}$ down to $15.5\,\mathrm{AU}$ can be reproduced by the model if 
the hypervolatile $\mathrm{CO}$ is driving activity. If the substantially more stable supervolatile $\mathrm{CO_2}$ is considered instead, \citet{meechetal17} find 
that the predicted brightness at $25\,\mathrm{AU}$ would be $\sim 5\,\mathrm{magn}$ fainter than the observed one, which corresponds to a two orders of 
magnitude smaller dust production rate. Therefore, $\mathrm{CO_2}$--driven distant activity is potentially relevant for Comet Boattini, but not for Comet C/2017~K2.

Although the activity of Comet C/2017~K2 may indeed be driven by the sublimation of condensed $\mathrm{CO}$, there are alternative options. One possibility 
is that $\mathrm{CO}$ is partially stored by an additional host medium that releases the $\mathrm{CO}$ at lower temperatures than possible for amorphous 
water ice. Based on comet abundances \citep{bockeleemorvanetal04}, the most likely host besides amorphous water ice would be condensed $\mathrm{CO_2}$ ice. Observations of protostars 
show that roughly $2/3$ of the $\mathrm{CO_2}$ is intimately mixed with water ice, but that the remainder forms a separate $\mathrm{CO:CO_2}$ system 
\citep{pontoppidanetal08}. Considering that the ESA Rosetta mission demonstrated that a fraction of ices in Comet 67P/Churyumov--Gerasimenko appear to have presolar origin 
\citep{calmonteetal16, martyetal17}, icy minor Solar System bodies could have inherited $\mathrm{CO:CO_2}$ mixtures from 
the interstellar medium.  Laboratory experiments show that $\mathrm{CO_2}$ readily traps hypervolatiles such as $\mathrm{CH_4}$ \citep{lunaetal08}, 
$\mathrm{N_2}$ \citep{satorreetal09}, and $\mathrm{CO}$ \citep{simonetal19}. Such substances are released from the $\mathrm{CO_2}$ ice above their individual sublimation 
temperatures (the process is called segregation or distillation), primarily at a temperature interval of $70$--$90\,\mathrm{K}$. As pointed out by \citet{jewittetal17}, the activity of  Comet C/2017~K2 took place 
at a heliocentric distance where a nucleus temperature of $60$--$70\,\mathrm{K}$ is expected. This makes segregation of $\mathrm{CO}$ out from a $\mathrm{CO_2}$ 
host a relevant process. Furthermore, \citet{luspaykutietal15} demonstrated that the production rates of $\mathrm{CO}$ and $\mathrm{C_2H_2}$ were correlated with 
that of $\mathrm{CO_2}$ in Comet 67P/Churyumov--Gerasimenko, and distinctively different from the production of water. Building on that study, \citet{gascetal17} 
showed that the production rates of $\mathrm{CH_4}$, $\mathrm{HCN}$, and $\mathrm{H_2S}$ also correlated with $\mathrm{CO_2}$, while $\mathrm{O_2}$ was 
correlated with water. This led \citet{gascetal17} to propose the presence of two main phases of separate condensed ices -- $\mathrm{H_2O}$ and $\mathrm{CO_2}$ -- each of 
them trapping a specific set of more volatile species.

The possibility that segregation of $\mathrm{CO:CO_2}$ mixtures is responsible for activity at extreme distances, as well as its potential role played 
in the $11\pm 1\,\mathrm{AU}$ activation distance of most comets, needs to be explored further. However, to address this issue it is first necessary 
to understand the level of segregation that may have taken place already in the Primordial Disk. This inspires to the second question.

\begin{trivlist}
\item Q\#2. What degree of $\mathrm{CO}$ release from $\mathrm{CO_2}$ entrapment is expected to take place in the Primordial Disk? Specifically, at 
what depth is the segregation front located?  
\end{trivlist}

As previously mentioned, the objects in the Primordial Disk at $15$--$30\,\mathrm{AU}$ could temporarily have been subjected to protosolar 
radiation equivalent to being located at $6$--$12\,\mathrm{AU}$ in the current Solar System. As seen from the works by \citet{jewitt09} and 
\citet{guilbertlepoutre12} this coincides with the conditions where crystallisation of amorphous water ice is expected to be an important 
driver of cometary activity. The observations of Comet C/2015~ER$_{61}$~(PANSTARRS) at $8.9\,\mathrm{AU}$ down to $4.8\,\mathrm{AU}$ by 
\citet{meechetal17b}, and their thermophysical modelling efforts to reproduce the observed coma brightness, indicate that the activity 
of this particular object is driven by sublimation of condensed $\mathrm{CO_2}$ ice. This inspires to the last question considered in the current work.
\begin{trivlist}
\item Q\#3. At what depths are the $\mathrm{CO_2}$ sublimation front and the crystallisation front of amorphous water ice expected to be 
located after thermal processing in the Primordial Disk? 
\end{trivlist}

An additional key goal of the current paper is to make the first in--depth presentation and application of the thermophysics modelling code 
``Numerical Icy Minor Body evolUtion Simulator'', or \texttt{NIMBUS}, written by the author. It has previously only been briefly mentioned 
and applied by \citet{davidssonetal21} and \citet{masieroetal21}.

The rest of the paper is organised as follows. The thermophysical code is described in Sec.~\ref{sec_model}. Readers who wish to fast--forward 
to the results are encouraged to first read the Sec.~\ref{sec_model} introduction, Sec.~\ref{sec_model_govern}, and Sec.~\ref{sec_model_illum}, 
for a minimum summary of the model. The model parameters used for the simulations in this paper are 
motivated in Sec.~\ref{sec_parameters}, the results are presented in Sec.~\ref{sec_results}, and 
they are discussed in Sec.~\ref{sec_discussion}. Finally, the conclusions are summarised in Sec.~\ref{sec_conclusions}.

\section{The thermophysical model} \label{sec_model}

The \emph{in situ} exploration of comet nuclei, initiated when spacecraft from Europe, the Soviet Union, and Japan explored 
Comet 1P/Halley in 1986, sparked a strong interest in the thermophysics of porous and sublimating ice/dust mixtures. It intensified 
during the $\sim 20$ years preparation for the ESA cornerstone mission Rosetta to Comet 67P/Churyumov--Gerasimenko in 2014--2016. 
A large number of thermophysical models were developed and were used to systematically explore the importance and behaviour of 
various physical processes in comets \citep[e.g.][]{weissmanandkieffer81, fanaleandsalvail84, kuhrt84, rickmanandfernandez86, prialniketal87, prialnikandbarnun87, prialnikandbarnun88, prialnikandbarnun90, mekleretal90, rickmanetal90, espinasseetal91, komleanddettleff91, komleandsteiner92, prialnik92, kossackietal94, kuehrtandkeller94, tancredietal94, benkhoffandhuebner95, oroseietal95, prialnikpodolak95, skorovandrickman95, capriaetal96, capriaetal01,  benkhoffboice96, podolakandprialnik96, enzianetal97, enzianetal99, shoshanietal97, markiewiczetal98, desanctisetal99, skorovetal99}. In parallel to model development, numerous laboratory experiments 
were carried out to further the understanding of thermophysical evolution and to validate models \citep[e.g.][]{barnunetal85, barnunetal87, benkhoffandspohn91, grunetal91, grunetal93, benkhoffetal95, kossackietal97}. 
The appearances of bright Comets C/1996~B2~(Hyakutake) and C/1995 O1 (Hale--Bopp), the discovery of the Kuiper belt \citep{jewittetal92, jewittluu93}, and NASA flybys of 
Comets 19P/Borrelly, 81P/Wild~2, 9P/Tempel~1, and 103P/Hartley~2 in 2001--2010 triggered further development of models, particularly considering larger body sizes, where gravitational compression and 
radiogenic heating become important \citep[e.g][]{gutierrezetal00, gutierrezetal01, desanctisetal01, desanctisetal07, capriaetal02, capriaetal09, choietal02, davidssonandskorov02b, davidssongutierrez04, davidssongutierrez05, davidssongutierrez06, merkprialnik03, merkprialnik06, podolakprialnik06, gonzalezetal08, gonzalezetal14, rosenbergandprialnik09, rosenbergandprialnik10, marboeufetal10, marboeufetal11, prialniketal08, saridandprialnik09, gortsasetal11, guilbertlepoutreetal11}. Work to understand the thermophysics of the Rosetta target comet is ongoing \citep[e.g.][]{desanctisetal15,kelleretal15, kossackietal15, huetal17, hofneretal17, mousisetal17, hoangetal20, 
davidssonetal21, davidssonetal21b}, and has partially challenged previous ideas about comet physics. This primarily concerns ongoing discussions about dust mantle fragmentation and dust coma formation \citep{gundlachetal15,jewittetal19}, as well as the storage medium for hypervolatiles \citep{gascetal17}. The inspiration to construct \texttt{NIMBUS} came from this vast body of work, and the code should be considered 
an independent implementation of what may be considered a ``standard comet nucleus thermophysical model''. It contains most, though perhaps not all, physical processes that are common to state--of--the--art 
thermophysical models. \texttt{NIMBUS} relies on and benefits from decades of efforts in the scientific community to better understand the thermophysics of porous media consisting of refractories and various types of ice. 
The only feature that may be considered new is the inclusion of $\mathrm{CO_2}$ as a host of more volatile substances (in this first implementation, exclusively CO) and the associated segregation process. 
The current paper attempts to describe \texttt{NIMBUS} in quite some technical detail, with the hope that this will benefit future users\footnote{It is the intention of the author to make the 
\texttt{NIMBUS} source code publicly available at GitHub.}.

\texttt{NIMBUS} considers a spherical body of radius $R$ consisting of a porous mixture of refractories and different types of ice. Currently, \texttt{NIMBUS} 
includes $\mathrm{H_2O}$, $\mathrm{CO_2}$, and $\mathrm{CO}$ but more species can be added in the future. The water ice may be amorphous and during heating it first transforms 
to cubic ice and then to hexagonal ice. Amorphous water ice may trap other species that are released gradually in three stages -- during crystallisation, 
the cubic--hexagonal transition, and sublimation. Also $\mathrm{CO_2}$ may trap other species except water, that are released through segregation prior to 
the onset of $\mathrm{CO_2}$ sublimation. All other volatiles may be condensed\footnote{The term ``condensed'' is here used for pure ice substances. Because they are not 
trapped in another medium, they are free to sublimate when the temperature is sufficiently high. \texttt{NIMBUS} considers all vapour--to--ice transitions to form 
clean ices, which is why the term ``condensed'' is used interchangeably with ``pure'' or ``free'' ice.} or trapped in $\mathrm{H_2O}$ or $\mathrm{CO_2}$ in some proportion.

\texttt{NIMBUS} makes no explicit assumptions about the geometric arrangements of matter on small local scales, e.~g., whether 
it is monolithic (with or without holes), consists of evenly distributed grains, or forms a lumpy medium where grains locally are concentrated 
into larger pebbles. However, a particular geometric model may be implicitly applied by the model physical relations chosen in order to deal with, e.~g., 
the porosity--correction of heat conductivity, the calculation of gas diffusivity, or evaluation of volume mass production rates. Such model physical   
relations are retrieved through function calls, and those functions may contain several different versions of such relations, published by 
different authors (and more can be added over time). It is up to the user to select the exact model physical relation to be applied, via an 
identification number that is submitted to the function. Therefore, the implied geometric arrangement of matter may change 
depending on the choice of model physical relation. This paper describes the relations chosen for the current simulations, but they are 
not hard--wired in \texttt{NIMBUS}.

\texttt{NIMBUS} tracks the conduction of heat and the diffusion of vapour both radially and latitudinally. Heat is transported through 
solid--state conduction, by radiative transfer, and during gas diffusion (advection). Sources of heat include solar irradiation, 
short-- and long--lived radionuclei, energy release during crystallisation of amorphous water ice, and latent energy released when vapour 
is recondensing. Sinks of heat include thermal infrared radiative loss of energy from the surface into space, and consumption of 
energy during ice sublimation, segregation (here, $\mathrm{CO}$ escaping entrapment in $\mathrm{CO_2}$ ice), and crystallisation 
(here, $\mathrm{CO}$ and $\mathrm{CO_2}$ released from amorphous water ice). The model body can be treated as a slow 
rotator (to fully account for daily variations in illumination conditions), or as a fast rotator (where each considered latitude slab 
receives the diurnally averaged radiative flux valid for that latitude). Any elliptic, parabolic, or hyperbolic orbit can be considered, 
and any spin axis orientation may be applied. Orbital elements and spin angles can change with time, allowing for, e.~g., consideration 
of spin axis precession and nutation.

The following sub--sections describe \texttt{NIMBUS} in detail. Specifically, they deal with governing equations and boundary 
conditions (Sec.~\ref{sec_model_govern}), composition and porosity (Sec.~\ref{sec_model_comp}), heat capacity and 
heat conductivity (Sec.~\ref{sec_model_ck}), volume mass production rates and latent heats (Sec.~\ref{sec_model_sublcond}), 
crystallisation and the cubic--hexagonal transition (Sec.~\ref{sec_model_cryst}), release of $\mathrm{CO}$ and $\mathrm{CO_2}$ 
from sublimating $\mathrm{H_2O}$ (Sec.~\ref{sec_model_CO_CO2_fromH2O}), release of $\mathrm{CO}$ from $\mathrm{CO_2}$ 
(Sec.~\ref{sec_model_COfromCO2}), gas diffusion (Sec.~\ref{sec_model_diffusion}), radiogenic heating (Sec.~\ref{sec_model_decay}), 
erosion (Sec.~\ref{sec_model_erosion}), the spatial grid, temporal resolution, orbit, and spin (Sec.~\ref{sec_model_grid}), 
illumination conditions (Sec.~\ref{sec_model_illum}), and implementation (Sec.~\ref{sec_model_implementation}). Finally, verification of the 
code is described in Sec.~\ref{sec_model_verification}.

\subsection{Governing equations and boundary conditions} \label{sec_model_govern}

The coupled differential equations for conservation of energy (Eq.~\ref{eq:01}), vapour mass (Eq.~\ref{eq:02}) and ice mass (Eq.~\ref{eq:03}) govern 
the thermophysical evolution of the body. For the symbols used throughout this work, see Tables~\ref{tab1a}--\ref{tab1b} (functions),
 Table~\ref{tab1c} (parameters), and Tables~\ref{tab2a1}--\ref{tab2b} (constants). The current version of \texttt{NIMBUS} considers $n_{\rm s}=6$ number of species for 
which the following labelling convention is applied: refractories ($k=1$), amorphous water ice ($k=2$), cubic water ice ($k=3$), 
hexagonal/crystalline water ice ($k=4$), carbon monoxide ($k=5$), and carbon dioxide ($k=6$). Future versions of \texttt{NIMBUS} will incorporate 
a larger number of species, but the current focus is on the top three in terms of abundance. Note that the index $k$ nominally is used 
to refer to species, but in the governing equations it is necessary to distinguish between species acting as hosts (denoted by $i$) and species 
that are guests (denoted by $j$) within hosts. 

\newpage
The energy conservation equation, formulated in a spherical geometry, reads 

\begin{strip}
\begin{equation} \label{eq:01}
\begin{array}{c}
\displaystyle \rho(\psi) c(T)\frac{\partial T}{\partial t}=\frac{1}{r^2}\frac{\partial}{\partial r}\left(\kappa(\psi,\,T)r^2\frac{\partial T}{\partial r}\right)+\frac{1}{r\sin l}\frac{\partial }{\partial l}\left(\kappa (\psi,\,T)\frac{\sin l}{r}\frac{\partial T}{\partial l}\right)-\sum_{i=4}^{n_{\rm s}}q_i(p_i,\,T)L_i(T)+\sum_{i=2}^{n_{\rm s}}\sum_{j=5}^{n_{\rm s}}\left( q'_i(T)\left\{H_i-F_{i,j}L_j(T)\right\}\right)\\
\\
\displaystyle  -\sum_{i=4}^{n_{\rm s}}g_i\left(\Phi_i(p_i,T,\psi)\frac{\partial T}{\partial r}-\frac{\Psi_i(p_i,T,\psi)}{r}\frac{\partial T}{\partial l}\right)+\mathcal{R}.\\
\end{array}
\end{equation}
\end{strip}
Dependencies on $\{r,\,l,\,t\}$ for all functions are implicit, and other dependencies on functions have been written 
explicitly (primarily to highlight coupling between Eqs.~\ref{eq:01}--\ref{eq:02}, omitting parameters and constants for clarity). 
From left to right the terms in Eq.~(\ref{eq:01}) denote: 1) the rate of change of internal energy (in units $\mathrm{[J\,m^{-3}\,s^{-1}]}$); 
2) radial heat conduction; 3) latitudinal heat conduction; 4) consumption of energy by net volume sublimation or release of energy 
by net condensation; 5) energy release during phase transitions and energy consumption during escape of an occluded species 
from a host medium; 6) radial and latitudinal convection; 7) energy release by radioactive decay.

The mass conservation equation of vapour reads,
\begin{equation} \label{eq:02}
\begin{array}{c}
\displaystyle \psi m_i\frac{\partial n_i}{\partial t}=-\frac{1}{r^2}\frac{\partial}{\partial r}\left(r^2\Phi_i(p_i,\,T) \right)-\frac{1}{r\sin l}\frac{\partial }{\partial l}\left(\Psi_i(p_i,\,T)\sin l\right)+\\
\\
\displaystyle q_i(p_i,\,T)+\sum_{j=2}^{n_{\rm s}}F_{j,i}q'_j(T),\\
\end{array}
\end{equation}
where $i\geq 4$. In Eq.~(\ref{eq:02}) the terms denote: 1) the rate of change of vapour density within pore spaces; 2) radial gas diffusion; 
3) latitudinal gas diffusion; 4) vapour production during sublimation or vapour removal during condensation; 5) release of vapour from a host ice.
 
The mass conservation equation for solid ices reads,
\begin{equation} \label{eq:03}
\begin{array}{c}
\displaystyle \frac{\partial \rho_i}{\partial t}=-q_i(p_i,\,T)+\tau_i(p_i,\,T),
\end{array}
\end{equation}
where $i\geq 2$. In Eq.~(\ref{eq:03}) the terms are: 1) the rate of change of ice bulk density; 2) the rate of removal by sublimation 
or formation by condensation; 3) the rate by which the ice is formed through phase transition of a precursor, or is destroyed by 
production of a successor.

\begin{table}
\begin{center}
\begin{tabular}{||l|l|l||}
\hline
\hline
Symbol & Description & Unit\\
\hline
$\mathbf{A}$ & Matrix, analytic gas diffusion method & $\mathrm{m\,K^{-1/2}}$\\
$A_{n-1,n}$ & Area of cell $n-1$ and $n$ interface &  $\mathrm{m^2}$\\
$A_{\rm cell}$ & Area of cell interface, generic  &  $\mathrm{m^2}$\\
$A_{\rm north}$, $A_{\rm south}$ & Area if northern/southern cell interfaces & $\mathrm{m^2}$\\
$\mathcal{A}$ & Total grain surface area within a volume & $\mathrm{m^2}$\\
$C_{235}$ & Initial molar fraction of $^{235}U$ in uranium & \\
$C_{238}$ & Initial molar fraction of $^{238}U$ in uranium & \\
$C_{\rm K}$ & Current number of potassium & \\
 & atoms in sample & \\
$C_{\rm K}^*$ & Initial--to--current molar ratio of $\mathrm{K}$\\
 & atoms in sample & \\
$D$ & Diffusivity of $\mathrm{CO}$ in $\mathrm{CO_2}$ ice & $\mathrm{m^2\,s^{-1}}$\\
$\delta E$ & Change of cell internal energy & $\mathrm{J}$\\
$F_{i,j}$ & Mass ratio: released guest $j$ & \\
 & to transformed host $i$ & \\
$H_{{\rm e},s}$ & Initial effect of radionuclide $s$ decay & Table~\ref{tab_X0_H} \\  
$H_k$ & Energy release during phase transition & $\mathrm{J\,kg^{-1}}$\\
$L_k$ & Latent heat of sublimation of species $k$ & $\mathrm{J\,kg^{-1}}$\\
$L_{\star}$ & Luminosity factor &\\
$M_{\rm H_2O}$ & Water ice mass per cell & kg\\
$M_k$ & Mass of species $k$ per cell & kg\\
$\mathcal{P}_{\rm e}$ & External pressure & $\mathrm{Pa}$ \\
$\mathcal{P}_{\rm g}$ & Gravitational pressure & $\mathrm{Pa}$ \\
$\mathcal{P}_{\rm m}$ & Compressive strength & $\mathrm{Pa}$ \\
$Q_k$ & Mass loss rate of species $k$ to space & $\mathrm{kg\,m^{-2}\,s^{-1}}$\\
$R_n$ & Radial cell boundary & $\mathrm{m}$\\
$\mathcal{R}$ &  Total radiogenic energy release & $\mathrm{J\,m^{-3}\,s^{-1}}$\\
$S$ & Solar flux at $1\,\mathrm{AU}$ & $\mathrm{J\,m^{-2}\,s^{-1}}$\\
$T$ & Temperature & K\\
$T_{\rm surf}$ & Surface temperature & $\mathrm{K}$\\
$\delta T$, $\Delta T$ & Temperature change & $\mathrm{K}$\\
$U$ & Total $\mathrm{CO}$ mass within a $\mathrm{CO_2}$ grain & $\mathrm{kg}$\\
$V$ & Cell volume & $\mathrm{m^3}$\\
$X_0$ & Initial mass fraction of radioactive &\\
 &  isotope $s$ in chondrite  & Table~\ref{tab_X0_H}\\ 
$Z$ & $\mathrm{CO}$ mass flux in $\mathrm{CO_2}$ & $\mathrm{kg\,m^{-2}\,s^{-1}}$\\
$\mathcal{Z}$, $\mathcal{Z}^{\star}$  & Zenith function & \\
\hline 
\hline
\end{tabular}
\caption{Functions (quantities that generally depend on one or several parameters and/or 
constants and that may vary with time or location) with descriptions and units. Upper case Latin.}
\label{tab1a}
\end{center}
\end{table}

\begin{table}
\begin{center}
\begin{tabular}{||l|l|l||}
\hline
\hline
Symbol & Description & Unit\\
\hline
$a_{\rm f}$ & Final core cell thickness & $\mathrm{m}$\\
$\mathbf{b}$ & Vector, analytic gas diffusion method & $\mathrm{m\,Pa\,K^{-1/2}}$\\
$c$ & Specific heat capacity of mixture & $\mathrm{J\,kg^{-1}\,K^{-1}}$\\
$c_k$ & Specific heat capacity of species $k$ & $\mathrm{J\,kg^{-1}\,K^{-1}}$\\
$c_0$, $c_{\jmath}$, $c_{\rm conv}$ & Initial, intermediate, final common ratio &\\
$d_{n-1,n}$ & Distance between cells $n-1$ and $n$ &  $\mathrm{m}$\\
$f_{\rm V,ref}$ & Volumetric fraction of refractories & \\
$f_{\rm V,vol}$ & Volumetric fraction of volatiles & \\
$g_k$ & Specific heat capacity of vapour & $\mathrm{J\,kg^{-1}\,K^{-1}}$\\
$h$ & Hertz factor & \\
$k_{\rm m}$ & Mixing rate coefficient & $\mathrm{s^{-1}}$\\
$n_k$ & Vapour number density of species $k$ & $\mathrm{molec\,m^{-3}}$\\
$n_{\rm r}$ & Number of radial cells & \\
$\mathbf{p}$ & Pressure array for mantle cells & $\mathrm{Pa}$ \\
$p_0$ & Fractal porosity & \\
$p_k$ & Partial gas pressure of species $k$ & Pa\\
$p_n$ & Pressure in cell $n$ & Pa\\
$p_{{\rm sat},k}$ & Saturation pressure of species $k$ & Pa\\
$q_k$ & Volume vapour mass production & \\ 
 & rate of species $k$ & $\mathrm{kg\,m^{-3}\,s^{-1}}$\\
$q_k'$ & Host mass transformation rate & \\ 
 & of species $k$ & $\mathrm{kg\,m^{-3}\,s^{-1}}$\\
$u$ &  $\mathrm{CO}$ bulk density within $\mathrm{CO_2}$ ice & $\mathrm{kg\,m^{-3}}$\\
$\tilde{u}$ & $\mathrm{CO}$ density within $\mathrm{CO_2}$ ice & $\mathrm{kg\,m^{-3}}$\\
$\tilde{u}_{\rm s}$ & $\mathrm{CO}$ density outside $\mathrm{CO_2}$ ice & $\mathrm{kg\,m^{-3}}$\\
$w$ & Newton--Raphson function (Eq.~\ref{eq:rev02}) &\\ 
\hline 
\hline
\end{tabular}
\caption{Functions (quantities that generally depend on one or several parameters and/or 
constants and that may vary with time or location) with descriptions and units. Lower case Latin.}
\label{tab1a2}
\end{center}
\end{table}

\begin{table}
\begin{center}
\begin{tabular}{||l|l|l||}
\hline
\hline
Symbol & Description & Unit\\
\hline
$\Gamma_{\rm s}$ & Thermal inertia based on $\kappa_{\rm s}$ & $\mathrm{J\,m^{-2}\,K^{-1}\,s^{-1/2}}$\\
 & & $=\mathrm{MKS}$\\
$\kappa$ & Total heat conductivity & $\mathrm{W\,m^{-1}\,K^{-1}}$\\
$\kappa_{\rm s}$ & Solid--state heat conductivity & $\mathrm{W\,m^{-1}\,K^{-1}}$\\
$\mu$ & Refractories--to--$\mathrm{H_2O}$ ice mass ratio & \\
$\rho$ & Local bulk density & $\mathrm{kg\,m^{-3}}$\\
$\rho_{\rm bulk}$ & Global bulk density & $\mathrm{kg\,m^{-3}}$\\
$\rho_{\rm c}$ & Bulk density of condensed $\mathrm{H_2O}$ ice & $\mathrm{kg\,m^{-3}}$\\
$\rho_{\rm h}$ & Bulk density of hexagonal $\mathrm{H_2O}$ ice & $\mathrm{kg\,m^{-3}}$\\
$\rho_{\rm ice}$ & Compact density of all ices & $\mathrm{kg\,m^{-3}}$\\
$\rho_k$ & Bulk density of species $k$ & $\mathrm{kg\,m^{-3}}$\\
$\rho_{\rm met}$ & Density of metals & $\mathrm{kg\,m^{-3}}$\\
$\rho_{\rm min}$,  & Density of minerals & $\mathrm{kg\,m^{-3}}$\\ 
$\rho_{\rm min}'$ & &\\
$\rho_{\rm ref}$,  & Density of refractories & $\mathrm{kg\,m^{-3}}$\\ 
$\rho_{\rm ref}'$ & &\\
$\tilde{\rho}_5$ & Bulk density of $\mathrm{CO}$ trapped in $\mathrm{CO_2}$ & $\mathrm{kg\,m^{-3}}$\\
 & within a large volume compared\\
 & to $\mathrm{CO_2}$ grain size & \\
$\tau_k$ & Transformation rate to/from $k$ & $\mathrm{kg\,m^{-3}\,s^{-1}}$\\
$\Phi_k$ & Radial mass flux rate of species $k$ & $\mathrm{kg\,m^{-2}\,s^{-1}}$\\
$\psi$ & Porosity & \\
$\Psi_k$ & Latitudinal mass flux rate of species $k$ & $\mathrm{kg\,m^{-2}\,s^{-1}}$\\
\hline 
\hline
\end{tabular}
\caption{Functions (quantities that generally depend on one or several parameters and/or 
constants and that may vary with time or location) with descriptions and units. Greek.}
\label{tab1b}
\end{center}
\end{table}

\begin{table}
\begin{center}
\begin{tabular}{||l|l|l||}
\hline
\hline
Symbol & Description & Unit\\
\hline
$a_{\rm t}$ & Targeted core cell thickness & $\mathrm{m}$\\
$\mathcal{B}$ & Biot number & \\
$d_{\rm co}$ & Co--declination & $\mathrm{^{\circ}}$\\
$d_{\rm surf}$ & Surface cell thickness & $\mathrm{m}$\\
$i$ & Index of host species & \\
$j$ & Index of guest species & \\
$\jmath$ & Newton--Rapson iteration index & \\
$k$ & Generic index of species & \\
$l$ & Latitudinal coordinate & rad\\
$\ell$ & Latitude cell index & \\
$m_{\rm g}$ & 67P non--water vapour loss & $\mathrm{kg}$\\
$m_{\rm w}$ & 67P water loss & $\mathrm{kg}$\\
$n$ & Radial cell index & \\ 
$n_{\rm max}$ & Top cell label, analytic & \\
 & gas diffusion method & \\
$n_{\rm s}$ & Number of species & \\
$P$ & Rotation period & s\\
$r$ & Radial coordinate & m\\
$R$ & Body radius & m \\
$s$ & Index of radionuclide & \\
$r_{\rm h}$ & Heliocentric distance & $\mathrm{AU}$\\
$t$ & Time & s\\
$t_{\rm c}$ & Disk clearing time & $\mathrm{yr}$\\
$t_{\rm noon}$ & Time of local midday & s\\
$z$ & Radial coordinate within a grain & $\mathrm{m}$\\
$\theta$ & Polar angle within a grain & $\mathrm{rad}$\\
$\mu_{\rm i}$ & 67P dust--to--all--ice mass ratio & \\
$\mu_{\rm v}$ & Coma dust--to-- water vapour mass ratio & \\
$\varphi$ & Azimuth coordinate within a grain & $\mathrm{rad}$\\
\hline 
\hline
\end{tabular}
\caption{Parameters (quantities that are not depending on other properties of the 
body and that may vary with time) with descriptions and units.}
\label{tab1c}
\end{center}
\end{table}

\begin{table}
\begin{center}
\begin{tabular}{||l|l|l|l||}
\hline
\hline
Symbol & Description & Value & Unit\\
\hline
$A$ & Bolometric  & 0.04 &\\
 & Bond albedo & &\\
$A_{\rm c}$ & Crystallisation pre-- & $1.05\cdot 10^{13}$ & $\mathrm{s^{-1}}$\\
 & exponential factor & &\\
$B_{\rm c}$ & Crystallisation  & $5370$ & $\mathrm{K}$\\
 & activation energy & & \\
$C_3$ & Cubic--to--hexagonal & $10^{-4}$ & $\mathrm{s^{-1}}$\\
 & transformation rate  & &\\
$C_{\rm lum}$ & Disk clearing speed & $921$ & $\mathrm{Myr^{-1}}$\\
$E_{\rm seg}$ & Segregation  & $3813$ & $\mathrm{K}$\\
 & activation energy &  &\\
$E_{{\rm one},s}$ & Energy release of  & Table~\ref{tab_radionuclides} & \\
 & single isotope & &\\
 &  $s$ decay  & &\\
$G$ & Newtonian gravity  & $6.672\cdot 10^{-11}$ & $\mathrm{m^3\,kg^{-1}\,s^{-2}}$ \\
 & constant & &\\
$H_2$ & Crystallisation  & $9\cdot 10^4$ & $\mathrm{J\,kg^{-1}}$\\
 & energy release & &\\
$H_3$ & Cubic--to--hexagonal  & 0 & $\mathrm{J\,kg^{-1}}$\\
 & energy release & & \\
$\mathcal{H}$ & $\mathrm{CO}$ mass transfer  & & $\mathrm{m\,s^{-1}}$\\ 
 & coefficient in $\mathrm{CO_2}$ & & \\
$L_{\rm p}$ & Pore length & $10^{-2}$ & $\mathrm{m}$\\
$L_{\odot}$ & Solar luminosity & $3.828\cdot 10^{26}$ & $\mathrm{W}$ \\
$\Delta M$ & 67P mass loss & $1.05\cdot 10^{10}$ & $\mathrm{kg}$\\
  &  & $\pm 3.4\cdot 10^9$ & $\mathrm{kg}$\\
$\mathcal{M}_{\rm H}$  & $\mathrm{H}$ molar mass &  $1.00794\cdot 10^{-3}$ & $\mathrm{kg\,mole^{-1}}$\\
$\mathcal{M}_4$ & $\mathrm{H_2O}$ molar mass & $1.8015\cdot 10^{-2}$ &  $\mathrm{kg\,mole^{-1}}$\\
$\mathcal{M}_5$ & $\mathrm{CO}$ molar mass & $2.80105\cdot 10^{-2}$ &  $\mathrm{kg\,mole^{-1}}$\\
$\mathcal{M}_6$ & $\mathrm{CO_2}$ molar mass & $4.40099\cdot 10^{-2}$ &  $\mathrm{kg\,mole^{-1}}$\\
$\mathcal{P}_5$ & Initial mass  & 0.524 & \\
 &  fraction of $\mathrm{CO}$  & & \\
 & that is condensed & & \\
$\mathcal{P}_6$ & Initial mass   & 0.717 & \\
 & fraction of $\mathrm{CO_2}$   & & \\
 & that is condensed &  & \\
$\mathcal{P}_5'$ & Initial mass   & 0.158 & \\
 & fraction of $\mathrm{CO}$   & &\\
 & trapped in $\mathrm{CO_2}$ & &\\
$R_{\rm g}$ & Universal gas  & 8.314510 & $ \mathrm{J\,g}$--$\mathrm{mole^{-1}\,K^{-1}}$\\
 & constant & &\\
$S_{\odot}$ & Solar constant & 1367 & $\mathrm{J\,m^{-2}\,s^{-1}}$\\
$T_{\rm sky}$ & Sky temperature & 0 &  $\mathrm{K}$\\
$\mathcal{T}$ & Time of perihelion  & & \\
 & passage & & \\
$V_{\rm imp}$ & Accretion impactor & 15 & $\mathrm{m\,s^{-1}}$ \\
 & velocity  & &\\
\hline 
\hline
\end{tabular}
\caption{Constants common to all simulations considered in this paper, their units and values. Numerical values are only provided for independent 
constants (for dependent ones, see their definitions in the text). Upper case Latin.}
\label{tab2a1}
\end{center}
\end{table}

\begin{table}
\begin{center}
\begin{tabular}{||l|l|l|l||}
\hline
\hline
Symbol & Description & Value & Unit\\
\hline
$a$ & Semi--major axis & 23 & $\mathrm{AU}$\\
$a_{\rm g}$ & Grain surface area & & $\mathrm{m^2}$\\
$e$ & Eccentricity & 0 & \\
$f_{\rm a}$ & Initial mass  &  1 & \\
 & fraction of $\mathrm{H_2O}$  & &\\
 & that is amorphous &  &\\
$f_{{\rm CI},s}$ & Current chondritic  & Table~\ref{tab_radionuclides} & \\
 & mass fraction  & &\\
 &  of element $s$ &  &\\
$f_{{\rm CI},s}'$ & Chondritic mass & Table~\ref{tab_radionuclides} & \\
 & fraction of element $s$  & & \\
&  in dry rock prior & &\\
 &  to aqueous alteration & &\\
$f_{\mathrm{ciso},s}$ & Current molar fraction  & Table~\ref{tab_radionuclides} & \\
& of radionuclide $s$  & &\\
 & in element & &\\
$f_{\rm CO}$ & Fraction of $\mathrm{CO}$ released  & 1 & \\
 & during crystallisation & &\\
$f_{\rm CO}'$ & Fraction of $\mathrm{CO}$ released    & & \\
 & during the cubic--to--  & 0 & \\
 & hexagonal transition & &\\
$f_{\rm CO_2}$ & Fraction of $\mathrm{CO_2}$ released  & 1 & \\
 & during crystallisation & &\\
$f_{\rm CO_2}'$ & Fraction of $\mathrm{CO_2}$ released  & & \\
 &  during the cubic--to-- & 0 & \\
 & hexagonal transition & &\\
$f_{{\rm elem},s}$ & Mass fraction of element  & Table~\ref{tab_radionuclides} & \\
 & in refractories & & \\
$f_{\rm H}$ & Chondritic hydrogen  & 0.021015 & \\
 & mass fraction & &\\
$f_{\mathrm{iso},s}$ & Initial molar fraction & Table~\ref{tab_radionuclides} & \\
 & of radionuclide $s$   & &\\
 & in element & &\\
$f_{\rm water}$ & Chondrite water  & 18.8 & $\mathrm{wt\%}$\\  
 & mass fraction & &\\
$\imath$ & Inclination & 0 & $^{\circ}$ \\
$k_{\rm B}$ & Boltzmann constant & $1.3806$ & \\
 &  & $\cdot 10^{-23}$ & $\mathrm{m^2\,kg\,s^{-2}\,K^{-1}}$\\
$m_{2-4}$ & $\mathrm{H_2O}$ molecule mass & $2.99\cdot 10^{-26}$ & kg\\
$m_5$ & $\mathrm{CO}$ molecule mass & $4.6513\cdot 10^{-26}$ & kg\\ 
$m_6$ & $\mathrm{CO_2}$ molecule mass & $7.3082\cdot 10^{-26}$ & kg\\
$m_{{\rm iso},s}$ & Isotopic mass of  & Table~\ref{tab_radionuclides} & \\
 & radionuclide $s$ & &\\
$m_{\rm u}$ & Atomic mass unit & $1.660538921$ & \\
 & & $\cdot 10^{-27}$ & $\mathrm{kg}$\\
$q$ & Perihelion distance & 23 & AU\\
$r_{\rm g}$ & Grain radius & $10^{-6}$ & $\mathrm{m}$\\
$r_{\rm p}$ & Pore radius & $10^{-3}$ & $\mathrm{m}$\\
$t_{\rm CAI}$ & Ca--Al--rich  & &\\
 & Inclusion & &\\
 &  formation time & &\\
$t_{\mathrm{h},s}$ & Half--life of  & Table~\ref{tab_radionuclides} & \\
 & radionuclide $s$ & &\\
$t_{\rm SS}$ & Solar System lifetime & $4.56\cdot 10^9$ & $\mathrm{yr}$\\
$t_0$ & \texttt{NIMBUS} time zero & &\\
$v_{\rm g}$ & Grain volume & & $\mathrm{m^3}$\\
\hline 
\hline
\end{tabular}
\caption{Constants common to all simulations considered in this paper, their units and values. Lower case Latin.}
\label{tab2a2}
\end{center}
\end{table}

\begin{table}
\begin{center}
\begin{tabular}{||l|l|r|l||}
\hline
\hline
Symbol & Description & Value & Unit\\
\hline
$\alpha$ & Spin axis right ascension & 0 &  $^{\circ}$\\
$\alpha_k$ & $p_{\rm sat}$ coefficient & Table~\ref{tab_sat} & \\
$\beta$ & Spin axis ecliptic latitude & $45$ & $^{\circ}$\\
$\beta_k$ & $p_{\rm sat}$ and $L$ coefficient & Table~\ref{tab_sat} &\\
$\gamma_k$ &  $p_{\rm sat}$ and $L$ coefficient & Table~\ref{tab_sat} &\\
$\delta$ & Spin axis declination & & \\
$\delta_k$ & $p_{\rm sat}$ and $L$ coefficient & Table~\ref{tab_sat} &\\
$\varepsilon$ & Emissivity & 0.9 & \\
$\zeta_4$ & $\mathrm{H_2O}$ degrees of freedom & 6 & \\
$\zeta_5$ & $\mathrm{CO}$ degrees of freedom & 5 & \\
$\zeta_6$ & $\mathrm{CO_2}$ degrees of freedom & 5 & \\
$\lambda$ & Spin axis ecliptic longitude & 0 & $^{\circ}$\\
$\mu_0$ & Initial refractories--to-- & 4 & \\
 & $\mathrm{H_2O}$ mass ratio & &\\
$\nu_5$ & Initial $\mathrm{CO/H_2O}$ ratio  & 0.155 & \\ 
 & by number & &\\
$\nu_6$ & Initial $\mathrm{CO_2/H_2O}$ ratio  & 0.17 & \\ 
 & by number & &\\
$\omega$ & Argument of perihelion & 0 &  $^{\circ}$\\
$\Omega$ & Longitude of the  & 0 &  $^{\circ}$\\
 & ascending node & &\\
$\xi$ & Tortuosity & 1 & \\
$\rho_{\rm Fe}$ & Iron density & $7100$ & $\mathrm{kg\,m^{-3}}$\\
$\rho_{\rm FeS}$ & Troilite density & $4600$ & $\mathrm{kg\,m^{-3}}$\\
$\rho_{\rm Fo}$ & Forsterite density & $3270$ & $\mathrm{kg\,m^{-3}}$\\
$\rho_{\rm imp}$ & Accretion impactor  & 300 & $\mathrm{kg\,m^{-3}}$\\
 & bulk density & &\\
$\rho_{\rm org}$ & Organics density & $1130$ & $\mathrm{kg\,m^{-3}}$\\
$\varrho_1$ & Refractories compact  & 3000 & $\mathrm{kg\,m^{-3}}$\\
 & density & &\\
$\varrho_2$ & Amorphous $\mathrm{H_2O}$ & 500 & $\mathrm{kg\,m^{-3}}$\\
 & compact density  & &\\
$\varrho_3$ & Cubic $\mathrm{H_2O}$  & 917 & $\mathrm{kg\,m^{-3}}$\\
 & compact density & &\\
$\varrho_4$ & Hexagonal $\mathrm{H_2O}$  & 917 & $\mathrm{kg\,m^{-3}}$\\
 & compact density & &\\
$\varrho_5$ & $\mathrm{CO}$ compact density & 1500 & $\mathrm{kg\,m^{-3}}$\\
$\varrho_6$ & $\mathrm{CO_2}$ compact density & 1500 & $\mathrm{kg\,m^{-3}}$\\
$\sigma_{\rm SB}$ & Stefan--Boltzmann  & $5.6705\cdot 10^{-8}$  & $\mathrm{W\,m^{-2}\,K^{-4}}$\\
& constant & &\\
$\tau_s$ & Decay $e$--folding scale & Table~\ref{tab_radionuclides} & \\
 &  of radionuclide $s$ & &\\
\hline 
\hline
\end{tabular}
\caption{Constants common to all simulations considered in this paper, their units and values. Greek.}
\label{tab2b}
\end{center}
\end{table}

The surface boundary condition of Eq.~(\ref{eq:01}),
\begin{equation} \label{eq:04}
\frac{S(t)(1-A)\mathcal{Z}(l,t)}{r_{\rm h}^2}=\sigma_{\rm SB}\varepsilon\left(T_{\rm surf}^4-T_{\rm sky}^4\right)-\kappa\frac{\partial T}{\partial r}\Big|_{r=R},
\end{equation}
envisions an infinitely thin layer responsible for absorbing solar radiation (and possibly cosmic background or Solar Nebula radiation) and emitting thermal radiation into space. 
The net difference between these fluxes equals a conductive energy flux that causes energy to either be added to, or removed from, an underlying layer of finite thickness (in practice the top grid cell). 
The common temperature of these two layers, $T_{\rm surf}$, is not uniquely provided by Eq.~(\ref{eq:04}), but is obtained from the combination of Eqs.~(\ref{eq:01}) and (\ref{eq:04}) 
applied to the finite layer. $T_{\rm surf}$ is therefore determined through the joint effects of all processes captured in Eq.~(\ref{eq:01}), such as sublimation of ices, radiogenic 
heating \emph{et cetera}, in addition to solar heating and radiative cooling. Most thermophysical models of comets explicitly include a sublimation term in the boundary 
condition, thereby mixing fluxes over the boundary surface with one specific volumetric energy sink while ignoring other processes accounted for in those models 
(such as heating by crystallisation or radioactive decay). Equation~(\ref{eq:04}) is a more stringently formulated boundary condition, that reinstates the expression found in older 
literature \citep[e.g.][]{mekleretal90, prialnik92}.

The surface boundary condition of Eq.~(\ref{eq:02}) is given by
\begin{equation} \label{eq:05}
\Phi_i\left(p_i,\,T\right)\Big|_{r=R}=Q_i,
\end{equation}
i.~e., the surface diffusion rate of species $i$ equals the production rate of that species. There is no exchange of solids with the exterior in \texttt{NIMBUS} 
and the flows of energy and mass vanish at $r=0$ (but see Sec.~\ref{sec_model_erosion} about \texttt{NIMBUSD}).

\subsection{Composition and porosity} \label{sec_model_comp}

The initial composition is specified by the mass ratio $\mu_0$ between refractories and water ice, the fraction of the water ice mass 
that is amorphous $f_{\rm a}$ (assuming no initial cubic ice), the abundance $\nu_k$ of species $k$ relative that of water 
by number, and the densities at full compaction $\varrho_k$. The total mass of water ice in a cell of volume $V$ and porosity $\psi$ is given by
\begin{equation} \label{eq:06}
M_{\rm H_2O}=(1-\psi)V\left(\frac{\mu_0}{\varrho_1}+\frac{f_{\rm a}}{\varrho_2}+\frac{1-f_{\rm a}}{\varrho_4}+\sum_{k=5}^{n_{\rm s}}\frac{\nu_km_k}{\varrho_i m_4}\right)^{-1}
\end{equation}
and the initial cell masses of species are therefore given by
\begin{equation} \label{eq:07}
\left\{\begin{array}{l}
\displaystyle M_1(t=0)=M_{\rm H_2O}\mu_0\\
\\
\displaystyle M_2(t=0)=f_{\rm a}M_{\rm H_2O}\\
\\
\displaystyle M_3(t=0)=0\\
\\
\displaystyle M_4(t=0)=(1-f_{\rm a})M_{\rm H_2O}\\
\\
\displaystyle M_k(t=0)=\frac{\nu_km_k}{m_4}M_{\rm H_2O}\hspace{2 cm}{\rm for\,}k\geq 5.\\
\end{array}\right.
\end{equation}
Note that these values change with time because of the physical processes described in Eq.~(\ref{eq:01})--(\ref{eq:03}), and 
as a consequence, changes to $\psi$ are calculated as well. The bulk density of the cell (see Eq.~\ref{eq:01}) is consequently
\begin{equation} \label{eq:08}
\rho(\psi,\,t)=\frac{1}{V}\left(\sum_{i=k}^{n_{\rm s}}M_k(t)+V\psi(t)\sum_{k=4}^{n_{\rm s}}m_kn_k(t)\right).
\end{equation}

Hyper-- and super--volatiles ($k\geq 5$) are initially distributed among different storage modes. In terms of mass fractions, $\mathcal{P}_k$ is 
condensed ice, $\mathcal{P}_k'$ is trapped in $\mathrm{CO_2}$ ice, and the remainder $1-\mathcal{P}_k-\mathcal{P}_k'$ is trapped in 
amorphous water ice. Note that $\mathcal{P}_6'=0$ ($\mathrm{CO_2}$ is not considered trapping itself). If there is $\mathrm{CO}$ trapped in 
amorphous water ice, \texttt{NIMBUS} allows for a fraction $f_{\rm CO}$ to be released during crystallisation (i.e., a fraction $1-f_{\rm CO}$ is 
being transferred to the newly formed cubic water ice). During the cubic--to--hexagonal transition, a fraction $f_{\rm CO}'$ of the original 
amount may be released, so that a fraction $1-f_{\rm CO}-f_{\rm CO}'$ is released once the water starts sublimating. Similar mass fractions 
$f_{\rm CO_2}$ and $f_{\rm CO_2}'$ are defined for $\mathrm{CO_2}$ that do not necessarily have to be equal to their $\mathrm{CO}$ counterparts. 
Water vapour that condenses is assumed to form pure crystalline water ice, even in the presence of $\mathrm{CO}$ and/or $\mathrm{CO_2}$ vapour, and 
regardless of temperature. During sublimation of water ice, clean crystalline water ice and ``dirty'' hexagonal ice (containing $\mathrm{CO}$ and/or $\mathrm{CO_2}$ 
remaining from the crystallisation and cubic--to--hexagonal processes) are assumed to be consumed in equal amounts, as far as possible.

Although the decay of radioactive species in reality has some effect on the overall mass budget, it is ignored in the context 
of calculating, i.~e., the bulk density and heat capacity of the body. The amount of radioactive nuclei are only accounted for 
in terms of the heating resulting from their decay, see Sec.~\ref{sec_model_decay}.

The porosity of the body can either be set to a constant initial value, or a radial porosity distribution $\psi=\psi (r)$ can be calculated 
by solving the hydrostatic equilibrium equation \citep[e.g.][]{henkeetal12} where the sum of the gravitational pressure 
\begin{equation} \label{eq:09}
\mathcal{P}_{\rm g}(r)=-4\pi G\int_r^R\frac{\rho(r')}{(r')^2}\left(\int_0^r\rho(r') (r')^2\,dr'\right)\,dr',
\end{equation}
and a potential external pressure $\mathcal{P}_{\rm e}$ is balanced by the pressure $\mathcal{P}_{\rm m}=\mathcal{P}_{\rm g}+\mathcal{P}_{\rm e}$ 
at which a granular material just resists compression below porosity $\psi$. At $\mathcal{P}_{\rm m}<10^5\,\mathrm{Pa}$ the relation $\mathcal{P}_{\rm m}=\mathcal{P}_{\rm m}(\psi)$ is 
taken as an average of the one measured for silica particles by \citet{guettlereta09} according to the omnidirectional version of their Eq.~(10), 
and the one measured for water ice particles by \citet{loreketal16}. The weight factors are taken as the volumetric fractions of ices, 
\begin{equation} \label{eq:10}
f_{\rm V,vol}=\frac{\sum_{k=2}^{n_{\rm s}}M_k/\varrho_k}{\sum_{k=1}^{n_{\rm s}}M_k/\varrho_k},
\end{equation}
and refractories, $f_{\rm V,ref}=1-f_{\rm V,vol}$. At  $\mathcal{P}_{\rm m}>10^5\,\mathrm{Pa}$ the $\mathcal{P}_{\rm m}=\mathcal{P}_{\rm m}(\psi)$ function is taken as 
that measured by \citet{yasuiarakawa09} for a $\mathrm{ref:vol}=29:71$ mixture at $T=206\,\mathrm{K}$ (see their Fig.~2).

The external pressure is set to the dynamic pressure
\begin{equation} \label{eq:10b}
\mathcal{P}_{\rm e}=\frac{1}{2}\rho_{\rm imp}V_{\rm imp}^2
\end{equation}
in order to mimic the compaction effect during accretion of impactors with density $\rho_{\rm imp}$ colliding at velocity $V_{\rm imp}$ during the formation of the body.

\subsection{Heat capacity and heat conductivity} \label{sec_model_ck}

The specific heat capacities applied in the nominal version of \texttt{NIMBUS} are measured in the laboratory as functions 
of temperature. $c_1(T)$ for refractories is represented by tabulated data for forsterite ($\mathrm{Mg_2SiO_4}$) as given by \citet{robieetal82}. 
All water ice phases are represented by crystalline water ice with
\begin{equation} \label{eq:11}
c_{2-4}(T)=7.49T+90
\end{equation}
as fitted by \citet{klinger81} to data from \citet{giauqueandstout36}. For carbon monoxide ice I apply 
\begin{equation} \label{eq:12}
c_5(T)=35.7T-187
\end{equation}
from \citet{tancredietal94} and for carbon dioxide ice ($c_6$) I apply the tabulated values from \citet{giauqueandegan37} in their Table~IV.

The specific heat capacity of vapour at constant volume is given by
\begin{equation} \label{eq:12b}
g_i=\frac{\zeta_ik_{\rm B}}{2m_i},
\end{equation}
applying the (translational and rotational) degrees of freedom $\zeta_4=6$ and $\zeta_5=\zeta_6=5$ (vibrational modes are not excited at the relevant temperatures).

The total specific heat capacity (see Eq.~\ref{eq:01})  is given by the mass--weighted average,
\begin{equation} \label{eq:13}
c(T)=\frac{\sum_{k=1}^{n_{\rm s}}M_k(t)c_k(T)+V\psi(t)\sum_{k=4}^{n_{\rm s}}m_kn_kg_k}{\sum_{k=1}^{n_{\rm s}}M_k(t)+V\psi(t)\sum_{k=4}^{n_{\rm s}}m_kn_k}.
\end{equation}

The heat conductivities of pure and compact species are taken as follows: refractories have $\kappa_1(T)$ equal to the values tabulated for the H5 ordinary chondrite 
Wellman by \citet{yomogidamatsui83}; amorphous water ice has
\begin{equation} \label{eq:14}
\kappa_2(T)=2.34\cdot 10^{-3}T+2.8\cdot 10^{-2}
\end{equation}
according to \citet{kuhrt84} based on a model by \citet{klinger80}; cubic and hexagonal water ice has
\begin{equation} \label{eq:15}
\kappa_{3-4}(T)=\frac{567}{T}
\end{equation}
according to \citet{klinger80}. Currently lacking better alternatives I here apply $\kappa_5=\kappa_6=\kappa_2$ \citep[see, e.g.,][]{capriaetal96, enzianetal97} for 
similar approximations).

The heat conductivity of a granular medium is strongly dependent on its porosity. The current simulations use the ``generation 2'' model of 
\citet{shoshanyetal02} to calculate the Hertz factor (the factor by which conductivity is reduced from that of compacted material due to porosity)
\begin{equation} \label{eq:16}
h(\psi)=\left(1-\frac{p_0(\psi)}{0.7}\right)^{({4.1p_0(\psi)+0.22})^2}
\end{equation}
where the fractal porosity is given by 
\begin{equation} \label{eq:17}
p_0(\psi)=1-\sqrt{1-\psi}.
\end{equation}
Note that this formula only is valid for $\psi<0.91$. For the current simulations a ceiling was applied so that $\psi>0.7$ would 
default to a Hertz factor $h(0.7)=0.013$. The solid state component of the heat conductivity is defined as
\begin{equation} \label{eq:18}
\kappa_{\rm s}(\psi,T)=h(\psi)\left(\frac{\sum_{i=1}^{n_{\rm s}}\kappa_i(T)M_i}{\sum_{i=1}^{n_{\rm s}}M_i}\right),
\end{equation}
i.~e., a mass--weighted average conductivity is applied for compacted material \citep[e.g.][]{prialnik92}, although there are alternative formulations \citep[e.g.][]{enzianetal97}. 
The conductivity is modified continuously as the temperature, porosity, and composition change during simulations. For reference, 
Table~\ref{tab4} shows the thermal inertia corresponding to the solid--state component, $\Gamma_{\rm s}=\sqrt{\rho(\psi)c(T)\kappa_{\rm s}(\psi,T)}$ for distinct 
values of $\{\psi,T\}$ for a $\mu_0=4$ mixture of dust and crystalline water ice.

\begin{table*}
\begin{center}
\begin{tabular}{||r||r|r|r|r|r|r|r|r|r|r||}
\hline
\hline
 & $T=20\,\mathrm{K}$ & $40\,\mathrm{K}$ &  $60\,\mathrm{K}$ & $80\,\mathrm{K}$ & $100\,\mathrm{K}$ & $120\,\mathrm{K}$ & $140\,\mathrm{K}$ & $160\,\mathrm{K}$ & $180\,\mathrm{K}$ & $200\,\mathrm{K}$\\
\hline
$\psi=0.2$  &  720 &    790 &    990 &   1300 &   1500 &   1700 &   1900 &   2000 &   2100 &   2200\\
$0.4$ &   440 &    490 &    610 &    770 &    950 &   1100 &   1200 &   1200 &   1300 &   1400\\
$0.6$ &   160 &    170 &    220 &    270 &    330 &    370 &    410 &    440 &    460 &    480\\
$0.7$ & 57 &     62 &     78 &   100 &    120 &    140 &    150 &    160 &    170 &    180\\
$0.8$   &   8.3 &      9.1 &     11 &     14 &     18 &    20 &     22 &     23 &     24 &     26\\
\hline 
\hline
\end{tabular}
\caption{Thermal inertia $\Gamma_{\rm s}\,\mathrm{[J\,m^{-2}\,K^{-1}\,s^{-1/2}]}$ as function of porosity $\psi$ and temperature $T$ 
for a $\mu_0=4$ mixture of refractories and crystalline water ice. These values apply for the specific heat capacities, conductivities, porosity--corrections, 
and compact densities of dust and water considered in this paper, and are provided as a guide to better understand the simulations.}
\label{tab4}
\end{center}
\end{table*}

\texttt{NIMBUS} also considers a contribution to heat conduction from radiative transfer \citep[see, e.g.][]{komleandsteiner92, enzianetal97}, 
thus the heat conductivity in Eq.~(\ref{eq:01}) is given by
\begin{equation} \label{eq:18b}
\kappa(\psi,\,T)=\kappa_{\rm s}(\psi,\,T)+4r_{\rm p}\varepsilon\sigma_{\rm SB}T^3,
\end{equation}
where $r_{\rm p}$ is the pore radius. For $r_{\rm p}=10^{-3}\,\mathrm{m}$, the radiative heat transfer component adds $\leq 2.3\,\mathrm{MKS}$ to 
the values in Table~\ref{tab4}, peaking at the highest $T$ and $\psi$. The radiative heat transfer component only becomes important at substantially 
higher temperatures that are relevant for dust mantles.

\subsection{Volume mass production rates and latent heats} \label{sec_model_sublcond}

Condensed $\mathrm{CO}$, $\mathrm{CO_2}$, and $\mathrm{H_2O}$ will experience net sublimation whenever the local 
partial gas pressure $p_k$ is below the saturation pressure $p_{{\rm sat},k}$ at the temperature in question. For a granular 
medium with grain radii $r_{\rm g}$ the volume vapour mass production rate is given by
\begin{equation} \label{eq:19}
q_k=\frac{3(1-\psi)}{r_{\rm g}}\sqrt{\frac{m_k}{2\pi k_{\rm B}T}}\left(p_{{\rm sat},k}-p_k\right)
\end{equation}
\citep{mekleretal90, prialnik92, tancredietal94} for $k\geq 4$. These production rates do not 
depend on the abundances of the species in question, i.e., multi--layer sublimation is assumed \citep[a zeroth--order process in context 
of the Polanyi--Wigner equation, e.g.,][]{suhasariaetal17}. Formally, Eq.~(\ref{eq:19}) breaks down when only 
(sub)monolayers of species--$k$ molecules remain on the grain surfaces of more refractory elements. However, the errors introduced 
by not switching to first--order sublimation at extremely low concentrations have no practical implications on the solutions, neither 
from a mass nor an energy point of view.  If the partial vapour pressure 
\begin{equation} \label{eq:20}
p_k=n_kk_{\rm B}T
\end{equation}
is higher than the saturation pressure then $q_k<0$ and the vapour will experience net condensation. It is assumed that amorphous water ice and cubic water 
ice do not sublimate ($q_2=q_3=0$). The saturation pressure of water vapour is certainly not zero below the temperatures at which crystallisation and 
the cubic--to--hexagonal transformation process are rapid ($\sim 137\,\mathrm{K}$ and $\sim 160\,\mathrm{K}$, respectively, see Sec.~\ref{sec_model_cryst}). 
Therefore, those species in principle do sublimate according to Eq.~(\ref{eq:19}). However, the time scales of significant net sublimation are long when deviations 
from saturation pressure are small, as is the case except at the very surface. Therefore, the errors introduced by these simplifications are small.

The saturation pressures are functions of temperature according to laboratory measurements fitted by analytical functions on the form
\begin{equation} \label{eq:20b}
\log_{10}p_{{\rm sat},k}(T)=\alpha_k+\frac{\beta_k}{T}+\gamma_k\log_{10}T+\delta_kT
\end{equation}
with coefficients in Table~\ref{tab_sat} provided by \citet{huebneretal06} for $\mathrm{H_2O}$ and $\mathrm{CO}$. 
Data from \citet{azregainou05} are used for $\mathrm{CO_2}$ instead of Eq.~(\ref{eq:20b}).\\

\begin{table}
\begin{center}
\begin{tabular}{||l|r|r|r|r||}
\hline
\hline
Species & $\alpha_k$ & $\beta_k$ & $\gamma_k$ & $\delta_k$\\
\hline
$\mathrm{H_2O}$ & 4.07023 & $-2484.986$ & 3.56654 & $-0.00320981$\\
$k=4$ & & & & \\
\hline
$\mathrm{CO}$ & 53.2167 & $-795.104$ & $-22.3452$ & 0.0529476\\
$k=5$ & & & & \\
\hline
$\mathrm{CO_2}$ & 49.2101 & $-2008.01$ & $-16.4542$ & 0.0194151\\
$k=6$ & & & & \\
\hline 
\hline
\end{tabular}
\caption{Coefficients from \protect\citet{huebneretal06} to be used in Eq.~(\ref{eq:20b}) to calculate saturation pressures of $\mathrm{H_2O}$ 
and $\mathrm{CO}$, and in Eq.~(\ref{eq:20c}) to calculate latent heats for $\mathrm{H_2O}$, $\mathrm{CO}$, and $\mathrm{CO_2}$.}
\label{tab_sat}
\end{center}
\end{table}

The temperature--dependent latent heats $L_k$ ($k\ge 4$) are on the form
\begin{equation} \label{eq:20c}
L_k(T)=\Big(-\beta_k\ln(10)+(\gamma_k-1)T+\delta_k\ln(10)T^2\Big)\frac{R_{\rm g}}{10^{-3}\mathcal{M}_k}
\end{equation}
with coefficients in Table~\ref{tab_sat} that are all taken from \citet{huebneretal06}.

\subsection{Crystallisation and the cubic--hexagonal transition} \label{sec_model_cryst}

I here define the functions $q_2'$ and $q_3'$ in Eqs.~(\ref{eq:01})--(\ref{eq:02}), as well as $H_2$, $H_3$, $F_{2,j}$, and $F_{3,j}$ for $j\geq 5$ 
in Eq.~(\ref{eq:01}). See Table~\ref{tab_H_F} for a summary of all $H_i$ and $F_{i,j}$ functions and values. Also $\tau_2$ and $\tau_3$ in 
Eq.~(\ref{eq:03}) are defined.

Amorphous water ice crystallise into cubic water ice at a rate
\begin{equation} \label{eq:23}
q'_2(t)=(M_2(t)/V)A_{\rm c}\exp(-B_{\rm c}/T)
\end{equation}
where $A_{\rm c}=1.05\cdot 10^{13}\,\mathrm{s^{-1}}$ and $B_{\rm c}=5370\,\mathrm{K}$ according to \citet{schmittetal89}. 
The energy release during crystallisation is taken as $H_2=9\cdot 10^4\,\mathrm{J\,kg^{-1}}$ \citep{ghormley68}. Part of this 
energy is being consumed by occluded species upon their release (see \citet{gonzalezetal08} for a discussion on this topic). 

Following \citet{enzianetal97}, Eq.~(\ref{eq:01}) assumes that this energy equals the latent heat during sublimation for the 
trapped species in question, explaining how the fifth term from the left in Eq.~(\ref{eq:01}) is defined. At $T=90\,\mathrm{K}$, where the 
crystallisation time--scale is $\sim 10^5\,\mathrm{yr}$, the applied latent heats are such that transition from an exothermic to an endothermic 
overall process takes place if the amorphous water ice contains $38\%\,\mathrm{CO}$ or $15\%\,\mathrm{CO_2}$ by number and the 
release is complete ($f_{\rm CO}=1$ or $f_{\rm CO_2}=1$). However, crystallisation may still be exothermic at higher concentrations 
of trapped species as long as $f_{\rm CO}<1$ or $f_{\rm CO_2}<1$.

The $F_{i,j}$--values in Table~\ref{tab_H_F} are obtained by calculating the mass ratio of guest to host, and are based on the 
requested abundances $A_k$, with appropriate corrections for the partitioning of a species among hosts ($\mathcal{P}_k$, $\mathcal{P}_k'$) 
as well as the fraction of hosted species that are released during a given transition ($f_{\rm CO}$, $f_{\rm CO_2}$, $f_{\rm CO}'$, and $f_{\rm CO_2}'$). 
Note that $q_i'F_{i,j}$ yields the volume production of a released guest species during the phase transition of its host ($\mathrm{[kg\,m^{-3}\,s^{-1}}]$), 
so that $-q_i'F_{i,j}L_i$ yields the energy consumption with appropriate units ($\mathrm{[J\,m^{-3}\,s^{-1}]}$) in Eq.~(\ref{eq:01}).

\begin{table*}
\begin{center}
\begin{tabular}{||l||l|l||}
\hline
\hline
Host $(i)$ & $H_i\,\mathrm{[J\,kg^{-1}]}$ & $\mathrm{CO}$ guest $F_{i,j=5}$ and $\mathrm{CO_2}$ guest $F_{i,j=6}$\\
\hline
\hline
Amorphous $\mathrm{H_2O}$  & $H_2=9\cdot 10^4\,\mathrm{J\,kg^{-1}}$ & $F_{2,5}=f_{\rm CO}A_5m_5(1-\mathcal{P}_5-\mathcal{P}_5')/f_{\rm a}m_4$\\
 ($i=2$) & & $F_{2,6}=f_{\rm CO_2}A_6m_6(1-\mathcal{P}_6)/f_{\rm a}m_4$\\
\hline
Cubic $\mathrm{H_2O}$  & $H_3=0$ & $F_{3,5}=A_5m_5(1-\mathcal{P}_5-\mathcal{P}_5')(1-f_{\rm CO}-f_{\rm CO}')/f_{\rm a}m_4$\\
 ($i=3$) & & $F_{3,6}=A_6m_6(1-\mathcal{P}_6)(1-f_{\rm CO_2}-f_{\rm CO_2}')/f_{\rm a}m_4$\\
\hline
Hexagonal $\mathrm{H_2O}$  & $H_4=0$ & $F_{4,5}=(1-f_{\rm CO}-f_{\rm CO}')A_5m_5(1-\mathcal{P}_5-\mathcal{P}_5')\rho_{\rm h}/f_{\rm a}m_4(\rho_{\rm c}+\rho_{\rm h})$\\  
($i=4$) & & $F_{4,6}=(1-f_{\rm CO_2}-f_{\rm CO_2}')A_6m_6(1-\mathcal{P}_6)\rho_{\rm h}/f_{\rm a}m_4(\rho_{\rm c}+\rho_{\rm h})$\\
\hline
Carbon monoxide  & $H_5=0$ & $F_{5,5}=0$\\
 ($i=5$) & & $F_{5,6}=0$\\
\hline
Carbon dioxide  & $H_6=0$ & $F_{6,5}=A_5m_5\mathcal{P}_5'/A_6m_6\mathcal{P}_6$\\
 $(i=6)$ & & $F_{6,6}=0$\\
\hline 
\hline
\end{tabular}
\caption{Host/guest terms in the governing equations, see Eqs.~(\ref{eq:01})--(\ref{eq:02}).}
\label{tab_H_F}
\end{center}
\end{table*}

No energy release is assumed to take place when cubic ice transforms to hexagonal ice ($H_3=0$). For simplicity the transformation is 
assumed to proceed at a fixed rate when the temperature exceeds $160\,\mathrm{K}$, thus 
\begin{equation} \label{eq:24}
q_3'=C_3\rho_3,
\end{equation}
assuming $C_3=10^{-4}\,\mathrm{s^{-1}}$, that corresponds to a half--life of cubic ice of $\sim 1.9\,\mathrm{h}$. This process is 
endothermic because of the consumption of latent heat.

Finally, the definitions of $\tau_2$ and $\tau_3$ in Eq.~(\ref{eq:03}) reflects the fact that amorphous water ice crystallises irreversibly at a rate $q_2$, 
while equation for cubic ice has both a source term (formation at $q_2'$) and a sink term (transformation to hexagonal water ice at $q_3'$),
\begin{equation} \label{eq:25}
\left\{\begin{array}{l}
\displaystyle \tau_2=-q_2'\\
\\
\displaystyle \tau_3=q_2'-q_3'.\\
\end{array}\right.
\end{equation}

\subsection{Release of $\mathbf{CO}$ and $\mathbf{CO_2}$ from sublimating $\mathbf{H_2O}$} \label{sec_model_CO_CO2_fromH2O}

Crystalline water consists of two phases that are tracked separately by \texttt{NIMBUS} -- pure condensed water ice, and (potentially) ``dirty'' 
hexagonal water ice. The former may be present from the start and/or forms when water vapour recondenses. It is pure in the sense that 
it does not host foreign species. The latter forms exclusively through the phase transition of cubic water ice, and may contain $\mathrm{CO}$ 
and/or $\mathrm{CO_2}$. The combined sublimation or condensation rate of the two equals $q_4$, and that contribution to the 
energy and mass conservation is fully accounted for by the  fourth terms in Eqs.~(\ref{eq:01}) and (\ref{eq:02}). For that reason, $H_4=0$, 
i.~e., the fifth term in Eq.~(\ref{eq:01}) does not include energy consumption caused by the sublimation of the hexagonal water ice 
host (it has already been included in the fourth term). However, it is necessary to account for the energy consumption of its entrapped guests.

Let $\rho_{\rm c}$ and $\rho_{\rm h}$ be the local bulk densities of condensed and hexagonal water, respectively. Note, that $\rho_4=\rho_{\rm c}+\rho_{\rm h}$. 
The transformation rate (equivalently, sublimation rate) of the hexagonal water ice host equals that of water itself,
\begin{equation} \label{eq:26}
q_4'=q_4,
\end{equation}
because sublimation is here considered a zeroth--order process. The expressions for $F_{4,5}$ and $F_{4,6}$ in Table~\ref{tab_H_F} 
specifies the guest/host mass ratio within hexagonal water ice, but also account for the fact that only a fraction $\rho_{\rm h}/(\rho_{\rm c}+\rho_{\rm h})$ 
of the available water contains trapped species. Therefore, $q_4'F_{4,j}$ provides the mass release ($\mathrm{[kg\,m^{-3}\,s^{-1}]}$) of $\mathrm{CO}$ ($j=5$) 
and $\mathrm{CO_2}$ ($j=6$).

The ice mass conservation equations for $\rho_{\rm c}$ and $\rho_{\rm h}$ read,
\begin{equation} \label{eq:27}
\left\{\begin{array}{l}
\displaystyle \frac{\partial\rho_{\rm c}}{\partial t}=-q_4\frac{\rho_{\rm c}}{\rho_{\rm c}+\rho_{\rm h}}\hspace{0.5cm}\mathrm{if}\,q_4>0\\
\\
\displaystyle \frac{\partial\rho_{\rm c}}{\partial t}=-q_4\hspace{0.5cm}\mathrm{if}\,q_4<0\\
\\
\displaystyle \frac{\partial\rho_{\rm h}}{\partial t}=q_3'-q_4\frac{\rho_{\rm h}}{\rho_{\rm c}+\rho_{\rm h}}\hspace{0.5cm}\mathrm{if}\,q_4>0\\
\\
\displaystyle \frac{\partial\rho_{\rm h}}{\partial t}=q_3'\hspace{0.5cm}\mathrm{if}\,q_4<0\\
\end{array}\right.
\end{equation}
The first two rows in Eq.~(\ref{eq:27}) show that sublimation ($q_4>0$) reduces the amount of condensed water ice at a rate 
that is a fraction $\rho_{\rm c}/(\rho_{\rm c}+\rho_{\rm h})$ of $q_4$, but that condensation ($q_4<0$) allocates the full water 
ice formation to the condensed phase. The two last rows in Eq.~(\ref{eq:27}) show that the amount of hexagonal ice only increases during 
the cubic--to--hexagonal transformation (at rate $q_3'$) because condensation ($q_4<0$) is not creating ``dirty'' hexagonal ice, 
while sublimation ($q_4>0$) proceeds at a rate that is a fraction $\rho_{\rm h}/(\rho_{\rm c}+\rho_{\rm h})$ of $q_4$. Adding rows 1 and 3 
(or 2 and 4), yields
\begin{equation} \label{eq:28}
\frac{\partial\rho_4}{\partial t}=\frac{\partial\rho_{\rm c}}{\partial t}+\frac{\partial\rho_{\rm h}}{\partial t}=-q_4+q_3'.
\end{equation}
Comparison with Eq.~(\ref{eq:03}) shows that Eq.~(\ref{eq:28}) is an identical statement and that
\begin{equation} \label{eq:29}
\tau_4=q_3',
\end{equation}
i.~e., that water ice (the sum of condensed and hexagonal phases) form as a combination of recondensation and cubic--to--hexagonal transformation, and 
is being destroyed through sublimation.

\subsection{Release of $\mathbf{CO}$ from $\mathbf{CO_2}$} \label{sec_model_COfromCO2} 

Because sublimation and condensation of $\mathrm{CO}$ are fully covered by the fourth terms in Eqs.~(\ref{eq:02}) and (\ref{eq:03}), 
and because $\mathrm{CO}$ is not considered a host neither of itself nor of any other species, then $H_5=F_{5,j}=0$. Because 
$\mathrm{CO_2}$ is assumed not to change structure and release energy during segregation ($H_6=0$), and is not considered a host of itself ($F_{6,6}=0$), 
the only remaining problem in the fifth terms of Eq.~(\ref{eq:01}) and (\ref{eq:02}) is to define the segregation rate $q_6'$ and $F_{6,5}$.

In order to derive an expression for $q_6'$, consider a spherical grain of $\mathrm{CO_2}$ ice that contains $\mathrm{CO}$ at 
local mass density $\tilde{u}=\tilde{u}(z,\,\theta,\varphi)$, where $\{z,\,\theta,\,\varphi\}$ are spherical coordinates within the grain. The evolution 
of $\tilde{u}$ with time is governed by Fick's second law,
\begin{equation} \label{eq:30}
\frac{\partial \tilde{u}}{\partial t}=\nabla\cdot \left( D\nabla \tilde{u}\right)
\end{equation}
where $D$ is the diffusivity (here referring to $\mathrm{CO}$ moving through $\mathrm{CO_2}$ ice). Integrating both sides of Eq.~(\ref{eq:30}) over the grain volume, 
assuming a spherical--symmetric distribution of 
$\mathrm{CO}$ within the grain, defining the mass flux
\begin{equation} \label{eq:31}
\mathbf{Z}=-D\frac{\partial \tilde{u}}{\partial z}\hat{z}
\end{equation}
and applying the Gau\ss~theorem, yields
\begin{equation} \label{eq:32}
\begin{array}{c}
\displaystyle \iiint \frac{\partial \tilde{u}}{\partial t}\,z^2\sin\theta\,dz\,d\theta\,d\varphi=\iiint\nabla\cdot \left( D\nabla \tilde{u}\right)\,z^2\sin\theta\,dz\,d\theta\,d\varphi\\
\\
\displaystyle \Rightarrow\frac{\partial u}{\partial t}=\iint (D\nabla \tilde{u})\cdot\hat{z}\,\sin\theta\,d\theta\,d\varphi\,\,\,\Rightarrow\\
\\
\displaystyle \frac{\partial u}{\partial t}=-Za_{\rm g}
\end{array}
\end{equation}
where $u$ is the total $\mathrm{CO}$ mass inside the grain, $Z=|\mathbf{Z}|$, and $a_{\rm g}$ is the grain surface area. 
The mass flux across the grain surface can also be written
\begin{equation} \label{eq:33} 
Z(t)=\mathcal{H}(\tilde{u}(r_{\rm g},t)-\tilde{u}_{\rm s})\approx\mathcal{H}\tilde{u}(r_{\rm g},t)
\end{equation}
where $\mathcal{H}$ is the mass transfer coefficient and $\tilde{u}_{\rm s}$ is the partial $\mathrm{CO}$ density in the gas that surrounds the grain. 
The gas density  $\tilde{u}_{\rm s}$ is here assumed to be negligible in comparison with the concentration $\tilde{u}(r_{\rm g},t)$ of $\mathrm{CO}$ 
molecules near the surface of the grain. If the Biot number is much smaller than unity,
\begin{equation} \label{eq:34}
\mathcal{B}=\frac{\mathcal{H}r_{\rm g}}{D}\ll 1,
\end{equation}
then the mass loss is sufficiently slow to remove internal abundance gradients, so that $\tilde{u}=u/v_{\rm g}$ everywhere, including at the surface ($\tilde{u}(r_{\rm g},t)=u(t)/v_{\rm g}$). 
Combining Eqs.~(\ref{eq:32})--(\ref{eq:34}) therefore yields,
\begin{equation} \label{eq:35}
\frac{\partial u}{\partial t}=-\frac{\mathcal{B}Da_{\rm g}}{v_{\rm g}r_{\rm g}}u.
\end{equation}

Equation~(\ref{eq:35}) can be generalised to a medium of volume $V$ containing of a large number of grains with combined surface area $\mathcal{A}$ 
with a total mass $U$ of trapped $\mathrm{CO}$,
\begin{equation} \label{eq:35b}
\frac{\partial U}{\partial t}=-\frac{\mathcal{B}D\mathcal{A}}{Vr_{\rm g}}U=-\frac{3(1-\psi)\mathcal{B}D}{r_{\rm g}^2}U.
\end{equation}
Denoting the bulk density of trapped $\mathrm{CO}$ within the entire volume by $\tilde{\rho}_5=U/V$, and recognising that $q_6'F_{6,5}=-\partial\tilde{\rho}_5/\partial t$
(the mass of released $\mathrm{CO}$ within unit volume and time), it yields
\begin{equation} \label{eq:36}
q_6'F_{6,5}=\frac{3(1-\psi)\mathcal{B}D}{r_{\rm g}^2}\tilde{\rho}_5.
\end{equation}

In order to proceed, $D$ must be evaluated. Applying Einstein's relationship \citep{cookeetal18},
\begin{equation} \label{eq:37}
D\approx \frac{1}{2}k_{\rm m}r_{\rm g}^2
\end{equation}
and recognising that the mixing rate coefficient $k_{\rm m}$ in the segregation process follows the Arrhenius equation according to 
laboratory experiments \citep[e.g.][]{obergetal09},
\begin{equation} \label{eq:38}
k_{\rm m}\propto\exp(-E_{\rm seg}/T),
\end{equation}
Eq.~(\ref{eq:36}) transforms to
\begin{equation} \label{eq:39}
q_6'F_{6,5}=\frac{3}{2}\tilde{\rho}_5(1-\psi)\Gamma_{\rm B}\exp\left(-\frac{E_{\rm seg}}{T}\right)
\end{equation}
where $\Gamma_{\rm B}$ is the proportionality constant (pre--exponential factor) in Eq.~(\ref{eq:38}) times $\mathcal{B}$, and $E_{\rm seg}$ is the activation energy of segregation. 
Recognising that the mass ratio between the trapped $\mathrm{CO}$ and the $\mathrm{CO_2}$ host is
\begin{equation} \label{eq:40}
F_{6,5}=A_5m_5\mathcal{P}_5'/A_6m_6\mathcal{P}_6=\tilde{\rho}_5/\rho_6,
\end{equation}
the expression for the segregation rate used by \texttt{NIMBUS} consequently is,
\begin{equation} \label{eq:41}
q_6'=\frac{3}{2}\rho_6(1-\psi)\Gamma_{\rm B}\exp\left(-\frac{E_{\rm seg}}{T}\right).
\end{equation}

Evaluating $\Gamma_{\rm B}$ and $E_{\rm seg}$ is a delicate problem. Laboratory studies of the behaviour of $\mathrm{CO}:\mathrm{CO_2}$ mixtures have been 
initiated fairly recently.  \citet{cookeetal18} studied the segregation of $\mathrm{CO}$ out of $\mathrm{CO_2}$ at low temperatures ($T=11$--$23\,\mathrm{K}$) 
and found a low activation energy $E_{\rm seg}=300\pm 40\,\mathrm{K}$, suggesting efficient diffusion along pore surfaces rather than bulk diffusion. However, the conditions 
within icy minor Solar System bodies are different from the interstellar conditions studied by \citet{cookeetal18}. Firstly, the relevant temperatures are higher and 
the porosity of $\mathrm{CO_2}$ ice decreases drastically with increasing temperature \citep{satorreetal09}. The smaller presence of pores at higher temperatures 
may strongly reduce the diffusivity and result in a much higher $E_{\rm seg}$--value. Secondly, the experiments are performed at ultrahigh vacuum in order to 
simulate interstellar conditions. However, the gas pressure in the interior of Solar System bodies may be significant, and the loss rate at a grain surface 
depends on the $\mathrm{CO}$ density difference between the grain and the surroundings (see Eq.~\ref{eq:33}). Although that difference was ignored in the 
derivation above, it could still have the effect of increasing the activation energy. \citet{simonetal19} studied the entrapment efficiency of $\mathrm{CO}$ in $\mathrm{CO_2}$ 
as function of temperature, ice layer thickness, and $\mathrm{CO}$ concentration during deposition. They found that $\mathrm{CO_2}$ entraps $\mathrm{CO}$ more 
efficiently than $\mathrm{H_2O}$ below the $\mathrm{CO_2}$ sublimation temperature. They also studied the desorption rate as function of temperature and 
found that $\mathrm{CO_2}$ released some $\mathrm{CO}$ near $T=40\,\mathrm{K}$ (corresponding to the $\mathrm{CO}$ sublimation temperature and likely representing 
desorption of $\mathrm{CO}$ deposited on the surface of the $\mathrm{CO_2}$ ice rather than authentic segregation). They found that low--level 
emission of $\mathrm{CO}$ took place at $45$--$70\,\mathrm{K}$, and that most $\mathrm{CO}$ was released during $\mathrm{CO_2}$ sublimation that peaked at $80\,\mathrm{K}$. 
It is interesting to compare these results with similar experiments for $\mathrm{CH_4}$ \citep{lunaetal08} and $\mathrm{N_2}$ \citep{satorreetal09} entrapped in $\mathrm{CO_2}$. 
Both species have a first desorption peak near $T=50\,\mathrm{K}$ (corresponding to  $\mathrm{CH_4}$ and $\mathrm{N_2}$ sublimation), a second broad desorption peak at 
$T=80$--$90\,\mathrm{K}$ (attributed to the removal of porosity in the $\mathrm{CO_2}$ host), and a third peak centred at $T=95$--$100\,\mathrm{K}$ during $\mathrm{CO_2}$ 
sublimation. The experiments of \citet{lunaetal08} and \citet{satorreetal09} were performed at a comparably high pressure of $10^{-5}\,\mathrm{Pa}$ 
which makes $\mathrm{CO_2}$ sublimate strongly at a temperature that was $\sim 20\,\mathrm{K}$ higher than in the experiment of \citet{simonetal19} that took place at 
$5\cdot 10^{-7}\,\mathrm{Pa}$. By doing so, \citet{simonetal19} potentially may have superimposed the two $\mathrm{CO}$ release peaks corresponding 
to $\mathrm{CO_2}$ densification and sublimation. It remains to be seen if $\mathrm{CO}$ displays three--peak segregation behaviour as do $\mathrm{CH_4}$ and $\mathrm{N_2}$ in 
segregation experiments at higher pressure. The pressure within Solar System minor icy bodies may become substantially higher than in these experiments. For example, 
the average $\mathrm{CO_2}$ pressure in the inner $20\,\mathrm{km}$ of a Hale--Bopp--sized model body during $\sim 15\,\mathrm{Myr}$ of steady--state in the 
early Solar System is $0.2\,\mathrm{Pa}$ (see Sec.~\ref{sec_results_HB}), while the average temperature is $108\,\mathrm{K}$ (i.~e., at such a pressure the temperature must reach even 
higher, to roughly $120$--$130\,\mathrm{K}$ to cause strong $\mathrm{CO_2}$ sublimation). In order to provide $\{\Gamma_{\rm B},\,E_{\rm seg}\}$--parameters 
relevant for modelling $\mathrm{CO:CO_2}$ segregation within Solar System bodies, such experiments should be performed at substantially higher $\mathrm{CO_2}$ pressures than 
currently available.

In the meanwhile, a simple and admittedly arbitrary method of estimating $\{\Gamma_{\rm B},\,E_{\rm seg}\}$ values was employed to 
enable the test simulations of this paper. Defining $\Delta t$ as the timescale of complete segregation ($q_6'F_{6,5}\Delta t=\tilde{\rho}_5$), 
inserting that into Eq.~(\ref{eq:39}), and logarithmising that equation, yields a linear relation with slope $E_{\rm seg}$, having $T^{-1}$ on the x--axis and $\ln(\Delta t)$ 
on the y--axis. Fixing two pairs of $\{T,\,\Delta t\}$ is therefore sufficient to determine the corresponding $\{\Gamma_{\rm B},\,E_{\rm seg}\}$ values. 
If $\mathrm{CO:CO_2}$ currently is present in icy minor Solar System bodies, such ice must have survived the intense radiation of the protosun. Postulating a clearing--time 
of $t_{\rm c}=1\,\mathrm{Myr}$ (at which the solar luminosity was $1.75$ times the current value, corresponding to a radiative equilibrium temperature of $66\,\mathrm{K}$), 
and demanding a segregation time--scale of $1\,\mathrm{Myr}$ in such conditions provided the first pair $\{T_1,\,\Delta t_1\}=\{66\,\mathrm{K},\,1\,\mathrm{Myr}\}$. Considering that the 
second segregation peak is near $80\,\mathrm{K}$, it is reasonable to attempt defining a time--scale of segregation at that temperature. \citet{simonetal19} find that 
segregation propagates at $0.3$ monolayers per minute, corresponding to full segregation in $\sim 10^5\,\mathrm{s}$ for a $1\,\mathrm{\mu m}$ thick ice layer ($\sim 500$ monolayers). 
However, that time--scale was measured at ultrahigh vacuum, and may very well correspond to sublimation rather than segregation. The interior of a Solar System body would be close 
to saturation, having a $\mathrm{CO_2}$ pressure at $T=80\,\mathrm{K}$ that exceeds that at $T=66\,\mathrm{K}$ by a factor $1.4\cdot 10^4$ according to Eq.~(\ref{eq:20b}) and Table~\ref{tab_sat}. 
Assuming that segregation is slower by a corresponding factor, it yields the second pair $\{T_2,\,\Delta t_2\}=\{80\,\mathrm{K},\,44\,\mathrm{yr}\}$. The corresponding Eq.~(\ref{eq:41}) 
parameters become $\{\Gamma_{\rm B},\,E_{\rm seg}\}=\{8\cdot 10^{11}\,\mathrm{s},\,3813\,\mathrm{K}\}$.

\subsection{Gas diffusion} \label{sec_model_diffusion}

The mass flux of vapour flowing through a porous medium depends on local temperature and pressure gradients, as well as on geometric 
parameters used to calculate the diffusivity (here referring to gas moving through the porous mixture of refractories and ices in the 
body interior). \texttt{NIMBUS} applies the Clausing formula \citep[e.g.][]{skorovandrickman95, davidssonandskorov02b}
\begin{equation} \label{eq:42}
\Phi_k(p_k,\,T,\,\psi)=-\frac{20L_{\rm p}+8L_{\rm p}^2/r_{\rm p}}{20+19L_{\rm p}/r_{\rm p}+3(L_{\rm p}/r_{\rm p})^2}\frac{\psi}{\xi^2}\sqrt{\frac{m_k}{2\pi k_{\rm B}}}\frac{\partial}{\partial r}\left(\frac{p_k}{\sqrt{T}}\right)\hat{r}.
\end{equation}
It is here envisioned that gas flows in cylindrical tubes of radius $r_{\rm p}$ and length $L_{\rm p}$. The formulation in 
Eq.~(\ref{eq:42}) is accurate for any $L_{\rm p}/r_{\rm p}$ ratio \citep{skorovandrickman95}, whereas expressions 
used in some thermophysical models only hold for infinitely long tubes. The tortuosity $\xi$ is the ratio of the path length 
through a series of tubes with different orientation, and the net path passed along the vertical. A similar expression is used 
for the latitudinal mass flux $\Psi_k$, except that the differential of $p_k/\sqrt{T}$ has to be replaced by 
$r^{-1}(\partial/\partial l)\hat{l}$.

\subsection{Radiogenic heating} \label{sec_model_decay}

It is possible to include the effect of radiogenic heating in \texttt{NIMBUS}, with the caveat that it does not allow 
for \emph{melting} of volatiles or dust. The code currently includes the long--lived radionuclides $^{40}\mathrm{K}$, $^{232}\mathrm{Th}$, $^{235}\,\mathrm{U}$, 
and $^{238}\,\mathrm{U}$, as well as the  short--lived radionuclides $^{26}\,\mathrm{Al}$ and $^{60}\,\mathrm{Fe}$. Although the simulations in the current 
work only includes the heating of long--lived radionuclides, the nominal abundances of short--lived radionuclides are here discussed as well for future reference.

\texttt{NIMBUS} uses four different parameters to define the abundances and physical properties of the radionuclides. These are: 1) the fraction of the refractories mass 
consisting of the chemical element in question, $f_{{\rm elem},s}$; 2) the fraction by number of a given element that consisted of the radioactive isotope at 
$t=0$ after CAI, $f_{{\rm iso},s}$; 3) the half--life of the radioactive isotope, $t_{\mathrm{h},s}$; 4) the energy released per atom during decay $E_{{\rm one},s}$. 
The index $s$ identifies the radionuclide as follows: $^{26}\,\mathrm{Al}$ ($s=1$), $^{60}\,\mathrm{Fe}$ ($s=2$), $^{40}\,\mathrm{K}$ ($s=3$), $^{232}\,\mathrm{Th}$ ($s=4$), 
$^{235}\,\mathrm{U}$ ($s=5$), and $^{238}\,\mathrm{U}$ ($s=6$).

The default abundances of the radionuclides are obtained as follows. The current elemental mass fractions in CI carbonaceous chondrites $f_{\rm CI}$ are taken from 
Table~3 in \citet{lodders03}, see Table~\ref{tab_radionuclides}. It is assumed that such material, prior to water melting and aqueous alteration of the CI chondrite parent 
body (that integrated water into the rock phase), consisted of a mass fraction $f_{\rm water}$ of water ice and a fraction $1-f_{\rm water}$ of dry rock. In order to calculate the 
elemental mass fractions  in the dry rock component of the CI carbonaceous chondrite parent body (that is the quantity relevant for \texttt{NIMBUS}) it is recognised that 
the current CI chondrite hydrogen mass fraction is $f_{\rm H}=0.021015$ \citep{lodders03} and it is assumed that all this hydrogen once was locked up in the water ice. 
The water mass fraction is then
\begin{equation} \label{eq:43}
f_{\rm water}=f_{\rm H}\frac{\mathcal{M}_4}{2\mathcal{M}_{\rm H}}\approx 18.8\,\mathrm{wt}\%.
\end{equation}
\citet{garenneetal14} performed thermogravimetric analysis of several carbonaceous chondrites, where the samples are heated and the amount of water loss from the rock phase 
is measured. They find $f_{\rm water}=27.5\%$ for CI chondrite Orgueil, an average $f_{\rm water}=13.6\%$ for three CR chondrites, and an average $f_{\rm water}=14.9\%$ for 16 CM chondrites. 
This indicates that Eq.~(\ref{eq:43}) is a reasonable estimate. The dry--rock mass fractions $f_{\rm CI}'=f_{\rm CI}/(1-f_{\rm water})$ are given in Table~\ref{tab_radionuclides}.

The current molar fractions of the isotope to that of their chemical elements $f_{{\rm ciso},s}$ in Table~\ref{tab_radionuclides} are taken from 
Table~6 in \citet{lodders03}. Note that the entries for $^{26}\mathrm{Al}$ and  $^{60}\mathrm{Fe}$ are zero because those radioisotopes are currently 
extinct. At $t=0$ after CAI those isotopes had fractional abundances by number of $f_{\rm iso, 1}=5.1\cdot 10^{-5}$ for $^{26}\mathrm{Al}$ and $f_{\rm iso,2}=1.6\cdot 10^{-6}$ for $^{60}\mathrm{Fe}$ 
\citep{henkeetal12}. The other $f_{{\rm iso},s}$--values are calculated from $f_{{\rm ciso},s}$ in the following. However, first it is necessary to specify the half--life $t_{\mathrm{h},s}$ that 
is related to the $e$--folding scale $\tau_s$,
\begin{equation} \label{eq:16d}
t_{\mathrm{h},s}=-\ln\left(\frac{1}{2}\right)\tau_s
\end{equation}
that are taken from \citet{henkeetal12} except the $^{26}\mathrm{Al}$ value that is from \citet{norrisetal83}. The $t_{\mathrm{h},s}$ values are given in Table~\ref{tab_radionuclides}.

\begin{table*}
\begin{center}
\begin{tabular}{||l|r|r|r|r|r|r||}
\hline
\hline
Quantity & $^{26}\mathrm{Al}$ & $^{60}\mathrm{Fe}$ & $^{40}\mathrm{K}$ & $^{232}\mathrm{Th}$ & $^{235}\mathrm{U}$ & $^{238}\mathrm{U}$ \\
 & $(s=1)$ & $(s=2)$ & $(s=3)$ & $(s=4)$ & $(s=5)$ & $(s=6)$\\
\hline
$f_{{\rm CI},s}$ &  $8.5\cdot 10^{-3}$ & $1.828\cdot 10^{-1}$ & $5.3\cdot 10^{-4}$ & $3.09\cdot 10^{-8}$ & $8.4\cdot 10^{-9}$ & $8.4\cdot 10^{-9}$\\
$f'_{{\rm CI},s}$ & $1.05\cdot 10^{-2}$ & $2.251\cdot 10^{-1}$ & $6.5\cdot 10^{-4}$ & $3.80\cdot 10^{-8}$ & $1.03\cdot 10^{-8}$ &  $1.03\cdot 10^{-8}$\\
$f_{{\rm ciso},s}$ & 0 & 0 & $1.1672\cdot 10^{-4}$ & 1 & $7.2\cdot 10^{-3}$ & $9.92745\cdot 10^{-1}$\\
$t_{\mathrm{h},s}\,\mathrm{[Myr]}$ & 0.705 & 2.6 & 1200 & 14000 &  690 & 4500\\
$f_{{\rm iso},s}$ & $5.1\cdot 10^{-5}$ &  $1.6\cdot 10^{-6}$ & $1.4682\cdot 10^{-3}$ & 1 &  0.2558 & 0.7442\\
$f_{{\rm elem},s}$ & $1.05\cdot 10^{-2}$ & $2.251\cdot 10^{-1}$ & $6.5344\cdot 10^{-4}$ & $4.8\cdot 10^{-8}$ & $2.78\cdot 10^{-8}$ & $2.78\cdot 10^{-8}$\\
$E_{{\rm one},s}\,\mathrm{[J]}$ & $4.9988\cdot 10^{-13}$  & $4.6367\cdot 10^{-13}$ &  $1.1103\cdot 10^{-13}$ & $6.4728\cdot 10^{-12}$ & $7.1137\cdot 10^{-12}$ &$7.6103\cdot 10^{-12}$\\
$m_{{\rm iso},s}\,\mathrm{[Da]}$ & 25.986891692 & 59.934071683 & 39.963998475 & 232.038055325 & 235.043929918 & 238.050788247\\
\hline 
\hline 
\end{tabular}
\caption{Radionuclide data -- see text for references. $f_{{\rm CI},s}$ is the \emph{current} CI carbonaceous chondrite mass fraction of the chemical element that the radioactive isotope $s$  
belongs to. $f_{{\rm CI}',s}$ is an attempt to correct $f_{{\rm CI},s}$ to the values valid for the dry rock component that presumably was mixed with water ice in the CI carbonaceous chondrite 
parent body prior to aqueous alteration. $f_{{\rm ciso},s}$ is the current isotope number fraction of the element. $t_{\mathrm{h},s}$ is the half--life.  $f_{{\rm iso},s}$ is the isotope number fraction 
of the element at $t=0$ and $f_{{\rm elem},s}$ is the mass fraction of the parent element in dry rock at that time. $E_{{\rm one},s}$ is the energy release during the decay of one isotope. $m_{{\rm iso},s}$ 
is the isotope mass.}
\label{tab_radionuclides}
\end{center}
\end{table*}

Let $C_{\mathrm{K}}$ be the current number of potassium atoms in a sample. Then the current number of $^{40}\mathrm{K}$ isotopes is $C_{\mathrm{K}}f_{\rm ciso,3}$ and the current number of $^{39}\mathrm{K}$ 
and $^{41}\mathrm{K}$ isotopes is $C_{{\rm K}}(1-f_{\rm ciso,3})$. The latter number of isotopes has remained constant over time, while the number of $^{40}\mathrm{K}$ isotopes at $t=0$ 
was $C_{\mathrm{K}}C_{\mathrm{K}}^*=C_{{\mathrm{K}}}f_{\rm ciso,3}\exp(t_{\rm SS}/\tau_{s=3})$ where $t_{\rm SS}=4.56\cdot 10^9\,\mathrm{yr}$ is the age of the Solar System. Then the $^{40}\mathrm{K}$ fraction at $t=0$ is obtained as
\begin{equation} \label{eq:16e}
\left\{\begin{array}{c}
\displaystyle f_{\rm iso,3}=\frac{C_{\mathrm{K}}^*}{1+C_{\mathrm{K}}^*-f_{\rm ciso,3}}\\
\\
\displaystyle C_{{\mathrm K}}^*=f_{\rm ciso,3}\exp(t_{\rm SS}/\tau_{s=3}).\\
\end{array}\right.
\end{equation}
For uranium I proceed in a similar way, and ignore isotopes other than $^{235}\mathrm{U}$ and $^{238}\mathrm{U}$ \citep[the third most common isotope, $^{234}\mathrm{U}$, constitutes merely $0.0058\%$ 
by weight in uranium;][]{goldinetal49}. Defining $C_{235}=f_{\rm ciso,5}\exp(t_{\rm SS}/\tau_{s=5})$ and $C_{238}=f_{\rm ciso,6}\exp(t_{\rm SS}/\tau_{s=6})$ as the original number of 
the isotopes in what currently is $1\,\mathrm{mole}$ of uranium, it is found that the primordial molar fractions were,
\begin{equation} \label{eq:16f}
\left\{\begin{array}{c}
\displaystyle f_{\rm iso,5}=\frac{C_{235}}{C_{235}+C_{238}}\\
\\
\displaystyle f_{\rm iso,6}=\frac{C_{238}}{C_{235}+C_{238}}.\\
\end{array}\right.
\end{equation}

The larger abundances of chemical elements in the past, owing to the presence of radioactive isotopes that now are partially or fully gone, 
are calculated by equating the current mass of stable isotopes in $1\,\mathrm{kg}$ chondritic rock, $f'_{\rm CI}(1-f_{{\rm ciso},s})$ 
with that in the past, $f_{\rm elem}(1-f_{{\rm iso},s})$. This assumes that the total chondritic rock mass has not changed, which is not entirely 
true (e.~g., a $^{26}\mathrm{Al}$ nucleus is somewhat heavier than its $^{26}\mathrm{Mg}$ daughter), but such minor modifications are here ignored. 
It also equates isotope molar and mass fractions, which is acceptable given other uncertainties, because the molecular mass differences among isotopes of the 
same element are small. If so,
\begin{equation} \label{eq:16g}
f_{{\rm elem},s}=\frac{f'_{\rm CI}(1-f_{{\rm ciso},s})}{(1-f_{{\rm iso},s})}.
\end{equation}
This works for aluminium, iron, and potassium ($s\leq 3$), that do have stable isotopes. As seen in Table~\ref{tab_radionuclides}, the differences between $f_{\rm elem}$ and $f'_{\rm CI}$ 
are very small, simply because the radioactive isotope fractions among those elements were always small. For thorium and uranium that do not have stable nuclei, the 
elemental fractions are calculated as
\begin{equation} \label{eq:16h}
\left\{\begin{array}{l}
\displaystyle f_{\rm elem,4}=f'_{\rm CI}\exp(t_{\rm SS}/\tau_{s=4})\\ 
\\
\displaystyle f_{\rm elem,5}=f_{\rm elem,6}=f'_{\rm CI}\left(C_{235}+C_{238}\right).\\
\end{array}\right.
\end{equation}
That explains the $f_{{\rm elem},s}$ values shown in Table~\ref{tab_radionuclides}. Finally, Table~\ref{tab_radionuclides} also lists the energy $E_{{\rm one},s}$ being 
released when one isotope decays, with values taken from \citet{henkeetal12} except for $^{26}\mathrm{Al}$ taken from \citet{castillorogezetal09}.

The literature often provides the applied mass fraction $X_0=f_{\rm elem}f_{\rm iso}$ of the radionuclide in the whole rock mass at $t=0$ \citep[e.g.][]{desanctisetal01, choietal02, guilbertlepoutreetal11}. Those values are calculated for the default \texttt{NIMBUS} composition and presented in Table~\ref{tab_X0_H} to ease a comparison. 
Table~\ref{tab_X0_H} also provides the effect $H_{{\rm e},s}=f_{{\rm elem},s}f_{{\rm iso},s}E_{{\rm kg},s}\exp(-1/\tau_s)/\tau_s$ produced at $t=0$, where $E_{{\rm kg},s}$ is the energy provided if 
$1\,\mathrm{kg}$ worth of the radioactive nuclide would decay at once ($\tau_s$ is here in seconds). This quantity is often reported in the literature as well \citep[e.g.][]{henkeetal12}.

\begin{table}
\begin{center}
\begin{tabular}{||l|r|r||}
\hline
\hline
Radionuclide & $X_0$ & $H_{{\mathrm{e}},s}\,\mathrm{[W\,kg^{-1}]}$\\
\hline
$^{26}\mathrm{Al}$ & $5.36\cdot 10^{-7}$ & $1.926\cdot 10^{-7}$\\
$^{60}\mathrm{Fe}$ & $3.6\cdot 10^{-7}$ & $1.399\cdot 10^{-8}$\\ 
$^{40}\mathrm{K}$ & $9.5934\cdot 10^{-7}$ & $2.826\cdot 10^{-11}$\\ 
$^{232}\mathrm{Th}$ & $4.8\cdot 10^{-8}$ & $1.272\cdot 10^{-12}$\\ 
$^{235}\mathrm{U}$ & $7.11\cdot 10^{-9}$ & $4.111\cdot 10^{-12}$\\
$^{238}\mathrm{U}$ & $2.07\cdot 10^{-8}$ & $1.944\cdot 10^{-12}$\\
\hline 
\hline 
\end{tabular}
\caption{Auxiliary radionuclide data -- see text for references. $X_0$ is the fraction of the rock mass consisting of the isotope at $t=0$. $H$ is the 
effect produced by $1\,\mathrm{kg}$ of rock at $t=0$.}
\label{tab_X0_H}
\end{center}
\end{table}

The default \texttt{NIMBUS} $X_0$ values for $\mathrm{Al}$, $\mathrm{K}$, and $\mathrm{Th}$ are intermediate between those of \citet{desanctisetal01} and \citet{choietal02} 
and closer to the latter. The \citet{guilbertlepoutreetal11} values are somewhat higher still. For $\mathrm{U}$ the \texttt{NIMBUS} $X_0$ are higher than both 
\citet{desanctisetal01} and \citet{choietal02}, but not as high as for \citet{guilbertlepoutreetal11}. The default \texttt{NIMBUS} $H_{{\rm e},s}$-values are 
somewhat higher for $\mathrm{Al}$, $\mathrm{K}$, and $\mathrm{U}$ but somewhat lower for $\mathrm{Fe}$ and $\mathrm{Th}$ compared to \citet{henkeetal12}.

The total radiogenic heating (see Eq.~\ref{eq:01}) is given by
\begin{equation}
\mathcal{R}=\rho_1\sum_{s=1}^6\frac{f_{\mathrm{elem},s}f_{\mathrm{iso},s}E_{\mathrm{one},s}}{\tau_s m_{{\rm iso},s}m_{\rm u}}\exp\left(-\frac{t}{\tau_s}\right)
\end{equation}
where the isotopic masses $m_{{\rm iso},s}$ are given in Table~\ref{tab_radionuclides}.

\subsection{Erosion} \label{sec_model_erosion}

If the near--surface gas pressure becomes sufficiently strong to overcome the local tensile strength, solid material will be 
torn off and ejected into the coma. The nominal version of \texttt{NIMBUS} does not take such erosion into account. However, an 
alternative version called \texttt{NIMBUSD} has been developed (``\texttt{D}'' for ``dust'', although this version of the code allows 
for erosion that ejects ices as well as refractories, that may contribute to an extended source of vapour in the coma). Because 
the different latitudes will evolve at different rates, the initially spherical nucleus will start to deform into a non--spherical shape. 
\texttt{NIMBUSD} keeps the original radial cell boundaries on a normalised axis (with zero at the centre and unity at the surface), 
but re--labels the cell boundary depths in response to the erosion. Physical quantities are interpolated to comply with the new grid 
after each ``erosion campaign'' (that typically is implemented when a certain fraction, e.~g., 10\% or 50\% of the top cell has been eroded). 
Because the erosion progresses at different speeds at each latitude, neighbouring slabs eventually get out of phase with each other. 
For that reason, the latitudinal diffusions of heat and vapour are switched off in \texttt{NIMBUSD}. Heavily eroding objects (e.~g., comets 
passing close to the Sun) are typically modelled on time--scales that are sufficiently short (a single parabolic passage, or a 
handful of apparitions for a short--period comet), that latitudinal heat and gas flows are negligible. \texttt{NIMBUSD} allows the 
user to define types of erosion rates, including: 1) observed rates stored on file as function of heliocentric distance; 2) a certain 
amount of solids is eroded in some proportion to the current outgassing rate of vapours; 3) a certain amount of solids is eroded in some proportion to 
the peak sub--surface net force caused by the gas pressure.

\subsection{Spatial grid, temporal resolution, orbit, and spin} \label{sec_model_grid}

The volume of the body is divided into grid cells. The radial distribution of grid cell boundaries is generated with various methods depending 
on the application, including geometric progression that allows for high spatial resolution near the surface, and equi--distant or equal--volume cells 
when coarser resolution is desirable. The latitudinal distribution of grid cell boundaries is uniform in $l$ or $\cos l$ depending on application. Once all 
cell boundaries have been set up, all cell volumes and the surface areas of the four walls associated with each cell are calculated analytically.

As an example, the grid--generation is here described for the geometric progression (radially) and equal--angle (latitudinally) case. 
The properties of the radial discretisation is determined by specifying the body radius $R$, the thickness of the outermost cell $d_{\rm surf}$ and 
a \emph{targeted} innermost cell thickness $a_{\rm t}$. The number of cells $n_{\rm r}$ are calculated as
\begin{equation} \label{eq:rev01}
\left\{\begin{array}{l}
\displaystyle c_{0}=\frac{R-a_{\rm t}}{R-d_{\rm surf}}\\
\\
\displaystyle n_{\rm r}={\rm round}\left(1+\frac{\log(d_{\rm surf}/a_{\rm t})}{\log(c_0)}\right)\\
\end{array}\right.\\
\end{equation}
where $c_0$ is an initial guess of the common ratio. The Newton--Raphson method is then used in order to 
find a more exact value $c_{\rm conv}$ for the common ratio, and the final inner most cell thickness $a_{\rm f}$,
\begin{equation} \label{eq:rev02}
\left\{\begin{array}{l}
\displaystyle w(c_{\jmath})=c_{\jmath}^{1-n_{\rm r}}+\left(\frac{R}{d_{\rm surf}}-1\right)c_{\jmath}-\frac{R}{d_{\rm surf}}\\
\\
\displaystyle w'(c_{\jmath})=(1-n_{\rm r})c_{\jmath}^{-n_{\rm r}}+\frac{R}{d_{\rm surf}}-1\\
\\
\displaystyle c_{\jmath+1}=c_{\jmath}-\frac{w(c_{\jmath})}{w'(c_{\jmath})}\\
\end{array}\right.\\
\end{equation}
which is iterated until convergence is reached, upon which $a_{\rm f}=d_{\rm surf}/c_{\rm conv}^{n_{\rm r}-1}$ is evaluated. The locations of 
the outer boundaries of the cells are  obtained as
\begin{equation} \label{eq:rev03}
\left\{\begin{array}{l}
\displaystyle R_1=a_{\rm f}\\
\\
\displaystyle R_n=R_{n-1}+a_{\rm f}c_{\rm conv}^{n-1},\,\,\,\,\,\,n=2,...,n_{\rm r}\\
\end{array}\right.
\end{equation}
For example, $R=1000\,\mathrm{m}$, $d_{\rm surf}=0.1\,\mathrm{m}$, and $a_{\rm t}=10\,\mathrm{m}$ yields $n_{\rm r}=464$, $a_{\rm f}=9.99501\,\mathrm{m}$ and $c_{\rm conv}=0.990104$.

If the desired number of latitudinal slabs is $n_{\rm l}$, the equal--angle variant trivially has northern boundaries at $l_{\ell+1}=l_{\ell}+\Delta l$ with 
$l_1=0$ and $\Delta l=\pi/n_{\rm l}$, in terms of the angular distance from the north pole. A cell with label $n$ radially and $\ell$ latitudinally has 
lower and upper surface areas given by
\begin{equation} \label{eq:rev04}
\left\{\begin{array}{l}
\displaystyle A_{n-1,n}=2\pi R_{n-1}^2\left(\cos l_{\ell-1}-\cos l_{\ell}\right)\\
\\
\displaystyle A_{n,n+1}=2\pi R_n^2\left(\cos l_{\ell-1} -\cos l_{\ell}\right),\\
\end{array}\right.\\
\end{equation}
with $R_0=0$ and $l_0=0$. Because \texttt{NIMBUS} is not geometrically three--dimensional, an arbitrary longitudinal width of a cell can be assigned. 
Here, that width is the full $2\pi$, which means that the cell shape resembles a torus. The northern and southern surface areas are given by
\begin{equation} \label{eq:rev05}
\left\{\begin{array}{l}
\displaystyle A_{{\rm north}}=\pi\sin l_{\ell-1}\left(R_n^2-R_{n-1}^2\right)\\
\\
\displaystyle A_{{\rm south}}=\pi\sin l_{\ell}\left(R_n^2-R_{n-1}^2\right).\\
\end{array}\right.\\
\end{equation}
The volume of the cell is given by
\begin{equation} \label{eq:rev06}
V=\frac{2\pi}{3}\left(\cos l_{\ell-1}-\cos l_{\ell}\right)\left(R_n^3-R_{n-1}^3\right).
\end{equation}

As described in more detail in Sec.~\ref{sec_model_implementation}, \texttt{NIMBUS} initially assigns species masses and internal energies 
to each cell, and then essentially operates by injecting new energy due to radioactivity and absorption of solar energy, and by moving mass and energy between 
cells (including escape of mass and radiative energy loss across the outer surface). Because the body is not modelled in three dimensions, special 
conditions prevail in cells that contain the polar axis (including all core cells). A \texttt{NIMBUS} core cell (with $1<\ell<n_{\rm l}$) only trades energy and mass with the core cells just north and 
south of it, and with the cell just exterior to it. All these cells share the same local hour. A real core would be simultaneously influenced by exchange with exterior cells 
having the full variety of local hours. However, as long as the time--scale for conducting energy from the surface to the core is much longer than the rotation period, 
the error in the core temperature will be negligible. Cells with $1<n<n_{\rm r}$ and $\ell=1$ trade energy and mass with cells immediately under and above, as well as with the cell 
just south of it (or, if $\ell=n_{\rm l}$, just north of it). Again, all these cells have the same local hour, but on a real body they have physical proximity to regions with very different 
local hours. In such regions, the quality of \texttt{NIMBUS} calculations would only be compromised if longitudinal lateral flows of energy and mass cannot be ignored. For these reasons, 
extreme slow--rotators should not be considered with \texttt{NIMBUS}, because they require a full three--dimensional treatment.

\texttt{NIMBUS} uses a dynamical time step that adjusts to the prevailing conditions. The dynamical time steps are controlled by forcing the 
change of internal energy at each step to fall within a certain percentile range, by making sure that the changes in gas partial pressures  
remain within certain limits, and that net vapour mass removal is not so large that saturation conditions cannot be re--established (because of 
insufficient amount of ice).

The orbit of the body is described by the standard elements: semi--major axis $a$, eccentricity $e$, inclination $\imath$, argument of perihelion $\omega$, 
the longitude of the ascending node $\Omega$, and the time of the perihelion passage $\mathcal{T}$ (in case $e=1$, then $a$ is replaced by the perihelion distance $q$). 
The orientation of the rotational axis of the body is given by the right ascension $\alpha$ and declination $\delta$ in the equatorial system, or by the longitude $\lambda$ and latitude $\beta$ in the ecliptic system. 
All these parameters can be considered functions of time in order to account for changes to the orbit and spin state during long--term simulations.

\subsection{Illumination conditions} \label{sec_model_illum}

The solar flux at $r_{\rm h}=1\,\mathrm{AU}$ equals the solar constant $S(t)=S_{\odot}$ when modelling the contemporary 
Solar System. However, when modelling the early Solar System \texttt{NIMBUS} accounts for changes of the protosolar luminosity 
according to the $1\,\mathrm{M_{\odot}}$ Hayashi-- and Henyey--tracks presented by \citet{pallastahler93}, $S(t)=L_{\star}(t)S_{\odot}$,  
where the luminosity factor $L_{\star}$ is a dimensionless number that should be multiplied to the current solar luminosity $L_{\odot}$ to obtain the luminosity $L_{\star}L_{\odot}$ at a given 
moment. Specifically, $L_{\star}=6.4$ on the ``birth--line'', that marks the end of main accretion and the start of the Solar Nebula. This is considered  $t_0$ or ``time zero'' in \texttt{NIMBUS}. 
It is likely that the formation of Calcium--Aluminium--rich Inclusions (CAI) at $t_{\rm CAI}$, generally considered the first Solar System solids that define the age of the 
Solar System \citep{macphersonetal95}, took place during the protostellar Class~0 stage \citep{tscharnuteretal09} because of its unique capacity 
of generating temperatures sufficiently high for CAI formation. If so, there is a time difference $t_0-t_{\rm CAI}>0$ of order 
$\sim 10^5\,\mathrm{yr}$ \citep{andremontmerle94}.  For simplicity, time is here measured with respect to $t_0$ instead of $t_{\rm CAI}$.

The luminosity factor falls to $L_{\star}=3.7$ in the first $3\cdot 10^5\,\mathrm{yr}$ and to $L_{\star}=1.7$ in the first $1\,\mathrm{Myr}$. 
At the end of the Solar Nebula lifetime (taken as $3\,\mathrm{Myr}$ based on the mean lifetime of gas disks in foreign solar--mass systems \citep{zuckermanetal95, haischetal01, siciliaaguilaretal06} 
the luminosity factor falls to $L_{\star}\approx 1$. Therefore, the time period during which giant planet 
formation and significant planetesimal growth in the outer Solar System takes place is characterised by substantially higher luminosities than 
in the current Solar System. The luminosity factor continues to drop and reaches a minimum of $L_{\star}\approx 0.55$ at $10\,\mathrm{Myr}$. 
At that point, marking the transition from the Hayashi--track to the Henyey--track, the luminosity factor starts to increase anew, reaching $L_{\star}\approx 0.87$ at 
the onset of hydrogen fusion at $30\,\mathrm{Myr}$. It is assumed that the luminosity factor then increases linearly to the current value in the following $4.56\,\mathrm{Gyr}$.

Prior to substantial dust agglomeration and planetesimal formation the disk midplane is likely to be shielded by the opaque disk. Thus, an option exists to reduce the
 illumination to a low level at $t_0$ and increase it exponentially to the unshielded value at a disk clearing time $t_{\rm c}$,
\begin{equation} \label{eq:05a}
S(t)=S_{\odot}L_{\star}(t)\min\left\{1,\,\exp\left(-C_{\rm lum}(t_{\rm c}-t)\right)\right\}
\end{equation}
where the coefficient $C_{\rm lum}$ determines the rate of change. If planetesimal formation is as rapid as suggested by models of the gravitational collapse of pebble 
swarms \citep{nesvornyetal10} formed in streaming instabilities \citep{youdingoodman05, johansenetal07} the 
clearing time could be as short as a few times $10^4\,\mathrm{yr}$. However, if the metallicity needs to increase to a certain value 
(through gas removal from the disk) in order for streaming instabilities to become efficient  \citep{johansenetal09}, $t_{\rm c}$ could be 
substantially larger. Slower planetesimal growth and a longer clearing time are also predicted in other planetesimal formation scenarios 
\citep{weidenschilling97, davidssonetal16}.

\texttt{NIMBUS} can operate with time--resolved rotation, where the cosine of the solar zenith angle is 
\begin{equation} \label{eq:05a2}
\mathcal{Z}(l,t)=\max\{0,\,\mathcal{Z}^{\star}(l,t)\}, 
\end{equation}
where
\begin{equation} \label{eq:05b}
\mathcal{Z}^{\star}(l,t)=\cos d_{\rm co}\cos l+\sin d_{\rm co}\sin l\cos\left(\frac{2\pi}{P}(t-t_{\rm noon})\right).
\end{equation}
Alternatively,  \texttt{NIMBUS} operates in fast--rotator mode were insolation is rotationally averaged for each latitude, 
\begin{equation} \label{eq:05c}
\mathcal{Z}(l,t)=\frac{1}{P}\int_0^P\max\{0,\,\mathcal{Z}^{\star}(l,t)\}\,dt.
\end{equation}
Equation~(\ref{eq:05a2}) is suitable for studying the detailed thermophysics of bodies on time--scales corresponding to a few orbits, 
while Eq.~(\ref{eq:05c}) enables simulations extending over $\mathrm{kyr}$--$\mathrm{Myr}$ timescales. Note that the co--declination $d_{\rm co}$ 
(the angle between the positive spin vector and the direction to the Sun) is a function of time because of orbital motion and/or spin orientation changes. 
Note that $S(t)$, given by $S_{\odot}$ or Eq.~(\ref{eq:05a}), and that $\mathcal{Z}(l,t)$, given by Eq.~(\ref{eq:05a2}) or Eq.~(\ref{eq:05c}), enter into  
Eq.~(\ref{eq:04}). The same geometrical grid can be applied regardless if Eq.~(\ref{eq:05a2}) or Eq.~(\ref{eq:05c}) is considered. Both provide 
a certain radiative flux at any given moment. The only difference is that Eq.~(\ref{eq:05a2}) contains a (potentially strong) diurnal variation, while Eq.~(\ref{eq:05c}) 
provides a constant flux (if integrated over an rotational period the total fluxes are the same).

\subsection{Implementation} \label{sec_model_implementation}

The \texttt{NIMBUS} numerical scheme is explicit, i.e., the properties at a current time step are calculated based on knowledge 
of the properties at the previous time step. Cells are assumed to be isothermal, i.~e., a single temperature is applied to the entire 
interior of a cell. It means that \texttt{NIMBUS} does not rely on temperature specification at grid points (cell interception points) 
and it does not attempt to interpolate temperatures between such grid points. Each grid cell carries a certain amount of mass for each 
$N_{\rm s}$ species (solids and vapours are stored separately) and a certain internal energy, and those masses and energies are updated 
over time by considering local sources and sinks as well as exchange with neighbouring cells through diffusion and transport processes. 
Temperatures are calculated from the internal energies, considering the local heat capacity, i.~e., if the energy change is $\delta E$, 
and the prevailing temperature is $T$, the temperature change $\delta T$ is obtained by solving the equation
\begin{equation} \label{eq:44}
\delta E=\sum_{k=1}^{N_{\rm s}}\left(M_k+n_km_kV\right)\int_T^{T+\delta T}c(T')\,dT'.
\end{equation}
Equation~(\ref{eq:44}) can be written as a simple function that can be solved analytically for $\delta T$ if approximating $c(T)$ by a 
locally linear function, which is an acceptable approximation because $\delta T$ is small compared to temperature ranges over 
which $c(T)$ changes significantly and non--linearly. Exchanges of masses and energies between cells are calculated based on 
the temperature and pressure gradients prevailing between neighbouring pairs of cells. For example, solid--state heat conduction is 
calculated by applying the Fourier law: 1) using the temperature difference between the cells to obtain $\Delta T$; 2) using the distance 
between the cell centres to obtain $d$; 3) using the average heat conductivity values for the two cells to obtain $\langle\kappa\rangle$; 
4) evaluating the energy exchange between the cells during a time step $\Delta t$ as $\langle\kappa\rangle\Delta tA_{\rm cell}\Delta T/d$, 
where $A_{\rm cell}$ is the surface area that the two cells have in common, through which the heat transfer takes place. The exact amount of 
energy removed from one cell is added to the other, leading to global energy conservation. The governing equations 
are fulfilled by modelling the underlying physical processes that give rise to these equations, rather than subjecting the equations themselves and 
their terms to mathematical manipulation (as would have been done in, e.~g., the finite element method).

The code first evaluates all changes to masses of solids and vapour consistent with the current time step, using the latest known set of temperatures and pressures. 
The code moves on to evaluate energy changes only if the mass changes are consistent with the following requirements: 1) no cell should lose all its $k$ vapour 
when diffusion net losses exceed gains of $k$ vapour from sublimation, segregation, or crystallisation; 2) local changes in gas pressure exceeds 
a certain user--defined threshold. If the mass changes are unacceptable, the time step is reduced and the evaluation starts from the beginning. 
Reducing the time step primarily has the effect of not allowing net mass loss (at a given rate) to proceed for so long that a cell runs out of gas. If the mass 
changes are approved, the energy changes are evaluated and compared with user--defined criteria. If the percentile changes are 
above a certain maximum threshold, the time step is reduced and the entire process starts from the beginning. Repeating the mass--change evaluation for 
the new smaller time--step is necessary, because many mass--changing mechanisms (e.~.g, net sublimation) lead to specific energy changes that need to be 
properly evaluated. If both the mass and energy changes have been approved, those changes are implemented and the code advances to the next time step. At this 
point, there is also a possibility to increase the time step, if the largest percentile energy change has fallen below a minimum threshold.

The amount of vapour in a grid cell is typically extremely small compared to the amount of the frozen volatile (often on the ppm--ppb level). 
The net gas diffusion fluxes are often large (particularly near the surface where temperature and pressure gradients are steep). Net mass loss at such high rates 
(if considered constant) could remove all available cell vapour in a very short time. Imposing the requirement that only a small percentage of the cell vapour mass can 
be removed before re--evaluation of the local conditions, would lead to extremely small time steps. One would find that the time--scale for re--establishing 
saturation conditions in the cell due to ice sublimation typically is even shorter, by orders of magnitude. An excessive amount of computing time would 
be spent on calculating small deviations from saturation pressure caused by gas diffusion, followed by guaranteed and practically instantaneous 
re--establishment of saturation conditions through ice sublimation (when $p_k<p_{{\rm sat},k}$) or vapour condensation (when $p_k>p_{{\rm sat},k}$). 
\texttt{NIMBUS} therefore assumes that cells that do contain ice of species $k$ have vapour of species $k$ at saturation pressure.

The magnitude of the mass flux rate of vapour $k$ between neighbouring cell pairs that both contain condensed ice $k$, is 
determined by the difference in saturation pressure between the cells according to Eq.~(\ref{eq:42}). As long as the two cells do not have the 
same temperature, the saturation pressure difference is non--zero. The amount of mass that a given cell loses or gains across a given boundary 
during the duration of the time--step is determined. This includes both radial and latitudinal flows. That mass is then subtracted from the losing cell 
and added to the gaining cell. This is done for all cells and all their mutual boundaries. The simple procedure of subtracting a given amount of mass from 
one cell and adding it to another guarantees mass conservation. At that point, most cells will no longer have vapour densities corresponding to saturation. 
The volume mass production rate (Eq.~\ref{eq:19}) is evaluated for the difference between saturation pressure and actual current pressure, whatever that 
difference might be. That provides a time--scale for re--establishment of saturation conditions. If this time--scale is shorter than the duration of the 
time--step, re--establishment of saturation condition is considered guaranteed. Re--introduction of saturation conditions in an under--dense cell 
means that the an amount of vapour is added to the gas phase (to reach saturation), and the same amount is subtracted from the ice phase. For an over--dense cell, the excess 
vapour is removed and the amount of ice increases accordingly. Again, the simple procedure of shifting specific masses between accounts guarantees mass conservation.

That allows for the usage of larger time steps where the total vapour mass change in a cell by outflow or inflow are calculated (typically being much larger than the current content of vapour), 
and where re--establishment of saturation conditions are made through a one--time larger net sublimation or condensation of ice. Such assumptions of local saturation 
conditions are common in thermophysical codes \citep[e.g.][]{espinasseetal93, oroseietal95, capriaetal96, desanctisetal99}. 
Doing so, there is only a need to use small time steps when: 1) a cell is about to run out of ice (making sure the demands for vapour does not exceed the availability of ice); 
2) when it is so cold (i.~e., the sublimation is so slow) that the ice is not capable of providing the requested amount of vapour in the time available; 3) or, in case 
condensed ice is lacking, injection of vapour from segregation and crystallisation are not sufficiently fast to prevent full net vapour removal by diffusion.

Saturation conditions for vapour $k$ are generally not prevailing in regions where ice $k$ is not present. That includes the dust mantle 
that is devoid of all ices, and regions above the $\mathrm{CO_2}$ and $\mathrm{CO}$ sublimation fronts. In order to calculate accurate vapour pressures in such 
regions, without necessarily having to resort to the time--consuming short time steps described above, \texttt{NIMBUS} calculate those pressures analytically 
in the radial direction according to the method described in the following. The lateral flows have substantially smaller temperature and pressure gradients, 
which means that those flows can be calculated with the nominal numerical procedure (without requiring too small time steps). That is to say, the latitudinal 
flows are calculated numerically, and the radial flows above the sublimation front are calculated analytically.

Consider the uppermost cell that contains plenty of ice $k$ (constituting the sublimation front of that element), ignoring cells closer to the surface that might 
contain very small amounts of ice $k$ deposited by temporary recondensation. Label the cells $n=[0,\,...\,n_{\rm max}]$ with $n=0$ being the previously 
mentioned icy cell and $n=n_{\rm max}$ being the surface cell. The method is applied to all latitudes $l$ and species $k\geq 4$ but the indices $\{l,\,k\}$ 
are not written explicitly below for brevity. The cell temperatures $T_n$ are considered known (those have just been updated during the previous time step), 
but the gas pressures $p_n$ for $n\geq 1$ (and the corresponding cell vapour masses $p_n\psi_nV_n/k_{\rm B}T_n$) are unknown. Cell $n=0$ is assumed to have local 
saturation pressure, $p_{n=0}=p_{{\rm sat},k}(T_{n=0})$. Additionally, a ``ghost cell'' in the coma just above the body surface is assumed to have $p_{n=n_{\rm max}+1}=0$. 
Steady--state is required, which means that the cell vapour masses do not change during the time step, i.~e., that the inflow of mass to a given cell is exactly balanced by the outflow. 
However, the flow rate changes from one time--step to the next. Using Eq.~(\ref{eq:42}), that mass transfer balance can be stated as
\begin{equation} \label{eq:45}
\begin{array}{l}
\displaystyle\left(\frac{p_{n-1}}{\sqrt{T_{n-1}}}-\frac{p_n}{\sqrt{T_n}}\right)\langle\psi\rangle_{n-1,n}\frac{A_{n-1,n}}{d_{n-1,n}}=\\
\\
\displaystyle \left(\frac{p_n}{\sqrt{T_n}}-\frac{p_{n+1}}{\sqrt{T_{n+1}}}\right)\langle\psi\rangle_{n,n+1}\frac{A_{n,n+1}}{d_{n,n+1}}\\
\end{array}
\end{equation}
where $A_{n-1,n}$ and $d_{n-1,n}$ are the common cell wall area and distance between cells, respectively, with the cell pairs in question specified by the sub--scripts. 
The average porosity, e.~g., $\langle\psi\rangle_{n-1,n}=(\psi_{n-1}+\psi_n)/2$, is applied for each cell pair, and because $\{L_{\rm p},\,r_{\rm p},\,\xi\}$ are assumed identical for all cells, 
those factors cancel on both sides. Equation~(\ref{eq:46}) can be re--arranged in three different ways,

\begin{equation} \label{eq:46}
\left\{\begin{array}{c}
\displaystyle -\left(\frac{\langle\psi\rangle_{n-1,n}}{\sqrt{T_n}}\frac{A_{n-1,n}}{d_{n-1,n}}+\frac{\langle\psi\rangle_{n,n+1}}{\sqrt{T_n}}\frac{A_{n,n+1}}{d_{n,n+1}}\right)p_n+\frac{\langle\psi\rangle_{n,n+1}}{\sqrt{T_{n+1}}}\frac{A_{n,n+1}}{d_{n,n+1}}p_{n+1}=\\
\\
\displaystyle -\frac{\langle\psi\rangle_{n-1,n}}{\sqrt{T_{n-1}}}\frac{A_{n-1,n}}{d_{n-1,n}}p_{n-1}\\
\\
\displaystyle \frac{\langle\psi\rangle_{n-1,n}}{\sqrt{T_{n-1}}}\frac{A_{n-1,n}}{d_{n-1,n}}p_{n-1}-\left(\frac{\langle\psi\rangle_{n-1,n}}{\sqrt{T_n}}\frac{A_{n-1,n}}{d_{n-1,n}}+\frac{\langle\psi\rangle_{n,n+1}}{\sqrt{T_n}}\frac{A_{n,n+1}}{d_{n,n+1}}\right)p_n+\\
\\
\displaystyle \frac{\langle\psi\rangle_{n,n+1}}{\sqrt{T_{n+1}}}\frac{A_{n,n+1}}{d_{n,n+1}}p_{n+1}=0\\
\\
\displaystyle \frac{\langle\psi\rangle_{n-1,n}}{\sqrt{T_{n-1}}}\frac{A_{n-1,n}}{d_{n-1,n}}p_{n-1}-\left(\frac{\langle\psi\rangle_{n-1,n}}{\sqrt{T_n}}\frac{A_{n-1,n}}{d_{n-1,n}}+\frac{\langle\psi\rangle_{n,n+1}}{\sqrt{T_n}}\frac{A_{n,n+1}}{d_{n,n+1}}\right)p_n=\\
\\
\displaystyle -\frac{\langle\psi\rangle_{n,n+1}}{\sqrt{T_{n+1}}}\frac{A_{n,n+1}}{d_{n,n+1}}p_{n+1}.\\
\end{array}\right.
\end{equation}

Now define an $n_{\rm max}\times 1$ vector $\mathbf{p}=\{p_1,\,p_2,\,...,\,p_{n_{\rm max}}\}$ containing the unknown pressures. 
Also define a sparse $n_{\rm max}\times n_{\rm max}$ matrix $\mathbf{A}$: 

\begin{equation} \label{eq:47}
\left\{\begin{array}{l}
\displaystyle \mathbf{A}(1,1)=-\left(\frac{\langle\psi\rangle_{0,1}}{\sqrt{T_1}}\frac{A_{0,1}}{d_{0,1}}+\frac{\langle\psi\rangle_{1,2}}{\sqrt{T_1}}\frac{A_{1,2}}{d_{1,2}}\right)\\
\\
\displaystyle \mathbf{A}(1,2)=\frac{\langle\psi\rangle_{1,2}}{\sqrt{T_{2}}}\frac{A_{1,2}}{d_{1,2}}\\
\end{array}\right.
\end{equation}

\begin{equation} \label{eq:47b}
\left\{\begin{array}{l}
\displaystyle \mathbf{A}(n,n-1)=\frac{\langle\psi\rangle_{n-1,n}}{\sqrt{T_{n-1}}}\frac{A_{n-1,n}}{d_{n-1,n}},\hspace{0.2cm}1< n<n_{\rm max}\\
\\
\displaystyle \mathbf{A}(n,n)=\frac{\langle\psi\rangle_{n-1,n}}{\sqrt{T_n}}\frac{A_{n-1,n}}{d_{n-1,n}}+\frac{\langle\psi\rangle_{n,n+1}}{\sqrt{T_n}}\frac{A_{n,n+1}}{d_{n,n+1}},\hspace{0.2cm}1< n<n_{\rm max}\\
\\
\displaystyle \mathbf{A}(n,n+1)=\frac{\langle\psi\rangle_{n,n+1}}{\sqrt{T_{n+1}}}\frac{A_{n,n+1}}{d_{n,n+1}},\hspace{0.2cm}1< n<n_{\rm max}\\
\end{array}\right.
\end{equation}

\begin{equation} \label{eq:47c}
\left\{\begin{array}{l}
\displaystyle \mathbf{A}(n_{\rm max},n_{\rm max}-1)=\frac{\langle\psi\rangle_{n_{\rm max}-1,n_{\rm max}}}{\sqrt{T_{n_{\rm max}-1}}}\frac{A_{n_{\rm max}-1,n_{\rm max}}}{d_{n_{\rm max}-1,n_{\rm max}}}\\
\\
\displaystyle \mathbf{A}(n_{\rm max},n_{\rm max})=\frac{\langle\psi\rangle_{n_{\rm max},n_{\rm max}+1}}{\sqrt{T_{n_{\rm max}}}}\frac{A_{n_{\rm max},n_{\rm max}+1}}{d_{n_{\rm max},n_{\rm max}+1}}.
\end{array}\right.
\end{equation}

All elements in $\mathbf{A}$ that have not been explicitly assigned in Eqs.~(\ref{eq:47})--(\ref{eq:47c}) are set to zero. With $\mathbf{A}$ formulated like this, the product $\mathbf{Ap}$ equals the 
left--hand sides of the expressions in Eq.~(\ref{eq:46}). Specifically, the two left--side terms in the first expression of Eq.~(\ref{eq:46}) are used for $\mathbf{A}(1,1)$ and $\mathbf{A}(1,2)$ in Eq.~(\ref{eq:47}), 
the three left--side terms in the second expression of Eq.~(\ref{eq:46}) are used for $\mathbf{A}(n,n-1)$, $\mathbf{A}(n,n)$, and $\mathbf{A}(n,n+1)$ with $2<n<n_{\rm max}$ in Eq.~(\ref{eq:47b}), and the 
two left--side terms in the third expression of  Eq.~(\ref{eq:46}) are used for $\mathbf{A}(n_{\rm max},n_{\rm max}-1)$ and $\mathbf{A}(n_{\rm max},n_{\rm max})$ in Eq.~(\ref{eq:47c}). In order to complete  
the equations in Eq.~(\ref{eq:46}), an $n_{\rm max}\times 1$ vector $\mathbf{b}$ needs to be defined, so that $\mathbf{Ap}=\mathbf{b}$. It has a single non--zero entry,
\begin{equation} \label{eq:48}
\mathbf{b}(1)=-\frac{\langle\psi\rangle_{0,1}}{\sqrt{T_0}}\frac{A_{0,1}}{d_{0,1}}p_{n=0}=-\frac{\langle\psi\rangle_{0,1}}{\sqrt{T_0}}\frac{A_{0,1}}{d_{0,1}}p_{{\rm sat},k}(T_0)
\end{equation}
corresponding to the right side of the first expression of Eq.~(\ref{eq:46}), because the right--hand side of the second expression in Eq.~(\ref{eq:46}) is explicitly zero (applied for elements $2<n<n_{\rm max}$ in $\mathbf{b}$), 
and $\mathbf{b}(n_{\rm max})=0$ as well, because it was decided to apply a zero pressure for the ghost cell, $p_{n=n_{\rm max}+1}=0$. Evidently, both $\mathbf{A}$ and $\mathbf{b}$ only 
contain geometric factors, porosities, and temperatures that are all known. It is therefore possible to calculate the unknown pressures,
\begin{equation} \label{eq:49}
\mathbf{p}=\mathbf{A}^{-1}\mathbf{b}.
\end{equation}

Applying a zero pressure for space at the body/coma interface is standard in many thermophysical models \citep[e.g.][]{mekleretal90, prialnik92, desanctisetal99, capriaetal09}. However, in reality the coma does have a pressure that is non--trivial to evaluate. This is because the near--surface coma is strongly deviating from thermodynamic equilibrium, 
i.~e., the molecular distribution function is not given by the Maxwell--Boltzmann distribution function within the so--called Knudsen layer \citep[e.g.][]{cercignani00, davidsson08}. 
\citet{davidssonandskorov04} solved the Boltzmann equation numerically, using the Direct Simulation Monte Carlo (DSMC) method, in order to calculate Knudsen layer properties for 
a range of body surface temperatures and near--surface temperature gradients. \texttt{NIMBUS} does not apply the near--surface density solutions of \citet{davidssonandskorov04} because 
those assumed presence of water ice up to the very surface, which is inconsistent with the presence of the dust mantle considered in Eq.~(\ref{eq:49}). Furthermore, the work of \citet{davidssonandskorov04} 
only concerned water sublimation, while \texttt{NIMBUS} also includes $\mathrm{CO_2}$ and $\mathrm{CO}$. However, \texttt{NIMBUS} can easily be upgraded to include a non--zero coma pressure 
by defining $\mathbf{b}(n_{\rm max})$ as the right--hand side of the third expression in Eq.~(\ref{eq:46}) and use a proper value for $p_{n_{\rm max}+1}$ for species $k$ obtained from Knudsen layer simulations.

Once the pressures $\mathbf{p}$ have been calculated, the corresponding total amount of vapour mass integrated over cells 1--$n_{\rm max}$ may have changed since the last time step. 
If the mass has increased, this mass enhancement is assumed to be provided by sublimating the corresponding amount of ice $k$ in cell $n=0$. If the new vapour mass is lower than that 
currently present, the excess is vented to space and contributes to $Q_k$. In certain conditions, the calculated steady--state pressures may exceed the local saturation pressures. 
This typically happens soon after sunset, when the interior is still warm and the saturation pressure of cell $n=0$ is relatively high, while the surface cells are cooling off and their 
saturation pressures drop drastically with temperature. If so, \texttt{NIMBUS} allows for the frost formation in those cells, by locally condensing the excess mass ($Q_k$ is reduced accordingly). 
At sunrise, the analytically calculated gas pressure typically is below the local saturation pressure, which triggers sublimation of the near--surface frost (increasing $Q_k$ by adding to 
the vapour arriving from larger depths).

This analytic method is not suitable for gas release due to crystallisation or segregation because those zones of release are thick compared to the single--cell sublimation fronts. 
A suitable $n=0$ cell is difficult to define, and pressures may be far from saturation. If vapour of species $k$ is released by segregation and/or crystallisation within a 
region above the sublimation front of $k$ ice, those contributions are added to the vapour mass radial profile set up by the analytical method. As long as those extra contributions 
keep the cell pressures below the local saturation pressures, they will only give rise to an elevated pressure compared to the pure analytical solution. In case the saturation pressure is 
exceeded, the excess will condense. If vapour of species $k$ is released by segregation and/or crystallisation below the sublimation front of $k$ ice, or if all condensed $k$ ice has been 
exhausted, that release takes place in an environment where no analytical solution exists (remembering that the validity of Eq.~(\ref{eq:49}) is limited to a particular near--surface 
region that lacks $k$ ice, and that the approach loses meaning when all $k$ ice in the body has been exhausted). In such cases, the accurate numerical evaluation takes over. 

If the local temperature and pressure gradients are shallow, comparably long time steps may 
still be afforded, because cells with a net loss rate would empty slowly. For example, radiogenically heated bodies typically have a large and almost isothermal core, while the temperature 
falls rapidly near the surface. If the core temperature is high enough to allow for $\mathrm{CO_2}$ release by slow crystallisation and $\mathrm{CO_2}$ ice sublimation, such vapour will 
diffuse upwards at a gentle pace. Most $\mathrm{CO_2}$ will recondense near the surface where it is sufficiently cold for $\mathrm{CO_2}$ freezing, and before the 
temperature and pressure gradients have become so steep that diffusion calculations would require a substantial reduction of the time step. In such situations simulations will proceed relatively fast. 
However, if a substantially more volatile species like $\mathrm{CO}$ is released during crystallisation in the same scenario, it will not encounter any region cold enough for condensation, 
and \texttt{NIMBUS} would have to deal with its diffusion across the very surface, where temperature and pressure gradients might be very steep. In such cases, the time step might 
become very small. In one numerical experiment, a $D=16\,\mathrm{km}$ body was studied right after the last amounts of condensed $\mathrm{CO}$ ice at its core were exhausted. 
The time it took to evacuate the $\mathrm{CO}$ gas that filled all pores from the core to the surface was measured. That complete removal took 6.8 years in the simulated world, and 
costed $3.3\,\mathrm{h}$ of CPU time. Prior to the removal of $\mathrm{CO}$ ice, the model object was advanced $23\,\mathrm{kyr}$ at the same cost. Because such 
drastic slow--downs may be calculationally prohibitive, \texttt{NIMBUS} comes with the option to omit detailed calculations of diffusion of vapour $k$ when ice $k$ is not present. Specifically,  
when segregation and/or crystallisation are solely responsible for production of $k$ vapour, gas may be vented directly to space. Such a simplification is acceptable 
because: 1) during multi--kyr simulations, a diffusion time--scale of a few years can be considered instantaneous and there is little value in resolving it temporally (on the minute level); 
2) the release mechanisms do not depend on the vapour pressure of element $k$, thus gradual diffusion or immediate venting does not change the way vapour $k$ is being produced; 
3) all energy release and consumption associated with the release of vapour $k$ are still calculated accurately, except that of advection that is very small compared to other sources and 
sinks of energy. However, the outgassing of vapour might be exaggerated by not accounting for the possibility of vapour diffusion towards the interior. Such inaccuracies are minimised when the 
temperature systematically increase with depth (as when radiogenic heating takes place), and when there is a background outward flow of vapour from a sublimation front at larger depths. 
If the user chooses not to activate the venting option, gas diffusion (including inward flow) is calculated according to Eq.~(\ref{eq:02}), and energy transport due to advection is accounted for in Eq.~(\ref{eq:01}).

\texttt{NIMBUS} stores snapshots of all the physical data it produces, in different ways depending on the type of body rotation: 1) for fast rotators, data is written to file at a fixed temporal resolution; 
2) for rotationally resolved bodies, data is stored with a certain rotation--angle resolution for selected full rotational periods. Such sequences can be used to reconstruct the total production 
rate of the full body, including all latitudinal, solar zenith angle, and day/night effects. Simulations can be stopped and re--started seamlessly on demand, because all 
information required for the continuation of a simulation are stored. Special programs have been developed to change, e.~g., the spatial resolution of a model and interpolate all physical 
quantities of a previous simulation onto a new grid. The files produced by that program can be read by \texttt{NIMBUS}, which allows for continued simulations under new conditions when necessary. 
There are also dedicated programs for plotting and analysing stored solutions. \texttt{NIMBUS} is implemented in MATLAB\textsuperscript{\textregistered} and also runs on GNU Octave.

\subsection{Code verification} \label{sec_model_verification}

The confirmation of the correctness of a computerised physics model is made by validation (comparison with measured data), 
and/or verification (comparison with other independently implemented computer codes that solve the same problem). Validation 
requires laboratory measurements of temperatures, gas densities, abundances \emph{et~cetera} as functions of depth and time, 
for ice/dust mixtures that are subjected to illumination cycles. As of now, no proper attempt to validate \texttt{NIMBUS} has been made.

\texttt{NIMBUS} has very simple basic numerical principles. Cells have a certain mass for each species and phase, and a particular internal energy. 
If a given mass is being removed from one cell, the exact same mass will be added to another phase of the same cell, added to a neighbouring cell, or ejected to space, 
depending on the considered process.  Energy transfer processes is a similar zero--sum game, with the added component that new energy is injected to the cell network 
by solar energy absorption and radiogenic heating. Because of this, mass and energy conservation are guaranteed to within machine accuracy by design, as long 
as there are no outright implementation errors. However, because machine accuracy is finite, small errors will accumulate with time. To exemplify the degree of 
mass and energy conservation, \texttt{NIMBUS} considered a $D=1\,\mathrm{km}$ body in conditions where $\mathrm{CO}$ sublimation was active but segregation, 
$\mathrm{CO_2}$ sublimation, crystallisation, and $\mathrm{H_2O}$ were dormant. The model was run for $6.2\,\mathrm{kyr}$ in the simulated world, during which $5\%$ of 
the $\mathrm{CO}$ was lost to space. The simulation consisted of $3.9\cdot 10^5$ time steps, i.~e., cycles of mass and energy swapping between cells. Comparing the difference in $\mathrm{CO}$ 
abundance of the body at the beginning and end, with the documented loss to space, the quantities differed by $0.01\%$. The difference in internal energy at the end compared to the 
beginning (including absorbed solar energy and lost thermal radiation), compared with the documented removal of energy by escaping vapour, differed by $0.05\%$. It is evident that 
coding errors are absent, and error accumulation is slow.

However, mass and energy conservation by themselves do not guarantee correctness. It is also important to verify that the transferred amounts have 
the correct magnitude. The basic correctness of the heat conductivity and radiogenic heat generation treatment was made by comparing \texttt{NIMBUS} 
simulations with an analytical solution \citep[e.g.][]{heveysanders06} to the energy conservation equation for a spherical homogeneous 
body that is heated radiogenically, cooled radiatively at its surface, and that has temperature--independent heat conductivity and heat capacity. 
In one particular case, using spatial and temporal resolutions typical of simulations in this paper, the $^{26}\mathrm{Al}$--driven heating of a 
$D=1\,\mathrm{km}$ body from $10\,\mathrm{K}$ at $t=0$ to a temperature peak at the centre of $4647\,\mathrm{K}$, followed by a drop to 
$1784\,\mathrm{K}$  at $t=25\,\mathrm{Myr}$, was carried out with maximum and mean errors in \texttt{NIMBUS} with respect to the analytical 
solution of $0.9\,\mathrm{K}$ and $0.5\,\mathrm{K}$, respectively. At a depth of $50\,\mathrm{m}$ below the surface, where temperature gradients are 
steep and change relatively fast with time, a peak of $2623\,\mathrm{K}$ and 
cooling to $275\,\mathrm{K}$ at $t=25\,\mathrm{Myr}$ was obtained with maximum and mean errors of $2.8\,\mathrm{K}$ and $1.3\,\mathrm{K}$. Although 
errors can be reduced further by considering higher spatial resolution and shorter time--steps, this level of uncertainty would be acceptable in most 
scientific applications.

\begin{figure*}
\centering
\begin{tabular}{cc}
\scalebox{0.45}{\includegraphics{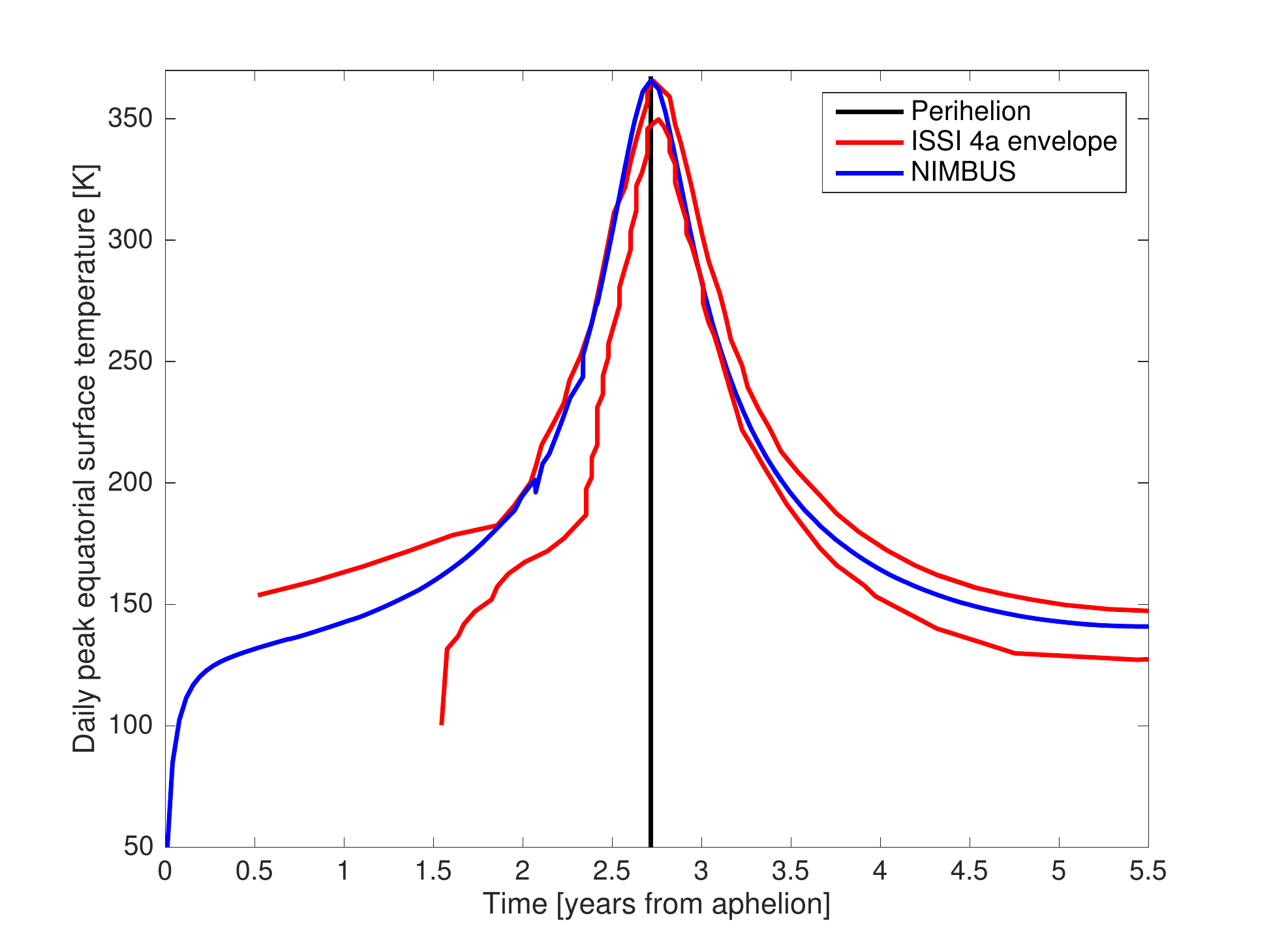}} & \scalebox{0.45}{\includegraphics{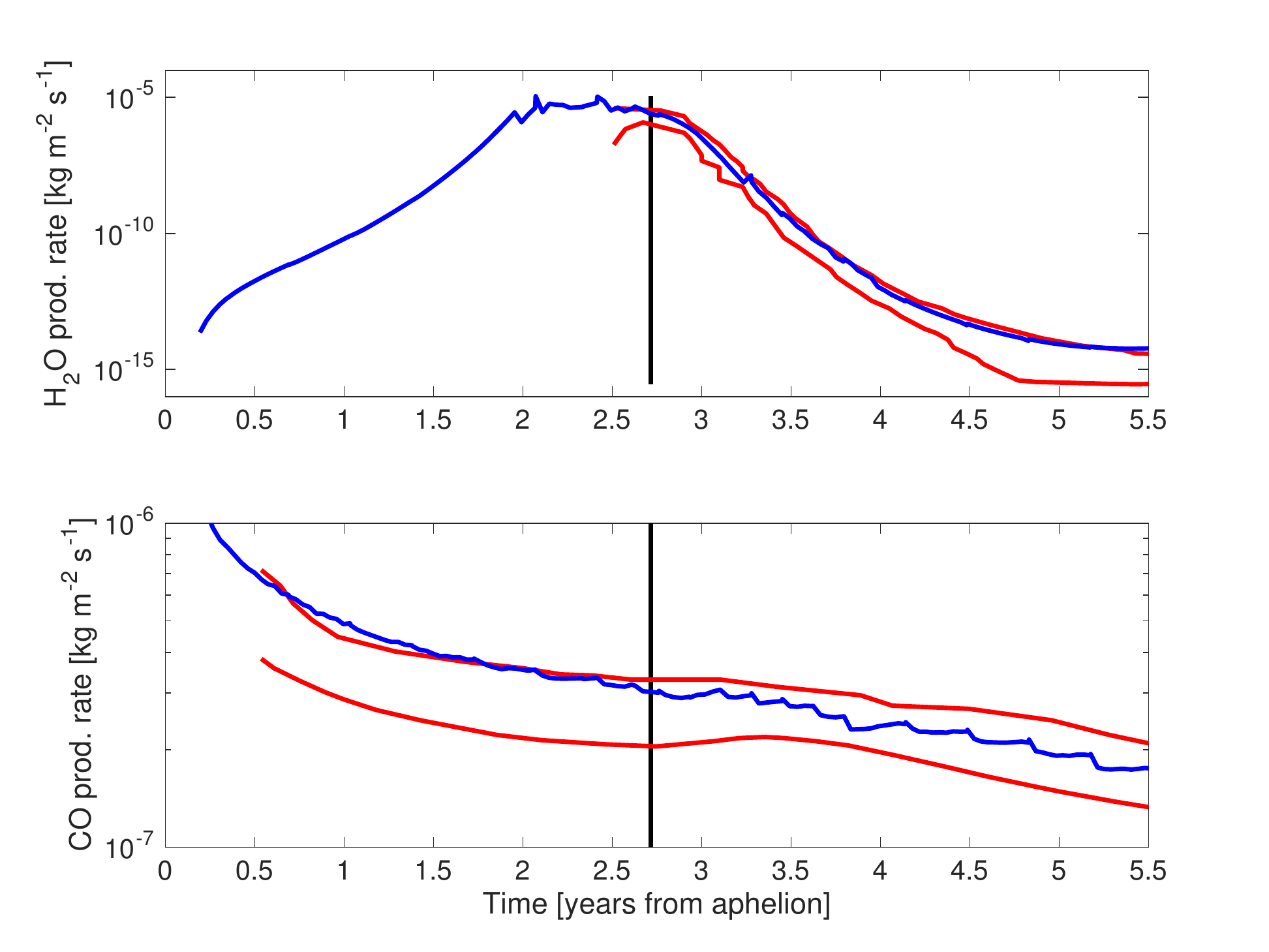}}\\
\end{tabular}
     \caption{Verification of the \texttt{NIMBUS} code against the ISSI~4a model of \protect\citet{huebneretal06}. The left panel shows 
the daily peak equatorial surface temperature throughout the first orbit (ISSI~4a model boundaries in red, and \texttt{NIMBUS} in blue). The right 
panel shows the daily peak equatorial production rates of $\mathrm{H_2O}$ (upper) and CO (lower).}
     \label{fig_verify}
\end{figure*}

\citet{huebneretal99, huebneretal06}, in the context of an International Space Science Institute 
(ISSI) Team in Bern, Switzerland, made detailed comparisons between five advanced thermophysical codes for a number of standard model cases. 
Their model ISSI~4a was selected for verification of \texttt{NIMBUS}.  That model considers a comet on a typical 
JFC orbit, that consists of a porous mixture of dust, water ice, and CO ice, assuming sublimation front withdrawal into the nucleus as the result of not considering 
dust mantle erosion \citep[for a detailed list of parameters, see][]{huebneretal06}. \texttt{NIMBUS} was set up to be as similar as possible 
to ISSI~4a \citep[which included some functions in Secs.~\ref{sec_model_ck}--\ref{sec_model_diffusion} being replaced by those used by][]{huebneretal06}. 
Despite being set up as similarly as possible, the five codes applied by \citet{huebneretal06} did not produce identical solutions. Figure~\ref{fig_verify} (left) shows 
the upper and lower enveloping boundaries of the equatorial daily peak surface temperatures for the five codes of \citet{huebneretal06} as red curves. 
It is not necessarily the same code that is responsible for the lowest or highest temperature throughout the orbit, but the red curves show the full range of 
the assemble of model solutions. The blue curve in the left panel of Fig.~\ref{fig_verify} shows the result of running \texttt{NIMBUS}, and as can be seen, 
it is consistent with the results of the other five models. The \texttt{NIMBUS} pre--perihelion temperatures most closely resemble the \citet{huebneretal06} 
algorithms~A \citep{enzianetal97} and C2 \citep{oroseietal99}, while the post--perihelion temperatures are closer to algorithm~D \citep{prialnik92}.

The upper right panel of Fig.~\ref{fig_verify} shows the equatorial daily peak $\mathrm{H_2O}$ production rate, with the two red curves 
enveloping the set of \citet{huebneretal06} models. For unknown reasons, \citet{huebneretal06} chose not to plot the first $\sim 2.5\,\mathrm{yr}$ 
of the orbit. Some of the \citet{huebneretal06} models display local fluctuations due to numerical instabilities that sometimes become as large as the 
difference between the five models. Those fluctuations were not reproduced in the upper right panel of Fig.~\ref{fig_verify}, but the red curves trace 
the upper and lower production rates for the numerically stable curves in regions where some models had problems. The \texttt{NIMBUS} solution (blue solid curve) 
closely follows the upper envelope curve, and is most similar to algorithm~C2. The lower right panel of Fig.~\ref{fig_verify} shows the equatorial daily peak 
$\mathrm{CO}$ production rate. \citet{huebneretal06} only showed the results of algorithms~B \citep{benkhoffboice96} and C2, 
being the upper and lower red curves in the lower right panel of Fig.~\ref{fig_verify}, respectively. The \texttt{NIMBUS} pre--perihelion solution closely 
follows algorithm~B, whereas the post--perihelion solution is intermediate between those of B and C2.

Obtaining identical numerical solutions to the same (highly non--linear) physical problem is not trivial because of differences in numerical schemes and certain 
variability in the treatment of physical processes, as explained in detail by \citet{huebneretal06}. It is clear from Fig.~\ref{fig_verify} that 
\texttt{NIMBUS} performs as well (or as poorly) as comparable advanced thermophysical modelling codes.

\section{Model parameters} \label{sec_parameters}

All model bodies studied here are assumed to have formed without short--lived radionuclides. That is done in order to isolate the thermal 
evolution caused by the unavoidable heating by the early Sun and by long--lived radionuclides. Therefore, the presented cases should 
be considered lower limits on the degree of processing (for the particular combinations of rock abundances and porosities considered here). 
Whether the real Primordial Disk objects contained short--lived radionuclides at the time of their formation is currently unknown. This 
uncertainty is because we do not yet know if the Solar Nebula contained short--lived radionuclids at birth, or if they were injected locally 
during formation, e.~g., as suggested in the supernova--triggered model by \citet{bossvanhala00} and \citet{bossetal08}. Until returned samples from the 
outer Solar System has settled this issue, it is motivated to consider models both with and without short--lived radionuclides: this paper focuses on the latter.

Three different body sizes have been considered, in order to better understand the effect of thermophysical evolution in the 
Primordial Disk for different types of bodies. The three body classes are named after similarly--sized objects in the current 
Solar System for easy reference. Although the simulations describe how objects of these sizes may have evolved in the early Solar System, the 
identifications with real objects should not be taken literally. Because of differences in formation distance, formation time, and subsequent evolution, the 
model bodies are not necessarily good representation of the current properties of their namesakes. The smallest body under study has a diameter of $D=4\,\mathrm{km}$ and 
is referred to as the ``67P/Churyumov--Gerasimenko'' case (or the 67P case for short). It is representative for the 
majority of JFCs, dynamically new comets, and small members of the Centaur and Kuiper Belt populations. Such bodies are not heated 
significantly by long--lived radionuclides, therefore, those models target the solar--driven evolution most directly. The intermediate 
body is an analogue to Comet C/1995 O1 (Hale--Bopp) with  $D=74\,\mathrm{km}$ \citep{szaboetal12}. The largest model body has 
$D=203\,\mathrm{km}$ which is similar in size to one of the largest Centaurs, (2060)~Chiron \citep{fornasieretal13}, as well as to 
the saturnian satellite Phoebe \citep{castillorogezetal12} that likely was captured \citep{johnsonlunine05} from the Primordial Disk.

The Hale--Bopp and Chiron/Phoebe cases consider bodies sufficiently large to make long--lived radionuclides important heat 
sources, although the  exact amount of heating depends on the abundance of refractories (radionuclides are only present in the rocky material). 
The refractories--to--$\mathrm{H_2O}$ mass ratio $\mu(t=0)=\mu_0$ is not well known for the bodies that once populated the Primordial Disk. It is easiest 
to measure for bodies so large that they cannot possibly have substantial macro porosity. Pluto and Charon, with well--determined sizes and masses 
thanks to the New~Horizons flyby, have $\mu=1.90\pm 0.04$ and $\mu=1.71\pm 0.07$ \citep{mckinnonetal17}, respectively. However, those 
may be upper limits to the original composition, considering that substantial amounts of volatiles may have been lost in the collision that 
formed the Pluto--Charon system. Several attempts to constrain $\mu$ for Comet 67P/Churyumov--Gerasimenko have been made. The 
total mass loss of the nucleus during the Rosetta mission is well known, $\Delta M=(10.5 \pm 3.4)\cdot 10^9\,\mathrm{kg}$ \citep{patzoldetal19}, and 
the total loss of water vapour $m_{\rm w}$ and other gases $m_{\rm g}$ can be estimated from in situ measurements. This yields the mass ratio of 
dust to water vapour $\mu_{\rm v}=(\Delta M-m_{\rm w}-m_{\rm g})/m_{\rm w}$ that is not necessarily identical to $\mu$ of the nucleus. Ice within 
large coma boulders may escape detection, which would lower $\mu$ with respect to $\mu_{\rm v}$.  Dust that release water vapour before returning to the nucleus 
as airfall \citep{thomasetal15a, thomasetal15b, kelleretal15, kelleretal17, davidssonetal21} 
could increase $\mu$ above $\mu_{\rm v}$. \citet{biveretal19} obtain $\mu_{\rm v}=2.63 \pm 0.19$ from gas measurements by MIRO, while \citet{combietal20} 
obtain $\mu_{\rm v}=1 \pm 1$ based on ROSINA data. Several estimates made with different methods and based on data from other Rosetta instruments are 
available (see review by \citet{choukrounetal20}). Based on COSIMA data,  \citet{choukrounetal20} obtain the mass ratio $0.2\leq\mu_{\rm i}\leq 3$ of refractories 
to all ices, which translates to  $0.3\leq\mu=(1+m_{\rm g}/m_{\rm w})\mu_{\rm i}\leq 4.1$ if using $m_{\rm g}$ and $m_{\rm w}$ from \citet{combietal20}, 
or $0.3\leq\mu\leq 5.2$ if using the values from \citet{biveretal19}. The lower limit is the ice--richest model of 67P published to date. The ice--poorest estimate is 
$\mu=8$ based on GIADA data \citep{fulleetal17}. \citet{davidssonetal21b} used \texttt{NIMBUS} to reproduce the water production curve of 67P and 
found that $\mu=1$ was necessary to make that possible. The albedo analysis of the interior of a boulder that was broken open by the Philae lander indicated $\mu=2.3^{+0.20}_{-0.16}$ \citep{orourkeetal20}. 
As a compromise, $\mu=4$ is applied in the current work. It is assumed that all water ice initially is amorphous.

\citet{gerakinesetal99} observed a diverse set of interstellar environments (massive protostars, sources near the Galactic Centre, and Taurus dark clouds) with 
the Infrared Space Observatory and found that the molar abundance of $\mathrm{CO_2}$ ice relative water had a narrow range of $10\%$--$23\%$. The 
average $\mathrm{CO_2}$ concentration was $[\mathrm{CO_2}]/[\mathrm{H_2O}]=0.17\pm 0.03$. Accordingly, it is here assumed that the amount of 
$\mathrm{CO_2}$ is $17\%$ relative $\mathrm{H_2O}$ by number.  \citet{gerakinesetal99} also find that $[\mathrm{CO_2}]/[\mathrm{CO}]=1.1$, hence 
it is assumed that $\mathrm{CO}$ is $15.5\%$ relative $\mathrm{H_2O}$ by number. In terms of the total body mass, the composition of all model bodies is 
$70.7\%$ refractories, $17.7\%$ water, $7.3\%$ carbon dioxide, and $4.3\%$ carbon monoxide (i.~e., the mass ratios of $\mathrm{CO_2}$ and $\mathrm{CO}$ 
relative water are 0.41 and 0.24, respectively).

The partitioning of $\mathrm{CO}$ and $\mathrm{CO_2}$ between entrapment in amorphous water ice and other storage modes aimed at 
making the crystallisation process energy--neutral. That is to say, the crystallisation energy should be entirely consumed when liberating the 
occluded $\mathrm{CO}$ and $\mathrm{CO_2}$. This (admittedly arbitrary) assumption was made in order to minimise the evolutionary 
changes triggered by the energy release during crystallisation, in order to isolate the effects of solar and radiogenic heating. Furthermore, it was 
required that $\mathrm{CO_2}$ should absorb $80\%$ of energy during crystallisation, and that the abundance of $\mathrm{CO}$ in condensed 
$\mathrm{CO_2}$ should be $20\%$ by number. This was made by considering the latent heat of $\mathrm{CO}$ at $T=70\,\mathrm{K}$ and of $\mathrm{CO_2}$ 
at $T=110\,\mathrm{K}$ (the highest and lowest temperatures, respectively, for which the \citet{huebneretal06} formulae formally are valid), which is the 
temperature interval where the bulk of crystallisation takes place (on the long time--scales  relevant for the considered problem). With those constraints, the $\mathrm{CO}$ 
partitions as follows: $52.4\%$ is condensed, $15.8\%$ is trapped in $\mathrm{CO_2}$, and $31.8\%$ is trapped in amorphous $\mathrm{H_2O}$. 
The $\mathrm{CO_2}$ is predominantly condensed ($71.7\%$) while the remainder ($28.3\%$) is trapped in amorphous $\mathrm{H_2O}$. The large contribution 
of $\mathrm{CO_2}$ during absorption of energy released during crystallisation was selected in order to place roughly half the $\mathrm{CO}$ in the condensed 
phase. If a 50/50 division of energy absorption between $\mathrm{CO}$ and $\mathrm{CO_2}$ is considered, only $\sim 2\%$ of the $\mathrm{CO}$ is been 
placed in the condensed phase, potentially biasing the $\mathrm{CO}$ loss time--scales towards too low values.

The hydrostatic code that solves Eq.~(\ref{eq:09}) requires that compacted densities and mass fractions are specified for ices and refractories. 
Ices have the compact density
\begin{equation} \label{eq:50}
\rho_{\rm ice}=\left(\frac{M_{\rm H_2O}}{\varrho_4}+\frac{M_5+M_6}{\varrho_6}\right)^{-1}\sum_{k=2}^{n_{\rm s}}M_k=1084\,\mathrm{kg\,m^{-3}}, 
\end{equation}
where the mass fraction of water in volatiles is $0.60$ according to the composition defined above, $\varrho_4=917\,\mathrm{kg\,m^{-3}}$ \citep{weast74}, and the densities 
of both $\mathrm{CO_2}$ and $\mathrm{CO}$ are taken as $\varrho_5=\varrho_6=1500\,\mathrm{kg\,m^{-3}}$ \citep{satorreetal09}. For a chondritic composition, the abundances 
of the 12 most common elements, according to \citet{lodders03}, are applied. Specifically: all Fe, S, and Ni are used to form metal/sulphides; all Mg, Si, Al, Ca, Na, and three O per 
Si are used to form rock; all C, N, and three H per C are used to form organics. The combination of metal, sulphides, and rock are here called ``minerals'', and the combination of 
minerals and organics is called ``refractories''. If so, there is a $93:7$ ratio of minerals:organics, and the minerals have $37.8\%$ metal/sulphide by mass. The density of metal/sulphide 
is taken as $\rho_{\rm met}=5850\,\mathrm{kg\,m^{-3}}$, which is the average of the densities for iron $\rho_{\rm Fe}=7100\,\mathrm{kg\,m^{-3}}$ and troilite 
$\rho_{\rm FeS}=4600\,\mathrm{kg\,m^{-3}}$ \citep{tesfayefirdutaskinen10}. The density of rock is taken as that of forsterite, $\rho_{\rm Fo}=3270\,\mathrm{kg\,m^{-3}}$ 
\citep{horai71}. If so, the density of minerals is $\rho_{\rm min}=3925\,\mathrm{kg\,m^{-3}}$. Taking the density of organics as that of ``spark tholins'', 
$\rho_{\rm org}=1130\,\mathrm{kg\,m^{-3}}$ \citep{horsttolbert13}, the density of refractories is $\rho_{\rm ref}=3367\,\mathrm{kg\,m^{-3}}$.

\begin{figure}
\begin{center}
     \scalebox{0.45}{\includegraphics{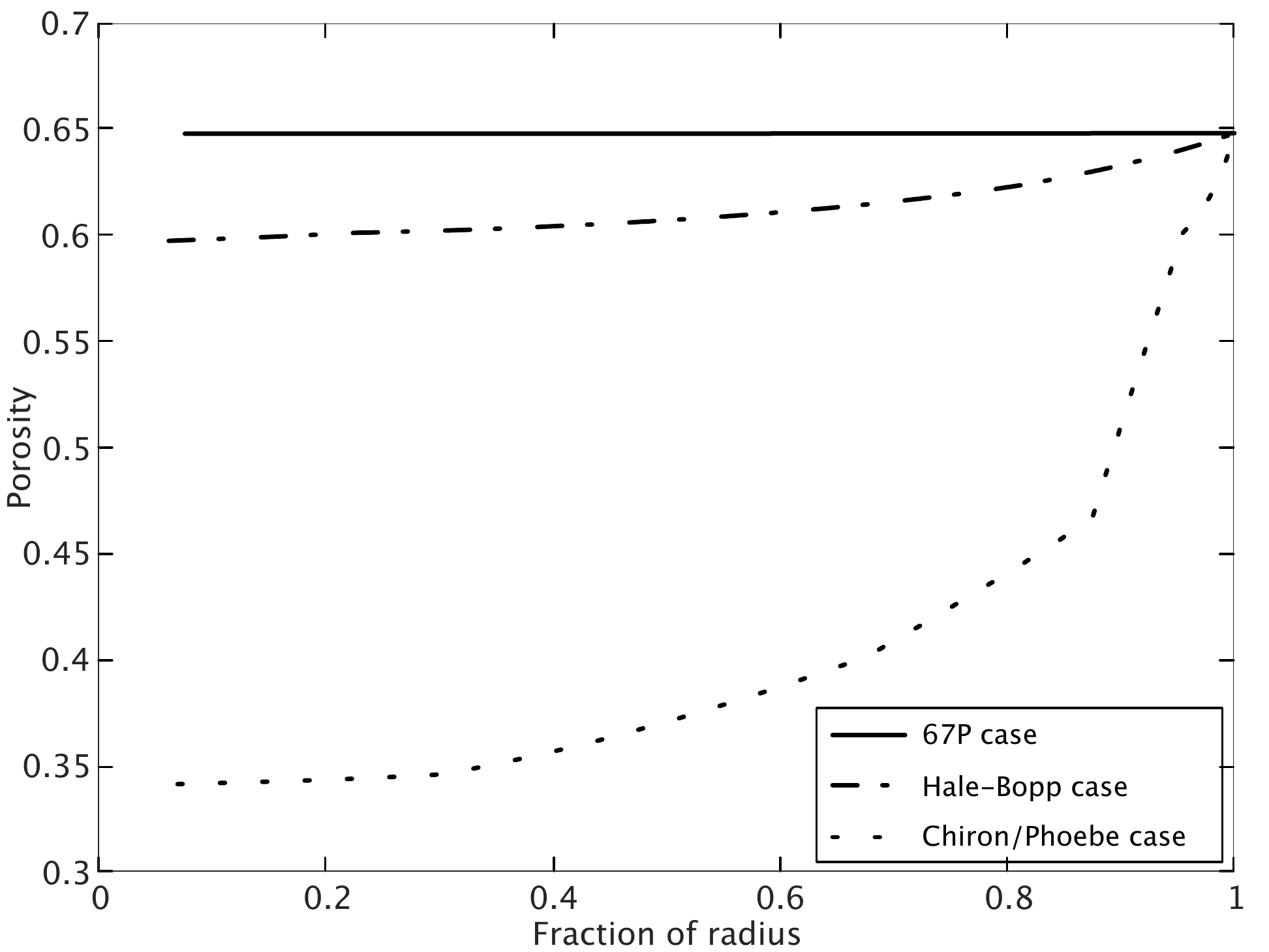}}
    \caption{Porosity as function of fractional body radius, obtained by hydrostatic equilibrium calculations, using compressive strengths 
of mixtures of silicate and water ice grains, and assuming dynamical pressure (helping self--gravity to compress the bodies) corresponding 
to a $15\,\mathrm{m\,s^{-1}}$ impact velocity during accretion \protect\citep[in order to give the 67P/C--G case the measured $535\,\mathrm{kg\,m^{-3}}$ 
bulk density of the comet;][]{preuskeretal15}.}
     \label{fig_porous}
\end{center}
\end{figure}

However, the composition of bodies in the outer Solar System is not necessarily chondritic. The 67P dust analysed by COSIMA is iron--poor and substantially more 
carbon--rich than chondritic material, if normalising both to the silicon abundance \citep{bardynetal17}. With metal/sulphide down to $28.9\%$ the mineral density 
is $\rho_{\rm min}'=3747\,\mathrm{kg\,m^{-3}}$. With a minerals:organics ratio drastically changed to $55:44$, the density of refractories drops to 
$\rho_{\rm ref}'=1835\,\mathrm{kg\,m^{-3}}$. The hydrostatic code was evaluated for $70.7\%$ refractories by mass, using $\rho_{\rm ref}'=1835\,\mathrm{kg\,m^{-3}}$, 
and $29.3\%$ ices with $\rho_{\rm ice}=1084\,\mathrm{kg\,m^{-3}}$. The initial total mass was set such that a $D=4\,\mathrm{km}$ body should have a bulk 
density $\rho_{\rm bulk}=535\,\mathrm{kg\,m^{-3}}$, as for 67P \citep{preuskeretal15}. Under these conditions, and in the absence of an external dynamic 
pressure (Eq.~\ref{eq:10b}), the self--gravity compresses the model body to a diameter $D=5.8\,\mathrm{km}$, having $\rho_{\rm bulk}=178\,\mathrm{kg\,m^{-3}}$, 
and an average porosity of $\psi=0.88$. In order to obtain the desired $D=4\,\mathrm{km}$ and $\rho_{\rm bulk}=535\,\mathrm{kg\,m^{-3}}$ it was necessary 
to apply an accretion velocity of $V_{\rm i}\approx 15\,\mathrm{m\,s^{-1}}$, resulting in $\psi=0.65$. The body is nearly homogeneous. The same composition and $V_{\rm i}$--value was 
applied to the Hale--Bopp and Chiron/Phoebe cases. The porosity profiles of the three model bodies are shown in Fig.~\ref{fig_porous}. Self--gravity effects are rather 
weak at $D=74\,\mathrm{km}$, lowering the core porosity to 60\% and yielding a bulk density $\rho_{\rm bulk}=585\,\mathrm{kg\,m^{-3}}$ and average porosity $\psi=0.62$. 
Self--gravity effects are stronger at $D=203\,\mathrm{km}$, lowering the core porosity to 34\% and yielding a bulk density $\rho_{\rm bulk}=836\,\mathrm{kg\,m^{-3}}$ and 
average porosity $\psi=0.45$. \citet{meechetal97} found that the observed coma of Chiron, if interpreted as a bound ballistic atmosphere, suggested 
$\rho_{\rm bulk}\stackrel{<}{_{\sim}} 1000\,\mathrm{kg\,m^{-3}}$. The radiometrically determined diameter of the primary in the binary Centaur (65489) Ceto/Phorcys system 
is $D=223\pm 10\,\mathrm{km}$ based on Herschel Space Observatory observations according to \citet{santossanzetal12}, i.e., a similar size to the Chiron/Phoebe case. Using the mass 
derived by \citet{grundyetal07}, the bulk density would be $\rho_{\rm bulk}=640^{+160}_{-130}\,\mathrm{kg\,m^{-3}}$ \citep{santossanzetal12}. However, \citet{grundyetal07} obtained 
a smaller Ceto diameter of $D=174\pm 17\,\mathrm{km}$ based on Spitzer Space Telescope observations, and consequently, a higher bulk density of 
$\rho_{\rm bulk}=1370^{+660}_{-320}\,\mathrm{kg\,m^{-3}}$. It is reassuring that the model $\rho_{\rm bulk}$ of the Chiron/Phoebe case is consistent with the estimate 
of \citet{meechetal97}, and is intermediate between the estimates of \citet{grundyetal07} and \citet{santossanzetal12} for a similarly--sized body. However, it should be remembered 
that Phoebe itself has a density $\rho_{\rm bulk}=1634\pm 46\,\mathrm{kg\,m^{-3}}$ \citep{matsonetal09}, which likely is caused by $^{26}\mathrm{Al}$ heating and compaction \citep{castillorogezetal12}.

Comets are often modelled with albedo $A=0.04$ and emissivity $\varepsilon=0.9$ and those values are applied here despite the possibility that near--surface ice may have 
made the Primordial Disk objects substantially brighter. The solar energy absorption in the current models could therefore be somewhat exaggerated, although 
solar wind sputtering and cosmic ray bombardment may have caused non--thermal removal of ice in the top few millimetres and a very low albedo.  Constituent grains in comet material 
are typically of order $\sim 1\,\mathrm{\mu m}$ in size \citep[e.g.][]{brownleeetal06, bentleyetal16}, hence $r_{\rm g}=1\,\mathrm{\mu m}$ is applied. Such grains 
are assumed to cluster into $\sim 1\mathrm{mm}$--sized pebbles \citep{blumetal17}, motivating the choice to use $r_{\rm p}=10^{-3}\,\mathrm{m}$ and $L_{\rm p}=10^{-2}\,\mathrm{m}$ 
(i.~e., diffusivity is calculated based on the assumptions that the cylindrical tube radii are similar to the diameters of pebbles, and that channels may be reasonably 
straight over distances of tens times their radii,  motivating the usage of $L_{\rm p}=10r_{\rm p}$ and $\xi=1$). The orbits of all model bodies are assumed to be located in the midst of 
the Primordial Disk ($a=23\,\mathrm{AU}$) and they are assumed to be circular and placed in the ecliptic plane. Fixed spin axes of $\{\lambda,\,\beta\}=\{0,\,45^{\circ}\}$ were applied, 
using an intermediate obliquity to prevent the polar regions to be permanent cold--spots (spin axes are not likely stable on the time--scales considered here, but this may be considered 
a reasonable average orientation).

The current simulations always considered a detailed diffusion treatment for $\mathrm{CO_2}$ and $\mathrm{H_2O}$. The same treatment was applied to $\mathrm{CO}$ as 
long as condensed $\mathrm{CO}$ ice existed within the model body. After condensed $\mathrm{CO}$ ice had been removed, $\mathrm{CO}$ vapour produced by 
segregation and crystallisation was vented directly to space.

\section{Results} \label{sec_results}

\subsection{The $\mathbf{D=4\,\mathrm{\bf km}}$ ``67P/Churyumov--Gerasimenko'' case} \label{sec_results_67P}

Three versions of the 67P case were considered, having disk clearing times of $t_{\rm c}=5\,\mathrm{kyr}$, $1\,\mathrm{Myr}$, and $3\,\mathrm{Myr}$. 
The orbital skin depth (i.~e., the depth over which the amplitude of temperature oscillations, caused by obliquity and orbital motion, are damped by an $e$--folding scale) 
was $\sim 30\,\mathrm{m}$ for the warmest body ($t_{\rm c}=5\,\mathrm{kyr}$) and $\sim 18\,\mathrm{m}$ for the coldest body ($t_{\rm c}=3\,\mathrm{Myr}$). 
In order to resolve the orbital skin depth and properly evaluate near--surface heat fluxes, near--surface radial grid cell thicknesses should be smaller than the skin depth by a 
factor of a few. In these simulations, the thickness of the uppermost grid cell was $5\,\mathrm{m}$, and cells grew to $50\,\mathrm{m}$ at the core by geometric progression, 
resulting in 102 radial cells. With 18 latitudinal slabs, the 67P case model body was spatially resolved by a total of 1,836 cells.

The $t_{\rm c}=5\,\mathrm{kyr}$ case exposed the model nucleus to the peak $6.4L_{\odot}$  illumination of the young protosun ($t_{\rm c}=0$ was avoided 
to allow for a gradual thermal adjustment to the harsh conditions). It was modelled for $1\,\mathrm{Myr}$. Under these conditions, half the 
condensed $\mathrm{CO}$ has been removed by $t=13.8\,\mathrm{kyr}$, $10\%$ is remaining at $t=41.1\,\mathrm{kyr}$, and all of it has 
been lost by $t=72.9\,\mathrm{kyr}$. This illustrates the gradual slow--down of the (close to exponential) loss rate with time: losing the last 10\% took almost as 
long as losing the first 90\%. The $\mathrm{CO_2}$ ice is not a safe haven for the $\mathrm{CO}$, because $90\%$ of the $\mathrm{CO}$ has 
segregated and left the model body by $t=87\,\mathrm{kyr}$.  Extremely small deposits ($0.005\%$ of the original amount) survive near the poles, 
in a layer 80--$160\,\mathrm{m}$ below the surface, but even those layers have lost $\sim 97\%$ of their original CO abundance. The sublimation of 
$\mathrm{CO_2}$ is rather substantial near the surface. The $\mathrm{CO_2}$ front withdraws $32\,\mathrm{m}$ below the surface in the first $0.2\,\mathrm{Myr}$ and 
then stops there. The degree of crystallisation is substantial near the surface, but $94.6\%$ of the global amorphous water ice reservoir remains at the end of the simulation. The degree of 
crystallisation is at least 10\% in the top $205\,\mathrm{m}$, and more than $80\%$ in the top $16\,\mathrm{m}$ (all water ice at the surface is cubic). 
Therefore, if comets formed right after the accretion of the envelope and the massive disk, the only plausible storage medium of 
hypervolatiles like $\mathrm{CO}$ would be amorphous water ice. Model results are summarised in Table~\ref{tab_results}.

\begin{figure*}
\centering
\begin{tabular}{cc}
\scalebox{0.2}{\includegraphics{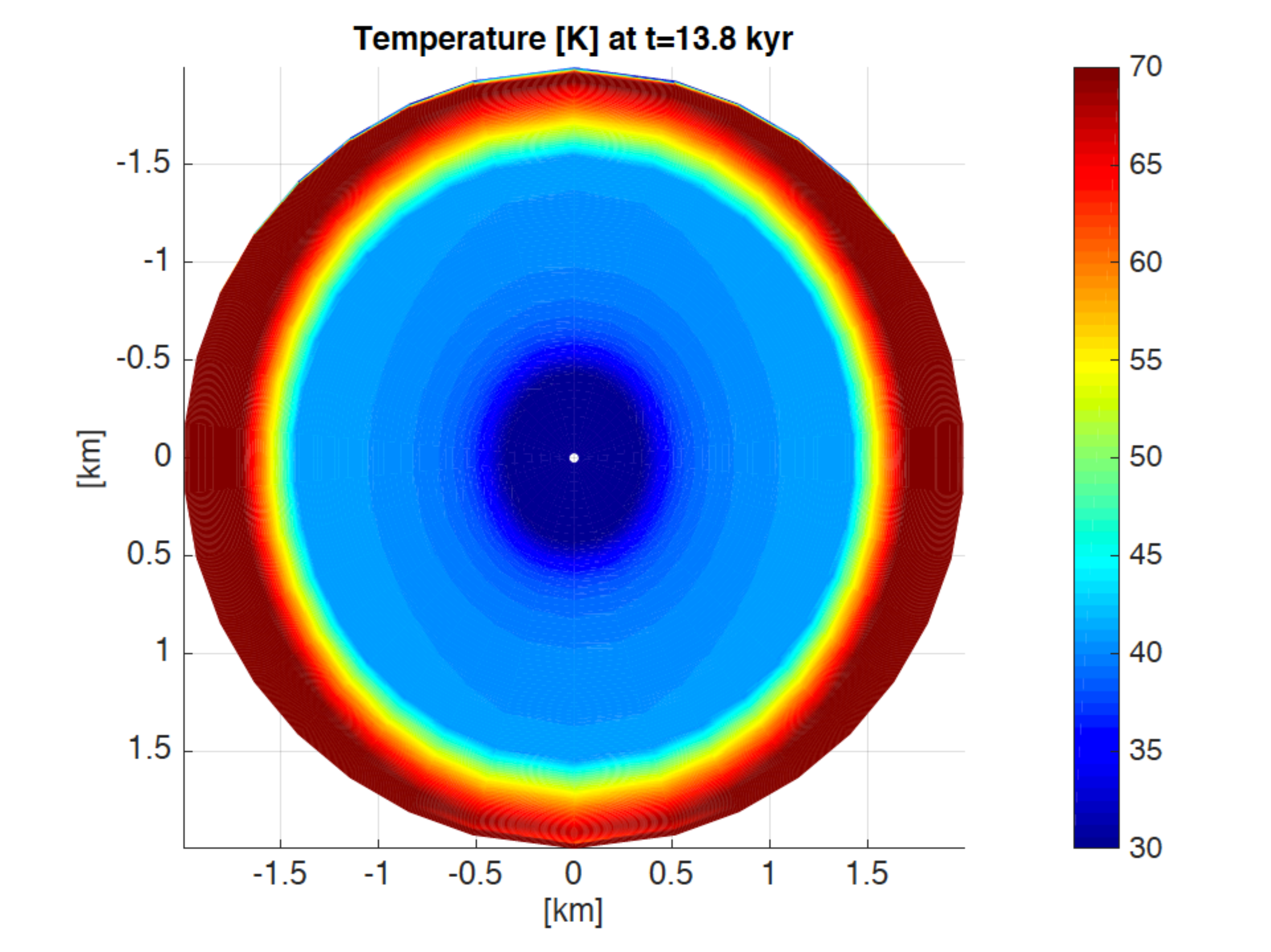}} & \scalebox{0.2}{\includegraphics{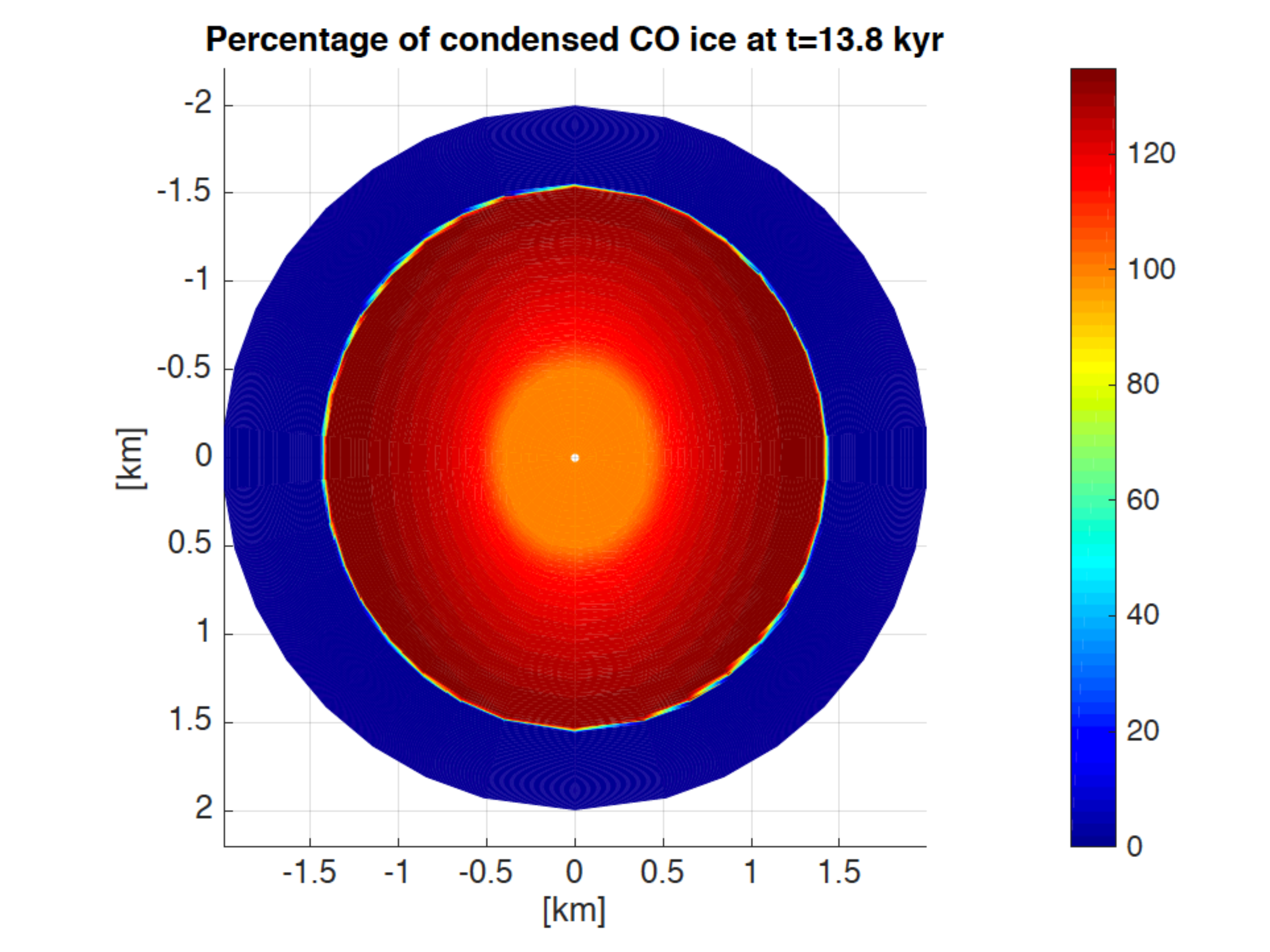}}\\
\scalebox{0.2}{\includegraphics{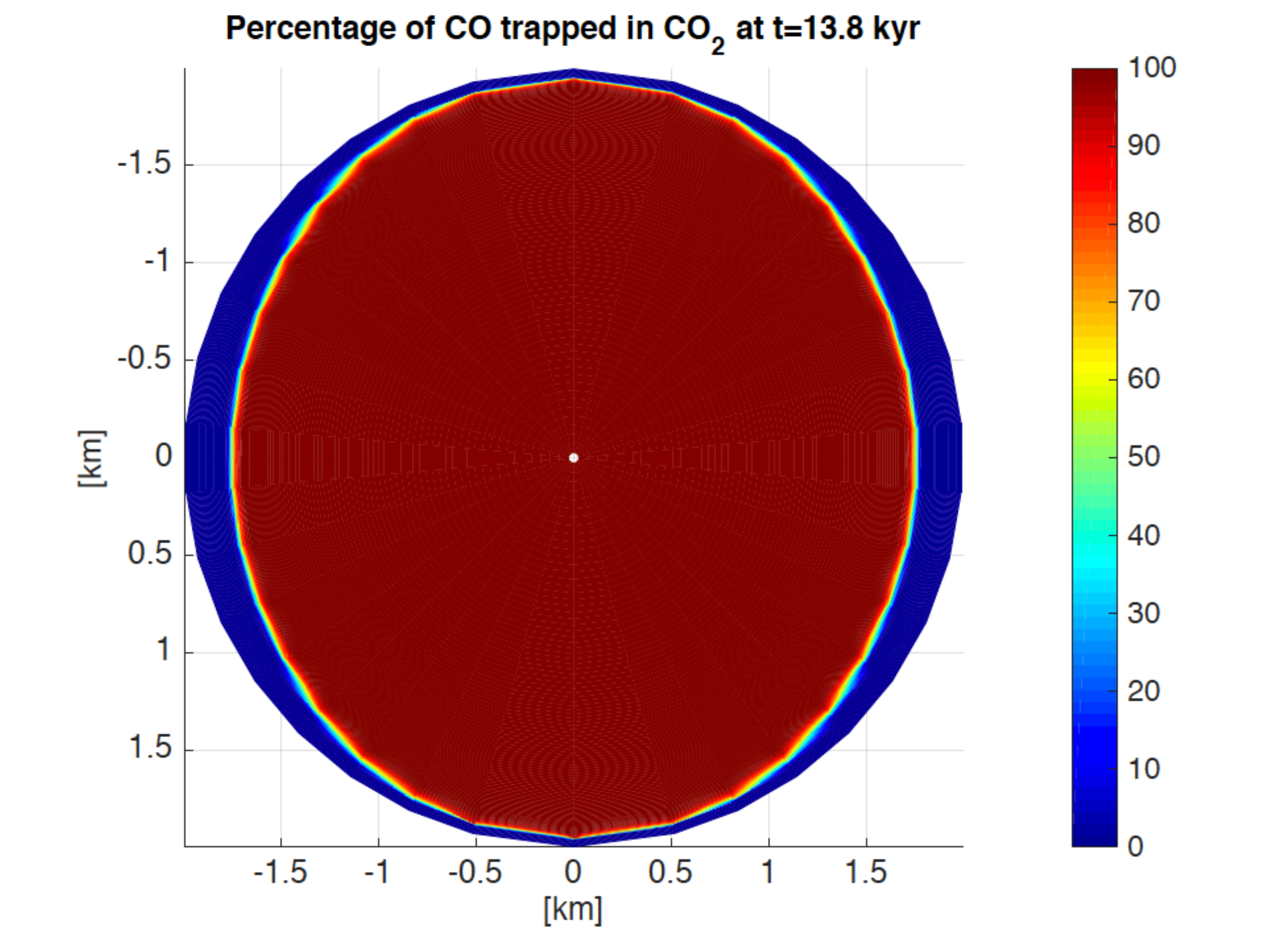}} & \scalebox{0.2}{\includegraphics{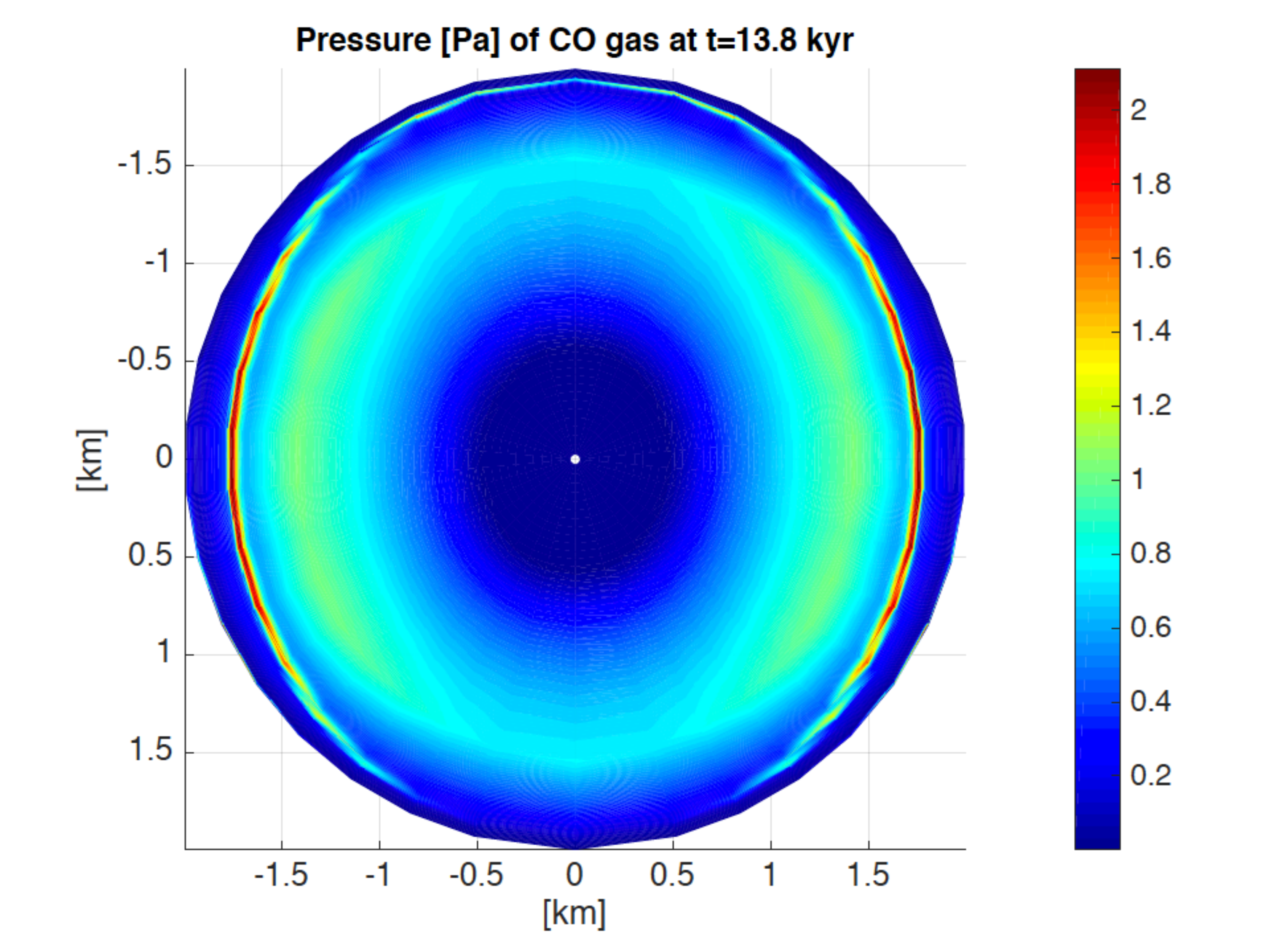}}\\
\end{tabular}
     \caption{The $D=4\,\mathrm{km}$ ``67P/Churyumov--Gerasimenko'' model case with $t_{\rm c}=5\,\mathrm{kyr}$ disk 
clearing is here shown at $t=13.8\,\mathrm{kyr}$. The panels show temperature (upper left), 
the abundance of CO ice (upper right; note that recondensation of downward--diffusing vapour temporarily 
has elevated the CO abundance below the sublimation front), the abundance of $\mathrm{CO:CO_2}$ mixtures (lower left), and the 
pressure of CO vapour (lower right).}
     \label{fig_case67P}
\end{figure*}

Figure~\ref{fig_case67P} shows temperature, concentrations of condensed and $\mathrm{CO_2}$--hosted $\mathrm{CO}$, as well as 
the $\mathrm{CO}$ gas pressure for the $t_{\rm c}=5\,\mathrm{kyr}$ 67P case at $t=13.8\,\mathrm{kyr}$. The temperature in the 
top few hundred meters is  $\sim 70\,\mathrm{K}$, which is too warm for condensed or $\mathrm{CO_2}$--hosted $\mathrm{CO}$ but 
sufficiently cool to preserve most $\mathrm{CO_2}$ and amorphous water ice. The $\mathrm{CO}$ sublimation front has withdrawn $570\,\mathrm{m}$ 
under the surface. Note that a substantial fraction of the core has a $\mathrm{CO}$ concentration that has been elevated by $\sim 30\%$ 
above the initial value because of downward diffusion and recondensation of $\mathrm{CO}$ vapour.  The region where $\mathrm{CO}$ has 
segregated out from $\mathrm{CO_2}$ is significantly thicker at the equator than at the poles ($187\,\mathrm{m}$ versus $38\,\mathrm{m}$). 
The $\mathrm{CO}$ pressure map reflects the location of the two fronts: a broader and lower pressure peak around the sublimation front, 
and a narrower and higher pressure peak around the segregation front.

The second 67P case was started with a fully opaque disk at $t=0.995\,\mathrm{Myr}$ that reached full transparency at $t_{\rm c}=1\,\mathrm{Myr}$. 
It was propagated to $t=1.7\,\mathrm{Myr}$. Because of the lower peak protosolar luminosity ($1.75L_{\odot}$ compared to 
$6.4L_{\odot}$ for the $t_{\rm c}=5\,\mathrm{kyr}$ case) the $\mathrm{CO}$ loss rate is lower. Nevertheless, $121\,\mathrm{kyr}$ after onset the 
model body has lost all condensed $\mathrm{CO}$. The peak temperature is sufficiently low to prevent significant segregation: 98\% of the 
$\mathrm{CO}$ trapped in the $\mathrm{CO_2}$ remained at the end of the simulation. The $\mathrm{CO:CO_2}$ mixture is still present at the 
surface, though the $\mathrm{CO}$ concentration has diminished by $\sim 13\%$. The losses of $\mathrm{CO_2}$ and amorphous water 
ice are insignificant, with both species being present in the top cell.

The third 67P case started at $t=2.995\,\mathrm{Myr}$, with the disk fully cleared up at $t_{\rm c}=3\,\mathrm{Myr}$, and it was first propagated 
until $t=4\,\mathrm{Myr}$. This is a time period during which the protosolar luminosity is expected to have been very similar to the current one. 
The model object lost all condensed $\mathrm{CO}$ in $169\,\mathrm{kyr}$, while $\mathrm{CO}$ segregation out from $\mathrm{CO_2}$, loss 
of $\mathrm{CO_2}$ through sublimation, and loss of amorphous water ice through crystallisation all were insignificant.

All 67P cases were run without activating the long--lived radionuclides. This can be motivated by considering Eq.~(11) of \citet{prialnikpodolak99} with parameters 
relevant for the current investigation, showing that the cooling rate exceeds the heating rate of long--lived radionuclides at diameters 
below $68\,\mathrm{km}$.

\subsection{The $\mathbf{D=74\,\mathrm{\bf km}}$ ``Hale--Bopp'' case} \label{sec_results_HB}

The Hale--Bopp case was started at $t=0$ and the disk cleared at $t_{\rm c}=5\,\mathrm{kyr}$. The body was spatially resolved by 109 radial cells 
(growing from $5\,\mathrm{m}$ thickness at the surface to $2\,\mathrm{km}$ at the core), and 18 latitudinal slabs. The model was 
propagated for $55.3\,\mathrm{Myr}$. Radiogenic heating by long--lived radionuclides is an important heat source 
for a $D=74\,\mathrm{km}$ body with $\mu=4$. The condensed $\mathrm{CO}$ was lost in $10.5\,\mathrm{Myr}$. 
The $\mathrm{CO}$ bound to $\mathrm{CO_2}$ is reduced to half on a $25\,\mathrm{Myr}$ time--scale and the 
amount was down to $13\%$ at the end of the simulation. Crystallisation was substantial as well, with half the 
amorphous water ice and the trapped $\mathrm{CO}$ and $\mathrm{CO_2}$ gone after $30\,\mathrm{Myr}$. 
At the end of the simulation, $32\%$ amorphous ice remained. The amount of condensed $\mathrm{CO_2}$ ice is 126\% of the 
amount present at the beginning of the simulation. The reason for that enhancement is that much of the $\mathrm{CO_2}$ 
that is driven out of amorphous water ice during crystallisation, unlike $\mathrm{CO}$, does not diffuse into space but 
recondenses within the nucleus. The $\mathrm{CO}$, $\mathrm{CO_2}$, and amorphous water abundances are final because 
a steady--state has been reach were the body is no longer evolving.

\begin{figure}
\begin{center}
     \scalebox{0.45}{\includegraphics{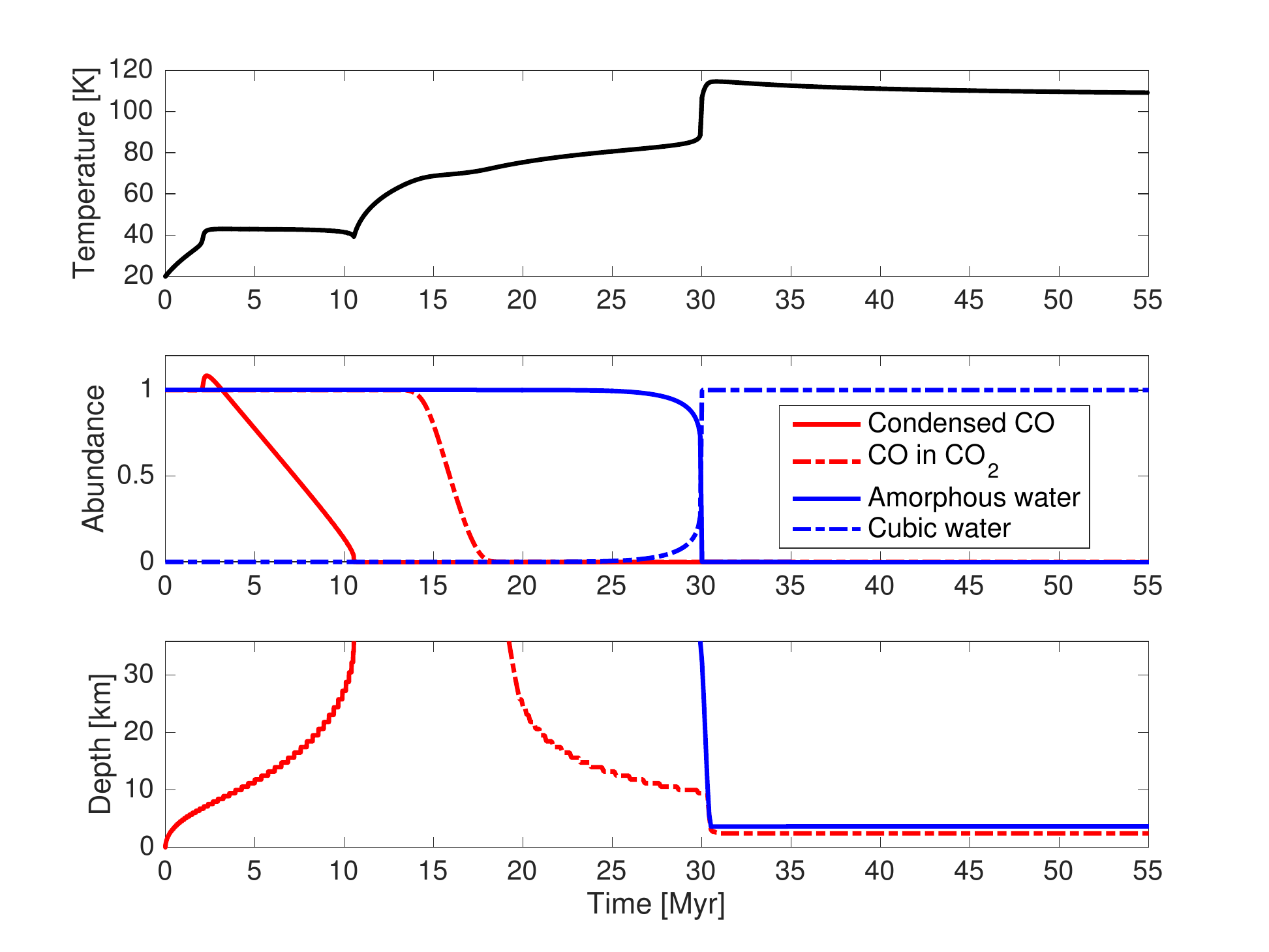}}
    \caption{The Hale--Bopp case. \emph{Upper panel:} The core temperature as function of time. \emph{Middle panel:} The core abundances 
of condensed $\mathrm{CO}$, $\mathrm{CO}$ trapped in $\mathrm{CO_2}$, and amorphous water ice, normalised to the local initial 
abundances of the respective species, versus time. The abundance of cubic water ice is shown as well, normalised to the local initial abundance of 
amorphous water ice. \emph{Lower panel:} The depths below the surface of the condensed $\mathrm{CO}$ sublimation front, the $\mathrm{CO}$ 
segregation front, and the crystallisation front, as functions of time, for the equator.}
     \label{fig_HB_time}
\end{center}
\end{figure}

The upper panel of Fig.~\ref{fig_HB_time} shows the core temperature as a function of time. The curve is best understood 
if compared with the core abundance, shown in the middle panel of Fig.~\ref{fig_HB_time}. The core is heating from its 
initial temperature $T_0=20\,\mathrm{K}$ during the first $2.2\,\mathrm{Myr}$. At that point, sublimation of $\mathrm{CO}$ 
at the core is initiated, establishing a steady temperature just below $43\,\mathrm{K}$ for the next $\sim 6\,\mathrm{Myr}$. 
During the following $\sim 2\,\mathrm{Myr}$ the core temperature falls somewhat and reaches a local minimum of $39\,\mathrm{K}$ 
at $t=10.5\,\mathrm{Myr}$ because of intensified $\mathrm{CO}$ sublimation. At this point all condensed $\mathrm{CO}$ is gone, and 
the temperature starts climbing rapidly to $69\,\mathrm{K}$ reached at $t=15\,\mathrm{Myr}$. At this point, $\mathrm{CO}$ segregation 
out of $\mathrm{CO_2}$ is initiated at the core. Because of the cooling effect the temperature stabilises anew for $\sim 3\,\mathrm{Myr}$ 
until the core segregation is completed at $t=18\,\mathrm{Myr}$. The core temperature rises to $80\,\mathrm{K}$ at $t\approx 25\,\mathrm{Myr}$. 
At this point, crystallisation is initiated, and the middle panel of Fig.~\ref{fig_HB_time} shows that the abundance of amorphous water starts to decline as the abundance 
of cubic water is rising. The abundances of $\mathrm{CO}$ and $\mathrm{CO_2}$ had been set so that the energy needed to 
liberate these species (taken as their latent heats of sublimation) should equal and neutralise the energy release of crystallisation 
(see Sec.~\ref{sec_parameters}). However, the saturation pressure of $\mathrm{CO_2}$ is extremely low at these temperatures, 
and much lower than the pressure of the released $\mathrm{CO_2}$ during crystallisation. As a consequence, the $\mathrm{CO_2}$ condenses 
and therefore releases its latent energy. This causes rapid heating and the temperature reaches $107\,\mathrm{K}$ 
at $t=30\,\mathrm{Myr}$. There is some further radiogenic net heating and the core temperature peaks at $T\approx 114\,\mathrm{K}$ at $t=30.8\,\mathrm{Myr}$. 
However, at this point a balance is reached between the radiogenic energy production rate and the thermal radiation cooling rate at the surface of 
the body. A steady--state temperature is reached at around $110\,\mathrm{K}$. There is no net $\mathrm{CO_2}$ sublimation, and it is 
also too cold for the cubic--to--hexagonal transition.

The lower panel of Fig.~\ref{fig_HB_time}  provides further insight into the mass loss processes. It shows the depths of the condensed 
$\mathrm{CO}$ sublimation front, the $\mathrm{CO}$ segregation front, and the crystallisation front, as functions of time. Condensed 
$\mathrm{CO}$ is being removed through combined protosolar and radiogenic heating, but the former process is dominating. Although the 
core abundance reduction of condensed $\mathrm{CO}$ is initiated at $t\approx 3.2\,\mathrm{Myr}$, the complete $\mathrm{CO}$ removal 
is a wave that spreads from the surface and downwards. When the core loses the last of its condensed $\mathrm{CO}$, the rest of the 
body is already free from pure $\mathrm{CO}$ ice. The segregation process, on the other hand, is a wave that spreads from the 
core and upwards. As previously mentioned, the core $\mathrm{CO_2}$ has lost all its $\mathrm{CO}$ at $t=18\,\mathrm{Myr}$. The 
region of $\mathrm{CO}$--free $\mathrm{CO_2}$ then expands outwards over the following $12\,\mathrm{Myr}$ until the 
wave grinds to a halt about $2.1\,\mathrm{km}$ under the surface. Likewise, crystallisation is a process that spreads from the 
core and outwards, albeit at a significantly faster rate -- the entire process takes $\sim 0.5\,\mathrm{Myr}$ and stops $3.6\,\mathrm{km}$ 
below the surface. Note that there are no movements of the segregation or crystallisation fronts at $t\geq 31\,\mathrm{Myr}$, showing 
that the body has reached a thermal steady--state. A warm interior and a substantially cooler thin near--surface region (explaining the survival of amorphous 
water ice and $\mathrm{CO}$--laden $\mathrm{CO_2}$ near the surface) is typical of radiogenically heated and radiatively cooled bodies 
\citep[e.g.][]{heveysanders06}.

 \begin{figure}
\begin{center}
     \scalebox{0.45}{\includegraphics{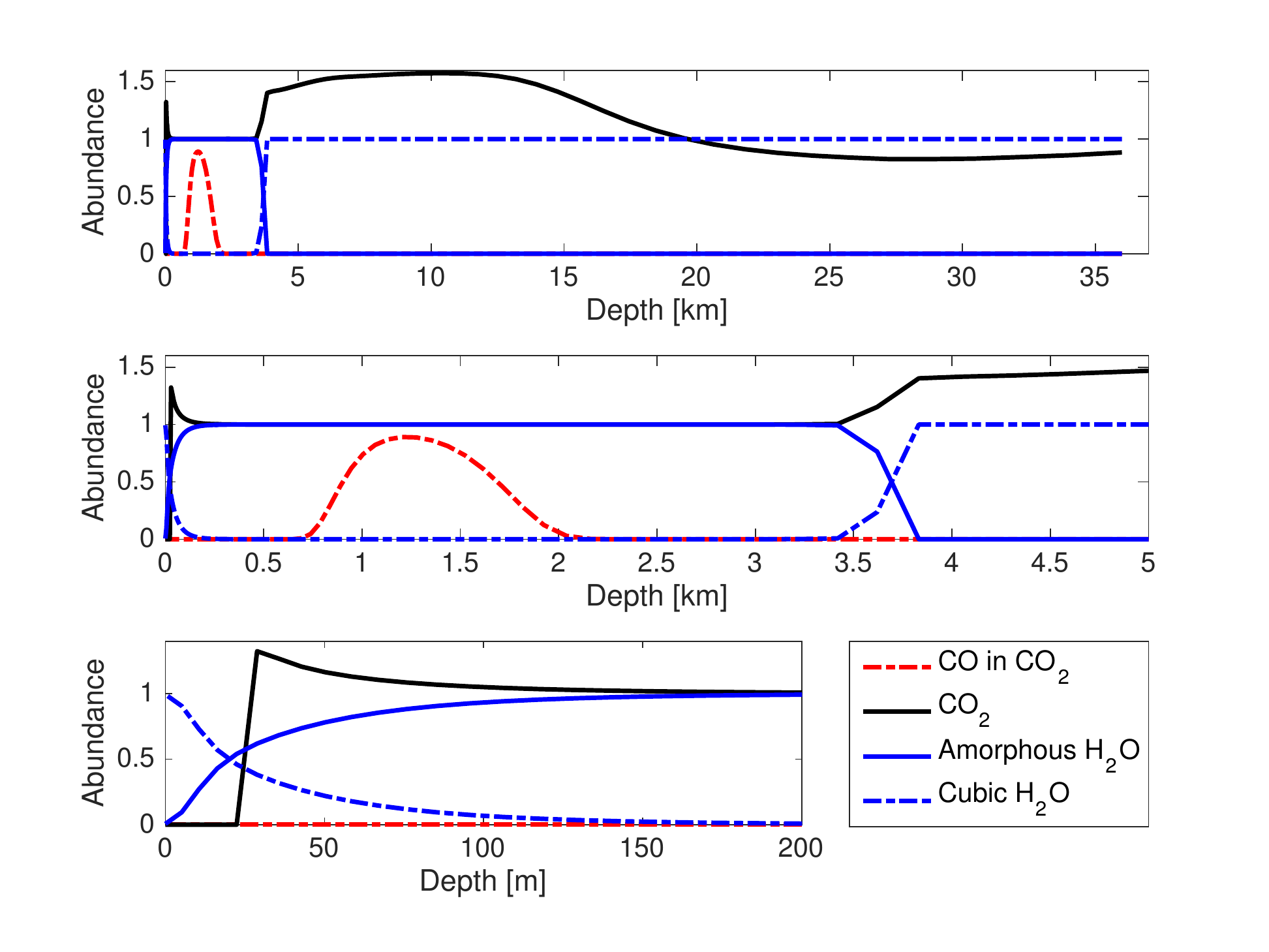}}
    \caption{The Hale--Bopp case. \emph{Upper panel:} The equatorial abundances versus depth of $\mathrm{CO}$ trapped in $\mathrm{CO_2}$, $\mathrm{CO_2}$ ice, 
amorphous water ice, normalised to their initial local abundances, at the end of the simulation. The abundance of cubic ice is shown as well, normalised 
to the initial local abundance of amorphous water ice. The upper panel shows the entire body. \emph{Middle panel:} The same as for the upper panel, 
but zooming on the upper $5\,\mathrm{km}$ of the body. \emph{Lower panel:} The same as for the upper panel, but zooming on the upper $200\,\mathrm{m}$ of the body.}
     \label{fig_HB_depth}
\end{center}
\end{figure}

Figure~\ref{fig_HB_depth} shows the abundances of $\mathrm{CO}$ trapped in $\mathrm{CO_2}$, $\mathrm{CO_2}$ ice, 
amorphous water ice, and cubic water ice, as functions of depth at the end of the simulation. The three panels show different levels of zoom on 
the surface region. Below a depth of $\sim 4\,\mathrm{km}$ the entire volume has crystallised and turned to cubic water ice. As already has 
been mentioned, the occluded $\mathrm{CO_2}$ driven out during this process has recondensed. However, Fig.~\ref{fig_HB_depth} shows 
that this recondensation has not necessarily been immediate and local. Instead, there has been a substantial upward diffusion of $\mathrm{CO_2}$ 
vapour before it recondensed, forming a $\leq 50\%$ mass enhancement $4$--$20\,\mathrm{km}$ below the surface. There has even 
been a small reduction of $\mathrm{CO_2}$ ice within a $15\,\mathrm{km}$--radius core, showing that some net sublimation of $\mathrm{CO_2}$ 
has taken place as well. That vapour has diffused upwards and contributed to the enhancement at depth $4$--$20\,\mathrm{km}$. 
Although the temperature at $2$--$4\,\mathrm{km}$ depth has not been sufficiently high to cause widespread crystallisation, it has been 
high enough to drive out $\mathrm{CO}$ from the $\mathrm{CO_2}$. The outermost $\sim 500\,\mathrm{m}$ of the body shows the effects 
of the early short--duration protosolar heating. Sublimation of $\mathrm{CO_2}$ ice has emptied the upper $\sim 20\,\mathrm{m}$. Note that 
a fraction of the liberated $\mathrm{CO_2}$ vapour has diffused downwards and created a somewhat elevated $\mathrm{CO_2}$ ice 
concentration down to a depth of $\sim 150\,\mathrm{m}$. Additionally, the entire region down to $\sim 150\,\mathrm{m}$ has experienced various 
levels of crystallisation.  At $\sim 20\,\mathrm{m}$ depth, the amount of amorphous water ice is about half of the original abundance, and it falls to 
zero at the surface. The only ice that would be visible at the surface of this Hale--Bopp analogue would be cubic water ice, at this stage of its existence. The 
protosolar heating has also driven out all $\mathrm{CO}$ from its $\mathrm{CO_2}$ host down to a depth of $\sim 700\,\mathrm{m}$. 
Because of the protosolar heating from above, and the radioactive heating from below, $\mathrm{CO}$--bearing $\mathrm{CO_2}$ ice is 
only present in a comparably thin layer located $0.7$--$2\,\mathrm{km}$ below the surface.

Note, that a later clearing time than $t_{\rm c}=5\,\mathrm{kyr}$ would have reduced and perhaps eliminated the crystallisation, 
segregation, and $\mathrm{CO_2}$ sublimation in the near--surface region. Also, if $\mu\ll 4$ the internal crystallisation, segregation, 
and $\mathrm{CO_2}$ re--distribution would become smaller and perhaps negligible. However, in this context it is worth remembering 
that C/1995 O1 (Hale--Bopp) is considered a dusty object, with $\mu\geq 5$ according to \citet{jewittmatthews99} and 
$\mu=5.1\pm 1.2$ according to \citet{dellorussoetal00}. Therefore, although the current simulation is not necessarily representative 
of all $D=74\,\mathrm{km}$--class bodies, it may describe the deep interior of Comet~Hale--Bopp itself fairly well.

\subsection{The $\mathbf{D=203\,\mathrm{\bf km}}$ ``Chiron/Phoebe'' case} \label{sec_results_CP}

Three versions of the Chiron/Phoebe case were considered, each having 91 spatial cells (growing from $5\,\mathrm{m}$ thickness at the surface to $8\,\mathrm{km}$ 
at the core) and 10 latitudinal slabs. The first considered the same time--dependent protosolar luminosity 
as the previously discussed models, starting at $t=0$, having a disk clearing at $t_{\rm c}=5\,\mathrm{kyr}$, and included long--lived radionuclides. 
A second model also included radiogenic heating but applied a fixed current--sun luminosity to better understand the sensitivity of the results 
to the heating sequence. A third model also considered current illumination conditions but removed the radiogenic heating. Such a model 
would be more representative of bodies having a substantially lower fraction of refractories than the other models considered here.

The radiogenically heated body exposed to protosolar heating was simulated during $40\,\mathrm{Myr}$. Compared to the Hale--Bopp case, 
the Chiron/Phoebe body has a $\sim 21$ times larger volume and a 2.7 times lower surface--to--volume ratio (i.~e., it is less capable of 
dissipating the radiogenic heat it produces, and is therefore heated to a higher temperature). Because of its lower porosity, the Chiron/Phoebe 
case body has a mass that is almost 30 times higher than the Hale--Bopp case body. Despite having 30 times more $\mathrm{CO}$ to 
get rid of, the time--scale for doing that is just $12.7\,\mathrm{Myr}$, marginally longer than the $10.5\,\mathrm{Myr}$ time--scale for the 
Hale--Bopp case. At the end of the simulation only $1.7\%$ of the $\mathrm{CO_2}$--trapped $\mathrm{CO}$ remains, and $7.2\%$ of 
the amorphous ice. The condensed $\mathrm{CO_2}$ is 137\% of the initial abundance because vapour released during crystallisation eventually 
recondenses with very small losses to space.

\begin{figure*}
\centering
\begin{tabular}{cc}
\scalebox{0.45}{\includegraphics{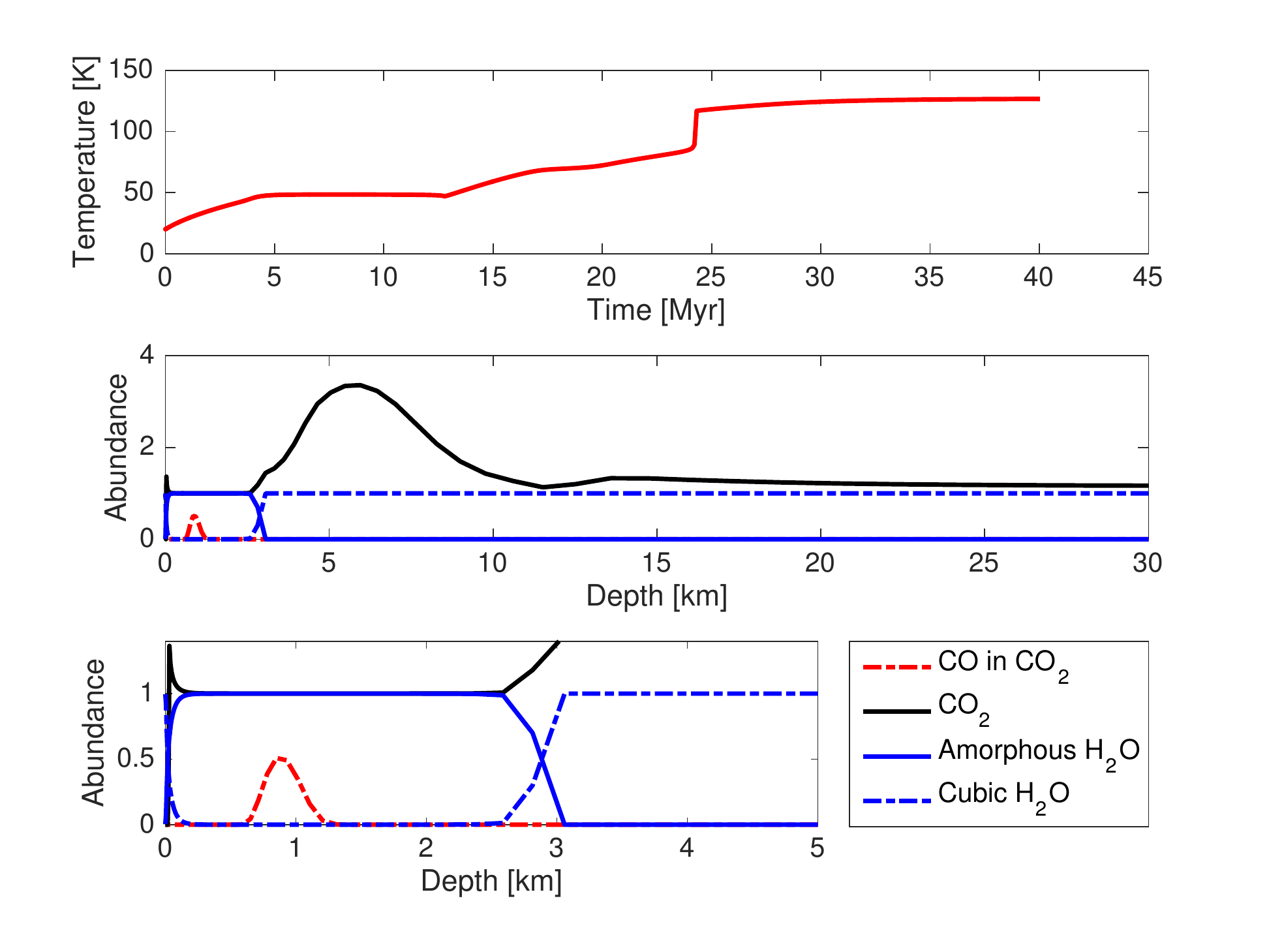}} & \scalebox{0.45}{\includegraphics{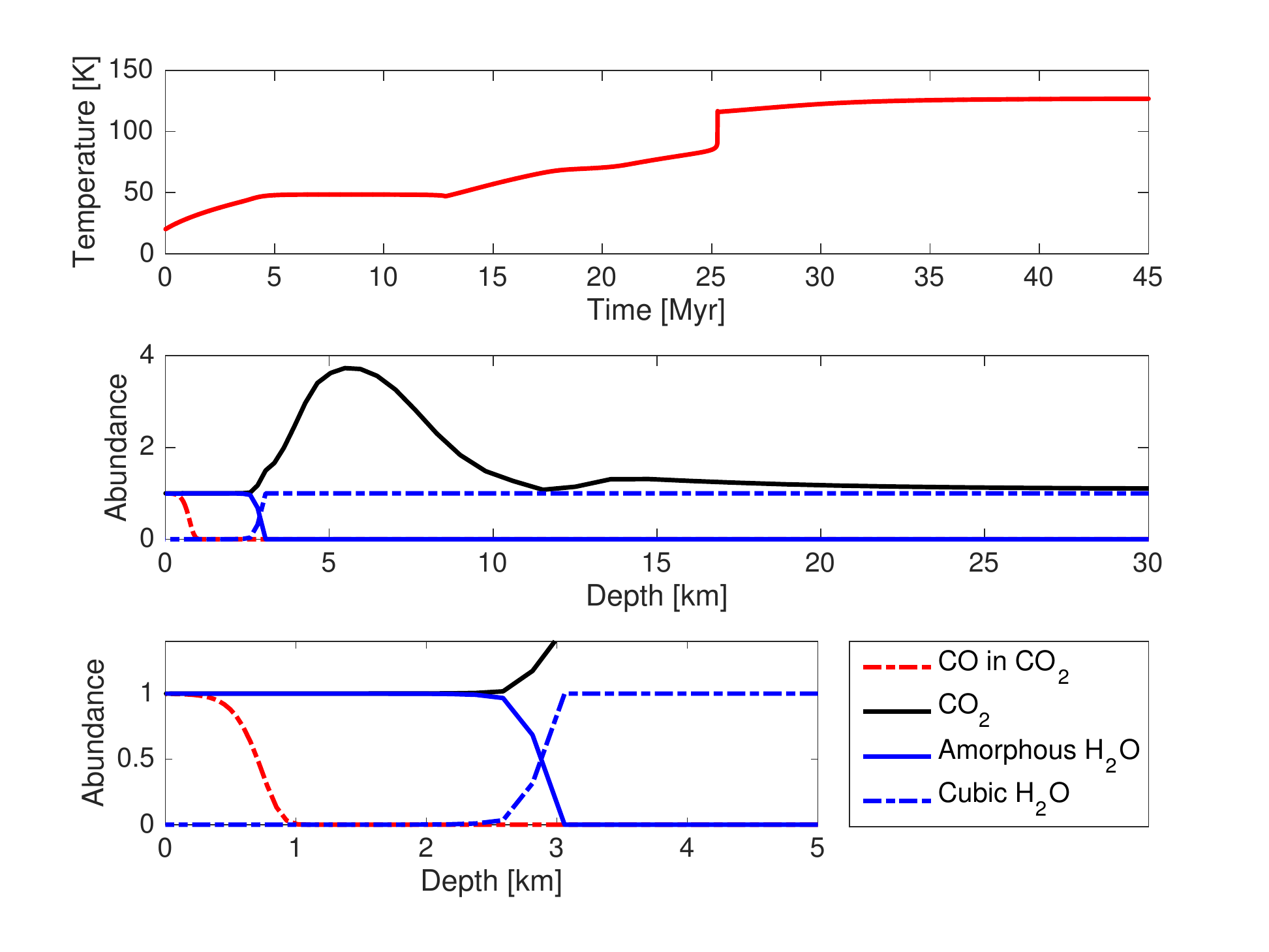}}\\
\end{tabular}
     \caption{Chiron/Phoebe cases, both including long--lived radionuclides. The left panels shows a $t_{\rm c}=5\,\mathrm{kyr}$ version with protosolar heating, 
the right panels assume the contemporary solar luminosity. The top panels show core temperature vs. time, the middle panels show 
equatorial composition vs. depth at the end of the simulation (the bottom panels are zooms of the middle ones).}
     \label{fig_cp}
\end{figure*}

The left panel of Fig.~\ref{fig_cp} shows the core temperature versus time, and the final distribution of ices as function of depth at 
the end of the simulation. The temperature curve is reminiscent of the Hale--Bopp case (Fig.~\ref{fig_HB_time}), but the values are 
higher. The plateau caused by $\mathrm{CO}$ ice sublimation is $\sim 48\,\mathrm{K}$ (compared to $43\,\mathrm{K}$ for 
the Hale--Bopp case). Because the $\mathrm{CO}$ vapour needs to diffuse over a longer distance, and because the lower core porosity 
reduces the diffusivity, the Chiron/Phoebe body evolve into having higher $\mathrm{CO}$ pressures and sublimation temperature than 
the Hale--Bopp case. The late steady--state temperature reaches $\sim 126\,\mathrm{K}$, compared to $110\,\mathrm{K}$ 
for the Hale--Bopp case because of its smaller surface--to--volume ratio. Due to relatively high internal temperatures, all amorphous ice 
below a depth of $\sim 3\,\mathrm{km}$ has crystallised (it is too cold for the cubic--to--hexagonal transformation). The $\mathrm{CO_2}$ 
vapour release during crystallisation has diffused upwards and recondensed in a zone located 3--$10\,\mathrm{km}$ below the surface, 
where the $\mathrm{CO_2}$ ice abundance has increased by up to a factor 3.4 above the initial abundance. The radiogenic 
heating has caused segregation of all $\mathrm{CO}$ out of $\mathrm{CO_2}$ ice below a depth of $\sim 1.2\,\mathrm{km}$.

Because of the intense protosolar heating in the first few Myr of the simulation the abundance of amorphous water ice falls from 
nearly the original value at a depth of $170\,\mathrm{m}$ to zero at the surface. The abundance of cubic water ice increases from 
zero to unity over the same near--surface region. The $\mathrm{CO_2}$ ice in the uppermost $\sim 30\,\mathrm{m}$ has sublimated. 
Furthermore, the $\mathrm{CO}$ has been segregated out of $\mathrm{CO_2}$ in the top $650\,\mathrm{m}$. That 
means that the only remaining $\mathrm{CO}$--bearing $\mathrm{CO_2}$ ice is found in a $\sim 600\,\mathrm{m}$ thick zone 
located roughly $1\,\mathrm{km}$ under the surface.

The second Chiron/Phoebe case, shown in the right panel of Fig.~\ref{fig_cp}, was propagated for $45\,\mathrm{Myr}$. By employing a 
fixed current--day solar luminosity, this model avoids the early scorching and is roughly equivalent of using $t_{\rm c}=3\,\mathrm{Myr}$. 
The time--scale for losing all condensed $\mathrm{CO}$ ($12.8\,\mathrm{Myr}$), the amounts of amorphous water ice ($7.1\%$), 
condensed $\mathrm{CO_2}$ ($137\%$), and $\mathrm{CO}$ trapped in $\mathrm{CO_2}$ (1.9\%) are very similar to that of the 
previously discussed Chiron/Phoebe case. That is because radiogenic heating is the major mechanism that drives the evolution. 
The only significant difference between the models are found at the surface: the entire upper $\sim 3\,\mathrm{km}$ layer is rich 
in amorphous water ice, the entire upper $\sim 1\,\mathrm{km}$ layer is additionally rich in $\mathrm{CO}$--bearing $\mathrm{CO_2}$, 
which means that amorphous water ice and $\mathrm{CO_2}$ ice are present at the very surface, both trapping $\mathrm{CO}$.

Finally, a Chiron/Phoebe case without heating by long--lived radionuclides was considered, assuming a current--day solar luminosity. 
As expected, the loss of condensed $\mathrm{CO}$ is comparably slow in this case. By $t=50\,\mathrm{Myr}$ about $42\%$ of the $\mathrm{CO}$ 
remained, falling to $25\%$ at $t=70\,\mathrm{Myr}$. At the end of the simulation ($t=100\,\mathrm{Myr}$) about $8\%$ of the condensed $\mathrm{CO}$ 
still remained (the time--scale of complete loss is estimated to be $200\,\mathrm{Myr}$). The model body only lost $3.3\%$ of the $\mathrm{CO}$ stored in 
$\mathrm{CO_2}$, it lost  $0.1\%$ of its $\mathrm{CO_2}$, and the amount of crystallisation was even smaller. 

\begin{table*}
\begin{center}
\begin{tabular}{||l|r|r|r|r|r|r|r||}
\hline
\hline
Model case & Diameter & $T_{\rm surf}$ & $T_{\rm core}$ & Cond. $\mathrm{CO}$ & $\mathrm{CO}$ in $\mathrm{CO_2}$ & Cond. $\mathrm{CO_2}$ & Amorp. $\mathrm{H_2O}$\\
\hline
67P, $t_{\rm c}=5\,\mathrm{kyr}$ & $4\,\mathrm{km}$ & $97.2\,\mathrm{K}$ & $81.9\,\mathrm{K}$ & $73\,\mathrm{kyr}$ & 0\% & 96.9\% & 94.6\%\\
67P, $t_{\rm c}=1\,\mathrm{Myr}$ & $4\,\mathrm{km}$ & $70.4\,\mathrm{K}$& $63.5\,\mathrm{K}$ & $121\,\mathrm{kyr}$ & 98.3\% & 99.9\% & 100\%\\
67P, $t_{\rm c}=3\,\mathrm{Myr}$ & $4\,\mathrm{km}$ & $60.9\,\mathrm{K}$ & $55.4\,\mathrm{K}$ & $169\,\mathrm{kyr}$ & 99.9\% & 100\% & 100\%\\
\hline
Hale--Bopp & $74\,\mathrm{km}$ & $97.1\,\mathrm{K}$ & $114.5\,\mathrm{K}$ & $10.5\,\mathrm{Myr}$ & 13.2\% & 126\% & 32.5\%\\
\hline
Chiron/Phoebe & $203\,\mathrm{km}$ & $97.3\,\mathrm{K}$ & $126.6\,\mathrm{K}$ & $12.7\,\mathrm{Myr}$ & 1.7\% & 136\% & 7.2\%\\
Protosun, LLR & & & & & & &\\
Chiron/Phoebe & $203\,\mathrm{km}$ & $61.3\,\mathrm{K}$ & $126.7\,\mathrm{K}$ & $12.8\,\mathrm{Myr}$ & 1.9\% & 137\% & 7.1\%\\
Current Sun, LLR & & & & & & &\\
Chiron/Phoebe & $203\,\mathrm{km}$ & $62.8\,\mathrm{K}$ & $43.6\,\mathrm{K}$ & $\sim 200\,\mathrm{Myr}$ & 100\% & 100\% & 100\%\\
Current Sun, no LLR & & & & & & &\\
\hline 
\hline
\end{tabular}
\caption{Results of the simulations, for models specified in the first two columns. The third and fourth column give peak equatorial surface temperatures, 
and peak core temperatures, respectively. The fifth column gives the time--scale of complete loss of condensed $\mathrm{CO}$.  The following columns 
provide the percentile abundances of $\mathrm{CO}$ trapped in $\mathrm{CO_2}$, free $\mathrm{CO_2}$ ice, and amorphous water ice at the 
end of the simulation compared to the initial amount. Note that free $\mathrm{CO_2}$ ice can reach above $100\%$ if $\mathrm{CO_2}$ vapour is released 
by amorphous water ice and it condenses. LLR: long--lived radionuclides.}
\label{tab_results}
\end{center}
\end{table*}

\section{Discussion} \label{sec_discussion}

The simulations presented in Sec.~\ref{sec_results} rely on \texttt{NIMBUS} and not \texttt{NIMBUSD}, i.~e., erosion of near--surface 
solids due to the gas flow is not accounted for. This can be motivated as follows. When the reduction from full to negligible opacity of the 
Solar Nebula takes a sufficient amount of time, the CO front withdraws 
underground so slowly that the outward force associated with the gas pressure is too weak to break up and eject the surface material. 
Once the surface is fully exposed to the protosolar early luminosity, the CO is already so far underground that pressure 
gradients are negligible compared to material cohesion. Regarding $\mathrm{CO_2}$, its volatility is so low, that it is unable 
to cause significant erosion even when being exposed to $6.4\,L_{\odot}$ at $r_{\rm h}=23\,\mathrm{AU}$ when present at the surface. 
To illustrate these points, a version of the $\{D=4\,\mathrm{km},\,t_{\rm c}=5\,\mathrm{kyr}\}$ was run with much finer spatial resolution 
($1\,\mathrm{cm}$ surface cells) to evaluate outward forces more accurately. The model was run for $60\,\mathrm{yr}$ (about half an orbit) 
from $t=1\,\mathrm{kyr}$ (when the effective luminosity was $0.16L_{\odot}$, corresponding to illumination at $r_{\rm h}=57\,\mathrm{AU}$ in the 
current Solar System). After $60\,\mathrm{yr}$, the CO front had withdrawn to $0.74$--$4.35\,\mathrm{m}$ depending on latitude, and the strongest force 
recorded was merely $0.02\,\mathrm{N}$. At that point, the model object was subjected to the full $6.4\,L_{\odot}$, causing a force spike of 
$15\,\mathrm{N}$ that rapidly decreased as the CO drew even deeper. The $15\,\mathrm{N}$ force is a generous upper limit, as the CO in reality would have withdrawn 
even deeper before Solar Nebula clearing at $t_{\rm c}=5\,\mathrm{kyr}$. The force due to $\mathrm{CO_2}$ sublimation from within the top cell 
peaked at $7\,\mathrm{N}$, and declined rapidly as well. Considering that dust mantle analogues with a porosity of $85\%$ have a tensile strength 
of $\sim 1\,\mathrm{kPa}$ according to laboratory measurements \citep{guettlereta09}, vapour escape without erosion and ejection of solids is a reasonable assumption. 

Furthermore, in their modelling of Centaur (2060)~Chiron, \citet{capriaetal00b} 
found that the gas flow only was capable of removing loose dust on the surface, that sustained weak activity at perihelion and waned after a few orbits (as sublimation and crystallisation 
fronts withdrew underground). With the water ice remaining intact at the low temperatures of Chiron, there was no mechanism in the model to liberate 
the grains and to replenish the surface deposit of loose dust. These simulations included CO ice and were made for heliocentric distances that 
overlap and somewhat subceeds the (effectively) $r_{\rm h}\geq 9.1\,\mathrm{AU}$ simulations in Sec.~\ref{sec_results} (Chiron has a perihelion distance of $8.5\,\mathrm{AU}$). 
One mechanism that possibly could remove water ice in a surface layer and create a weakly cohesive mantle of dust is solar wind sputtering. 
In laboratory measurements of that process, \citet{munteanetal16} found that a $\sim 1.5\,\mathrm{cm}$ layer of water ice could be 
removed in objects located at $\mathrm{42\,\mathrm{AU}}$ during the Solar System lifetime, constituting an upper limit on the thickness of dust mantles formed accordingly, for objects in 
the Scattered Disk that later become Centaurs. A global $1.5\,\mathrm{cm}$ layer of dust (with an assumed density of $1500\,\mathrm{kg\,m^{-3}}$) 
would be able to sustain the observed dust production rate of Chiron \citep[$\sim 1\,\mathrm{kg\,s^{-1}}$;][]{luujewitt90b} for $\sim 10^5\,\mathrm{yr}$. 
Other Centaurs have dust production rates of up to order $10^2$--$10^3\,\mathrm{kg\,s^{-1}}$ \citep{jewitt09}, that could be sustained for at most 
$10^2$--$10^3\,\mathrm{yr}$ (considering that most Centaurs are smaller than Chiron). A working hypothesis is therefore that Centaurs (the best analogues 
of Primordial Disk objects) do not experience large--scale erosion, but merely removal of minor unreplenishable deposits of surface dust. This further motivates the 
assumption in Sec.~\ref{sec_results} that erosion can be ignored.

Based on the \texttt{NIMBUS} simulations it is possible to address the questions formulated in Sec.~\ref{sec_intro}. 
The questions are here re--stated for convenience.

\begin{trivlist}
\item Q\#1. Under what conditions is loss of condensed $\mathrm{CO}$ ice complete or partial in the Primordial Disk? 
If loss of condensed $\mathrm{CO}$ ice is complete, what are the time--scales for that loss in bodies of different size in the Primordial Disk? 
\end{trivlist}

According to the simulations in Sec.~\ref{sec_results}, the loss of condensed CO proceeds at an exponentially declining 
rate. If the bodies remain at $23\,\mathrm{AU}$ sufficiently long, the process runs to completion. The current simulations therefore 
do not support the hypothesis of \citet{desanctisetal01}, who suggested that a core of condensed CO might be preserved if 
the body reaches a state where sublimation is balanced by condensation. Such a quasi--stationary condition does not arise, at 
least not for the combination of model parameters applied in the current work. One scenario that would allow for a long--lived 
reservoir of $\mathrm{CO}$ ice is the existence of an internal compacted shell or core with sufficiently low porosity to prevent 
diffusion from that region (at least until the temperature and gas pressure become high enough to forcefully create channels through 
which vapour can diffuse). However, considering that self--gravity is not capable of removing porosity even in $200\,\mathrm{km}$--class 
bodies (see Fig.~\ref{fig_porous}) it is hard to imagine that such compacted interiors would be common, particularly for JFC--sized objects. 
The time--scale for complete loss of $\mathrm{CO}$ ice is about 70--$170\,\mathrm{kyr}$ for a $D=4\,\mathrm{km}$ body, depending on 
the clearing time of the Solar Nebula (Table~\ref{tab_results}). The exact number depends on the actual CO abundance, conductivity, and 
diffusivity of the body, but a loss time--scale of order $0.1\,\mathrm{Myr}$ is realistic for JFC--sized bodies. Bodies with sizes ranging 
between those of Hale--Bopp and Chiron have CO ice loss  time--scales of roughly 10--$13\,\mathrm{Myr}$ when being relatively dust--rich ($\mu=4$). 
This time--scale is comparable to the shortest proposed lifetime of the Primordial Disk \citep[e.~g., $\sim 15\,\mathrm{Myr}$ according to][]{nesvornyandmorbidelli12}. 
If the Primordial Disk indeed was dispersed that early, it is likely that bodies in the $D=70$--$200\,\mathrm{km}$ 
range could preserve a fraction of their original CO ice content, at least if they were less dusty than considered in the current simulations. Because the loss is 
partially driven by radiogenic heating, the long--term survival of CO ice in such bodies depends largely on what heliocentric distance they are scattered to. 
Bodies ending up in the Oort Cloud would obtain sufficiently low near--surface temperatures for CO vapour diffusing from the deep interior to condense 
near the surface (in a fashion akin to the near--surface $\mathrm{CO_2}$ concentrations seen in Figs.~\ref{fig_HB_depth} and \ref{fig_cp}). Bodies ending up 
in the Kuiper Belt would likely lose all condensed CO, considering the work by \citet{desanctisetal01} and \citet{choietal02} in combination with the 
results in Sec.~\ref{sec_results}. Sufficiently large and dust--poor bodies ending up in the Scattered Disk may have been able to conserve some CO ice 
if their semi--major axes are sufficiently large. However, the apparent lack of activity beyond $\sim 12\,\mathrm{AU}$ \citep{bockeleemorvanetal01, jewittetal08, jewitt09, lietal20} suggests that such bodies are not common among Centaurs (and by inference, the Scattered Disk) or Kuiper Belt objects. 
The most likely scenario is therefore that $D\leq 200\,\mathrm{km}$ bodies indeed lost their CO ice, which places constraints on the Primordial Disk lifetime: it could 
not have had a lifetime shorter than 10--$13\,\mathrm{Myr}$ (unless the initial CO abundance was significantly lower than considered here). If the bodies 
contained less radionuclides than assumed in this work, the Primordial Disk lifetime increases accordingly. If the actual dust--to--water--ice mass ratio $\mu$ of 
Centaurs could be determined accurately (being the former Primordial Disk large members that can be reached by spacecraft with less difficulty), thermophysical simulations  
of the type considered in this paper could be used to determine the shortest allowable Primordial Disk lifetime. If the Primordial Disk disruption was associated with the 
Late Heavy Bombardment, all bodies with $D\leq 200\,\mathrm{km}$ would have lost all their CO ice prior to disruption (by $200\,\mathrm{Myr}$, or after at most half of 
the Primordial Disk lifetime), even in the unrealistic scenario that they are dust--free and experienced no radiogenic heating at all.

Observations of debris disks around young stars sometimes reveal emission of CO gas from within those structures. Confirmed cases in systems of known 
age include NO~Lup \citep[$2\pm 1\,\mathrm{Myr}$;][]{lovelletal20}, TWA~7  \citep[$10\,\mathrm{Myr}$;][]{matraetal19}, HD~138813 \citep[$11\,\mathrm{Myr}$;][]{liemansifryetal16, halesetal19}, 
HD~129590 \citep[10--$16\,\mathrm{Myr}$;][]{kraletal20}, HD~131835 \citep[$16\,\mathrm{Myr}$;][]{mooretal15, liemansifryetal16, halesetal19}, HD~181327 \citep[$23\pm 3\,\mathrm{Myr}$;][]{marinoetal16}, 
$\beta$~Pictoris \citep[$23\pm 3\,\mathrm{Myr}$;][]{matraetal17b}, and Fomalhaut \citep[$440\pm 40\,\mathrm{Myr}$;][]{matraetal17}. This CO is not considered primordial by these 
authors (in the sense of having remained in gas form since the time of formation), but originates from exocomets that store the CO as ice. These CO disks often extend up to 
100--$250\,\mathrm{AU}$ from the central star, beyond the CO snow line for most systems. These authors therefore suggest that CO ice from the interior of exocomets is being 
exposed during collisional cascades in these debris disks, and is vaporised through UV desorption. That may certainly be the case, but the simulations in Sec.~\ref{sec_results} suggest 
that an additional CO gas source could be radiogenically heated planetesimals, perhaps in combination with cratering that penetrates surface regions where CO might recondense, thus 
allowing for venting of CO gas from deeper and warmer regions.  Such a source of $\mathrm{CO}$ is potentially preferable in systems where the gas disk extends beyond the debris 
disk (i.e., presence of CO vapour in regions without detectable evidence of high collisional activity), and could explain why gas--rich disks are not preferentially associated with dust--rich debris disks \citep{liemansifryetal16}. 
Bodies in the 70--$200\,\mathrm{km}$ class with $\mu=4$ can sustain CO release for 10--$13\,\mathrm{Myr}$, but that time--scale can easily be extended if the bodies are less dusty 
(i.~e., the CO release is slowed down by a smaller radiogenic power) or simply have larger size. 

It is clear that Primordial Disk bodies with sizes of a few kilometres stand no chance of still containing primordial CO ice at the time the disk 
was dispersed and some of them were placed in the Oort Cloud. The notion that dynamically new comets are pristine in the sense that they 
still carry all species they formed with and have not experienced any form of global--scale processing, is therefore not necessarily correct. 
If a dynamically new comet contains $\mathrm{CO}$ ice there are 
at least two possible reasons: 1) the comet experienced heating strong enough to cause segregation and/or crystallisation just prior to 
its ejection, and the released $\mathrm{CO}$ condensed near the surface when the object reached sufficiently large heliocentric distances; 
2) the comet did not originate from the Primordial Disk. The first scenario requires that the object spent some time substantially closer to 
the Sun than $23\,\mathrm{AU}$ prior to ejection, because both $\mathrm{CO:CO_2}$ mixtures and amorphous ice are stable at that distance. 
The viability of that scenario must be tested in future simulations, because it is not clear whether the CO has time to escape before sufficiently large 
distances are reached, or if the accumulated amount of condensed CO is sufficiently large to drive observed levels of activity. Furthermore, if 
this was a common and efficient process, one might have expected a larger number of active comets at large distances. One version of the second scenario concerns 
planetesimals that formed within the giant planet zone and were moved to the Oort Cloud during 
the first $\sim 0.1\,\mathrm{Myr}$ after the Solar Nebula cleared up (even shorter time--scales may be necessary due to the higher level of 
protosolar illumination compared to $23\,\mathrm{AU}$). The work of \citet{brasseretal07} shows that an object with a perihelion near Jupiter 
needs to reach $a\approx 10\,\mathrm{AU}$ to avoid orbit circularisation by Solar Nebula gas, and it needs $a\stackrel{>}{_{\sim}}60$--$200\,\mathrm{AU}$ 
(depending on inclination) to experience changes to $a$ during close encounters with Jupiter that are of a size similar to $a$ itself (i.~e., enabling a rapid 
increase of $a$ that allows stars in the birth cluster and galactic tides to raise the perihelion out from the planetary region and incorporate it into the Oort Cloud). 
In order to successfully create an Oort Cloud orbit, this process must necessarily take place before the Solar Nebula gas drag has time to move the perihelion 
away from Jupiter's orbit. That time--scale is $\sim 10\,\mathrm{kyr}$ if $\imath\approx 0^{\circ}$ but approaches $100\,\mathrm{kyr}$ if $\imath\approx 30^{\circ}$ 
\citep{brasseretal07}. At Saturn, $a>20$--$100\,\mathrm{AU}$ is needed to avoid circularisation, $a\geq 300$--$2000\,\mathrm{AU}$ is needed to 
initiate large orbital perturbations, and Oort Cloud insertion must take place in less than $0.1$--$1\,\mathrm{Myr}$. For Neptune, the insertion must take 
place within $1$--$10\,\mathrm{Myr}$. The time--scale of Oort Cloud injection by Jupiter is shorter than the CO loss time for $D=4\,\mathrm{km}$ bodies, therefore constituting a 
mechanism that could create a subset of Oort Cloud comets that actually contain CO ice. The time--scales for Saturn, Uranus, and Neptune are longer, 
but it must be remembered that once $a$ has reached values high enough to initiate the Oort Cloud insertion sequence, a very small fraction of the time is 
spent at distances small enough to allow for efficient $\mathrm{CO}$ loss. It is therefore possible that all giant planets contributed to populating the Oort Cloud 
with some objects rich in CO ice. However, the fraction of such objects in the Oort Cloud could be small. The number of planetesimals in the giant planet region 
that avoided accretion may have been low, and of these merely 1--$4\%$ end up in the Oort Cloud according to the simulations of \citet{brasseretal07}. 
Furthermore, objects with $D\stackrel{<}{_{\sim}}2\,\mathrm{km}$ never reach the Oort Cloud because the Solar Nebula gas drag is too strong. Comet C/2017~K2 
could be a relatively rare example of a body that formed among the giant planets and was inserted into the Oort Cloud long before the planetary gravitational instability 
took place and the Primordial Disk was disrupted. If that is the case, it also shows that the giant planets had grown sufficiently large to feed the Oort Cloud 
before the Solar Nebula in that region became transparent, otherwise Comet C/2017~K2 would not have contained $\mathrm{CO}$ ice. A second version of 
the second scenario is that some, perhaps a majority, of Oort Cloud objects were captured from foreign stellar system in the solar birth cluster \citep{levisonetal10}. 
If that is the case, the current simulations offer little insight into their histories.

As mentioned in Sec.~\ref{sec_intro},  segregation of CO out from $\mathrm{CO_2}$ could be another potential reason for 
unusually distant activity,  in addition to, or instead of, $\mathrm{CO}$ ice. The second group of questions concern $\mathrm{CO:CO_2}$ mixtures.

\begin{trivlist}
\item Q\#2. What degree of $\mathrm{CO}$ release from $\mathrm{CO_2}$ entrapment is expected to take place in the Primordial Disk? 
Specifically, at what depth is the segregation front expected to be located?  
\end{trivlist}

According to Sec.~\ref{sec_results} simulations, small JFC--like objects could have segregated fully if the Solar Nebula cleared very early, 
while almost complete preservation takes place if $t_{\rm c}\geq 1\,\mathrm{Myr}$. $D=70\,\mathrm{km}$--class objects preserve roughly 
$\sim 10\%$ while $D=200\,\mathrm{km}$--class objects preserve about $\sim 1\%$ $\mathrm{CO_2}$--trapped $\mathrm{CO}$, respectively, 
when $\mu=4$ (Solar Nebula clearing time matters little at these sizes). If radiogenic heating is insignificant, preservation is essentially complete. 
Admittedly, exact numbers depend on the preliminary $\{\Gamma_{\rm B},\,E_{\rm seg}\}$--values assigned in Sec.~\ref{sec_model_COfromCO2}. 
However, the $\{D=4\,\mathrm{km},\,t_{\rm c}=5\,\mathrm{kyr}\}$ object reaches peak temperatures of $82\,\mathrm{K}$ and $97\,\mathrm{K}$ at 
the core and on the surface, respectively, and the laboratory measurements of \citet{lunaetal08} and \citet{satorreetal09} show that no 
hypervolatile can survive within $\mathrm{CO_2}$ at such temperatures. The $\{D=4\,\mathrm{km},\,t_{\rm c}=1\,\mathrm{Myr}\}$ object never 
exceeds $64\,\mathrm{K}$ at the core, where $\mathrm{CO_2}$ entrapment is very stable. With surface temperatures reaching $70\,\mathrm{K}$ the 
level of segregation in the surface region could be substantial, and perhaps more so than the current simulations suggests. Yet, it seem clear that 
small objects exposed to the protosun within the first million years could display a very wide range of $\mathrm{CO:CO_2}$ abundances, while later 
exposure likely allows for primordial abundance--levels. Small objects with sufficiently large $t_{\rm c}$ would have $\mathrm{CO:CO_2}$ mixtures 
throughout their volumes, including the surface. The 70--$200\,\mathrm{km}$--class models studied here would have $\mathrm{CO:CO_2}$ mixtures 
confined to the upper $1$--$2\,\mathrm{km}$ (because of radiogenic heating if as dusty as $\mu=4$), and potentially lack $\mathrm{CO:CO_2}$ in 
the upper few hundred meters if exposed to the protosun very early.

The suggestion that Comet 67P contains intimate $\mathrm{CO_2}:\mathrm{CO}$ mixtures \citep{gascetal17} excludes 
an early Primordial Disk exposure to intense protosolar radiation. The need for an extended opaque era may imply a late formation time for comets as well, considering 
that planetesimal formation and Solar Nebula clearing necessarily are related. In this context, ``late'' means $\stackrel{>}{_{\sim}} 1\,\mathrm{Myr}$ after CAI 
(potentially earlier), which still is consistent with early giant planet formation. If cometary $\mathrm{CO_2}:\mathrm{CO}$ mixtures are confirmed, and 
our knowledge of the segregation process is improved, the lack of $\mathrm{CO}$ ice and presence of $\mathrm{CO_2}$--trapped $\mathrm{CO}$ may 
provide important means of dating the clearing of the Solar Nebula. It should also be remembered, that 67P/C--G may have formed in the outermost 
regions of the Primordial Disk \citep[considering its unusually high fraction of deuterated water;][]{altweggetal15}. Comets formed closer to the Sun could 
have experienced earlier formation and Solar Nebula clearing.

\begin{figure*}
\centering
\begin{tabular}{cc}
\scalebox{0.45}{\includegraphics{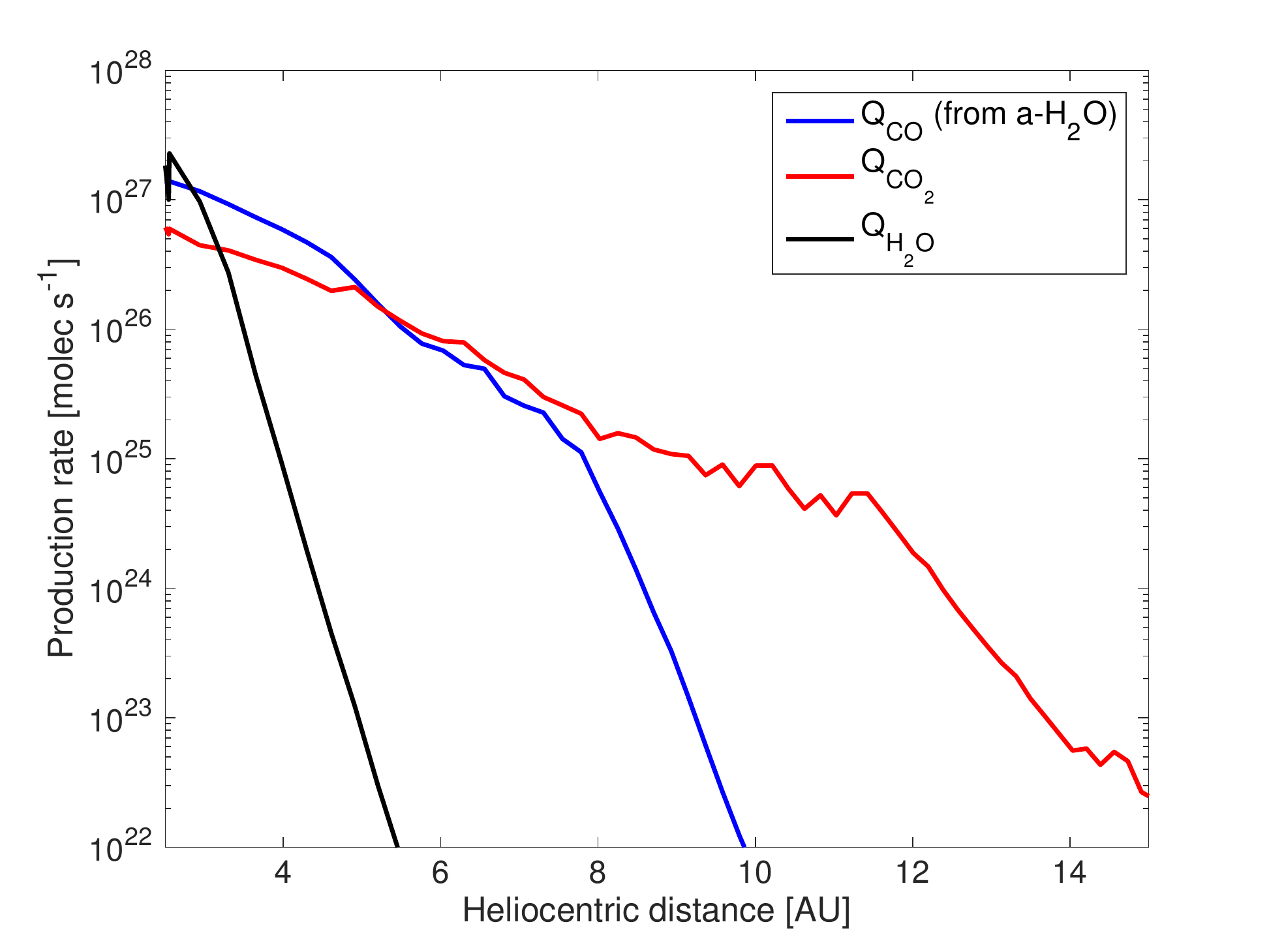}} & \scalebox{0.45}{\includegraphics{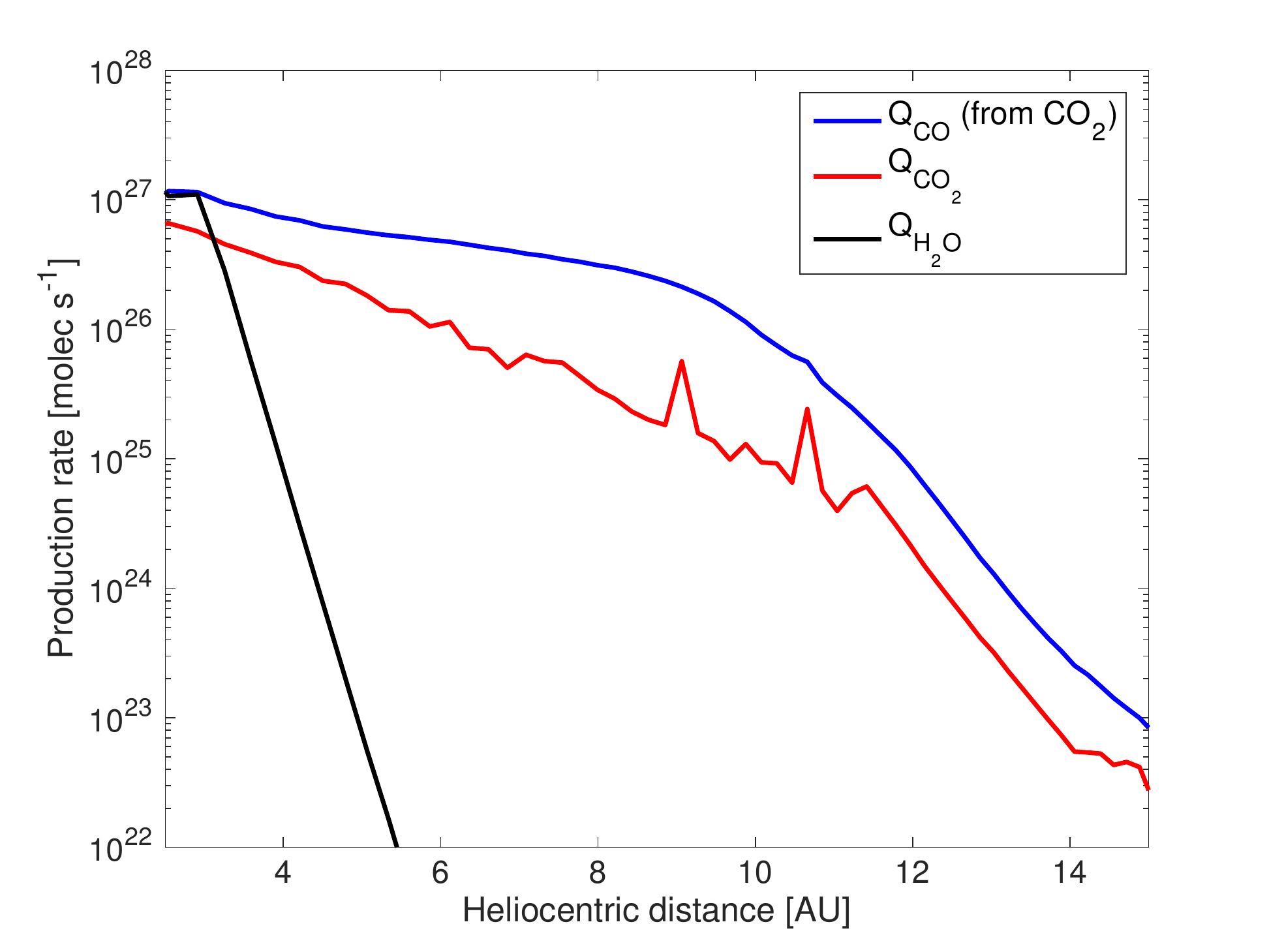}}\\
\end{tabular}
     \caption{Inbound production rates of $\mathrm{CO}$, $\mathrm{CO_2}$, and $\mathrm{H_2O}$ for a $D=4\,\mathrm{km}$ dynamically new comet with $q=0.25\,\mathrm{AU}$ 
and $e=1$, modelled for $\leq 20\,\mathrm{AU}$ and plotted in the $2.5$--$15\,\mathrm{AU}$ region. This model object had $\mu=1$, $15.5\%$ CO, and $17\%$ $\mathrm{CO_2}$ by number relative water. 
The left panel shows a case where all $\mathrm{CO}$ initially is trapped in amorphous water ice. The right panel shows a case where all $\mathrm{CO}$ initially is trapped in $\mathrm{CO_2}$ 
ice. \texttt{NIMBUSD} was used for these particular simulations, and it was assumed that the solid material was eroded at half the rate compared to local total production rates of all volatiles. 
The assumed erosion rate is arbitrarily chosen, but the qualitative behaviour of the solutions are not strongly dependent on the exact erosion rate value.}
     \label{fig_CO_source}
\end{figure*}

If intimate $\mathrm{CO_2}:\mathrm{CO}$ mixtures indeed are present in comets, they could play an important role in distant activity of comet nuclei and Centaurs.
To illustrate this point, Fig.~\ref{fig_CO_source} compares two models of a dynamically new comet: one with clean $\mathrm{CO_2}$ where 
$\mathrm{CO}$ is traditionally trapped in amorphous water ice (left), and another where the amorphous $\mathrm{H_2O}$ is clean and the 
$\mathrm{CO}$ is trapped in $\mathrm{CO_2}$ (right). These models were run with \texttt{NIMBUSD}, i.e., dust mantle erosion was included. The latter object behaves qualitatively similar to observed comets. 
It reaches activation around $11\pm 1\,\mathrm{AU}$ (compare with the discussion in Sec.~\ref{sec_intro}), where the $\mathrm{CO}$ and $\mathrm{CO_2}$ productions 
have reached 1--$10\%$ of the strong activity near $2\,\mathrm{AU}$. At a somewhat larger distance, e.~g., at $14\,\mathrm{AU}$, the activity is 
2--3 orders of magnitude lower than at $12\,\mathrm{AU}$. At $12\,\mathrm{AU}$, the $\mathrm{CO}$ production rate is $\sim 4$ times higher than 
that of $\mathrm{CO_2}$. The combination of the two might drive the dust production that often is observed in the $11\pm 1\,\mathrm{AU}$ region, 
although telescopes yet do not have the sensitivity to routinely measure the $\mathrm{CO}$ and $\mathrm{CO_2}$ directly at such distances, to explore their absolute and relative abundances. 
Segregation should be capable of driving the $\mathrm{CO}$ production observed in Centaurs, with almost the same intensity throughout the $6$--$10\,\mathrm{AU}$ region. 
The equatorial segregation front in the model of Fig.~\ref{fig_CO_source} (right) is merely $\sim 4\,\mathrm{cm}$ below the surface at $8\,\mathrm{AU}$, and such 
shallow depths could perhaps be maintained over long periods of time if the erosion rate is sufficiently high. If so, segregation could explain the sunward $\mathrm{CO}$ activity observed in 
Centaurs 29P/Schwassmann--Wachmann~1 and 174P/Echeclus \citep{jewitt09, wierzchosetal17}.

Figure~\ref{fig_CO_source} (right) also shows that significant and extremely remote activity \citep[e.~g., a dust production of $240\pm 110\,\mathrm{kg\,s^{-1}}$ for 
C/2017~K2 at $23.7\,\mathrm{AU}$;][]{huietal18} may not be possible to maintain through segregation after all, unless such activity represents the ``$\sim 50\,\mathrm{K}$ 
emission peak'' seen in segregation experiments \citep{lunaetal08, satorreetal09}. At the moment, the best explanations for the activity of C/2017~K2 appear to be an 
origin in the giant planet region or a foreign stellar system, or alternatively, minor near--surface $\mathrm{CO}$ condensates formed during ejection of an actively segregating/crystallising 
object originating from the Primordial Disk. However, the latter alternative is less compelling because of the perceived difficulty of reconciling a probably common process with 
the unusual distantly active comets.

The more classical model in  Fig.~\ref{fig_CO_source} (left) also becomes active around $11\pm 1\,\mathrm{AU}$. However, somewhat surprisingly considering the discussion 
in Sec.~\ref{sec_intro}, the driver is not $\mathrm{CO}$ released from crystallising amorphous water ice. Instead, sublimating $\mathrm{CO_2}$ is responsible for the 
activity. At $12\,\mathrm{AU}$ the $\mathrm{CO}$ production from crystallisation is $7\cdot 10^5$ times weaker than that of segregation, and the $\mathrm{CO}$ 
production rate is $2\cdot 10^5$ times below that of $\mathrm{CO_2}$. The $\mathrm{CO}$ production from crystallisation becomes comparable to $\mathrm{CO_2}$ 
production at $\sim 8\,\mathrm{AU}$ and the model object needs to get within $\sim 5\,\mathrm{AU}$ before $\mathrm{CO}$ exceeds $\mathrm{CO_2}$. 
\citet{guilbertlepoutre12} found that crystallisation becomes an important source of activity at $11\pm 1\,\mathrm{AU}$. Her thermophysical model did not consider 
sublimation of any volatile, but solved the heat flow problem within a rotating spherical (and, in practice, asteroidal) body at different latitudes, for different spin axis orientations. 
The thermophysical model contained a module similar to Eq.~(\ref{eq:23}), allowing her to calculate crystallisation rates as function of depth, latitude, and time. In the absence of sublimation 
cooling, the temperatures become comparably high, which explains why crystallisation reaches full force as distant as $11\pm 1\,\mathrm{AU}$ from the Sun. However, 
when $\mathrm{CO_2}$ ice is included in a thermophysical model, the \texttt{NIMBUSD} simulations in Fig.~\ref{fig_CO_source} show that the cooling from $\mathrm{CO_2}$ 
is substantial. With $\mathrm{CO_2}$, the temperature does not become high enough to allow for substantial crystallisation until the object is to within $\sim 8\,\mathrm{AU}$ 
of the Sun. When discussing the $11\pm 1\,\mathrm{AU}$ activation barrier for Centaurs, \citet{jewitt09} dismissed $\mathrm{CO_2}$ with an argument similar to that 
for $\mathrm{CO}$, e.~g., his calculations showed that $\mathrm{CO_2}$ should be active throughout the planetary region while Centaurs are not. However, the simulations in 
Fig.~\ref{fig_CO_source} show that the $\mathrm{CO_2}$ outgassing drops at least two orders of magnitude between $12\,\mathrm{AU}$ and $14\,\mathrm{AU}$, 
becoming completely insignificant further away. Potential reasons for the discrepancy is that \texttt{NIMBUSD} includes heat conduction, rotation (including night--time cooling), 
and a finite diffusivity for vapour from sub--surface ice, that \citet{jewitt09} did not consider in his work. I therefore propose that the activation distance around $11\pm 1\,\mathrm{AU}$, 
observed in numerous long--period comets and Centaurs, is caused by the onset of $\mathrm{CO_2}$ sublimation, potentially in association with $\mathrm{CO}$ release during 
segregation, and not necessarily by crystallisation.

The last question concerned $\mathrm{CO_2}$ ice and amorphous water ice:

\begin{trivlist}
\item Q\#3. At what depths are the $\mathrm{CO_2}$ sublimation front and the crystallisation front of amorphous water ice expected to be 
located after thermal processing in the Primordial Disk? 
\end{trivlist}

Bodies that are exposed to the most intense protosolar heating lose all $\mathrm{CO_2}$ ice in the upper $\sim 30\,\mathrm{m}$ and 
the degree of crystallisation is at least $\sim 50\%$ in that layer. Complete preservation of amorphous water ice is only expected below a depth of 
$\sim 170\,\mathrm{m}$. Although small bodies would be amorphous throughout the rest of their interiors, sufficiently large and dusty objects would 
only have amorphous ice in the upper $\sim 3\,\mathrm{km}$ because of radiogenic heating. Such bodies would also have substantially enhanced 
$\mathrm{CO_2}$ ice abundances at $5$--$20\,\mathrm{km}$ depth because of upward displacement. Bodies with $t_{\rm c}\geq 1\,\mathrm{Myr}$ 
would have $\mathrm{CO_2}$ ice and amorphous water ice up to their very surfaces (while the internal crystallisation and $\mathrm{CO_2}$ displacements 
would still be present in the sufficiently radioactive ones). Whether the $\mathrm{CO_2}$ and crystallisation fronts are located at the surface, or at 
depths of, e.~g., $1\,\mathrm{m}$, $5\,\mathrm{m}$, or $10\,\mathrm{m}$ would have a tremendous influence on their capacity to produce 
$\mathrm{CO}$ and $\mathrm{CO_2}$ when entering the inner Solar System for the first time as new arrivals from the Oort Cloud. The observed $\mathrm{CO/H_2O}$ 
coma abundance ratio varies by $\sim 3$ orders of magnitude among comets passing close to the Sun, and the $\mathrm{CO_2/H_2O}$ coma abundance ratio varies
 by $\sim 2$ orders of magnitude \citep{ahearnetal12}. Although such differences routinely are interpreted as reflecting variability in the bulk composition of 
comet nuclei, another possibility is that the abundance ratio diversity is caused by differences in front depths (the larger the depth of the source of vapour, the smaller the mass flux 
that reaches the surface). That would imply that at least parts of the Solar Nebula (in what would become the Primordial Disk) cleared during the first $\sim 0.2\,\mathrm{Myr}$, 
when the protosolar radiation was strong enough to cause near--surface $\mathrm{CO_2}$ sublimation and crystallisation. If so, cometary coma $\mathrm{CO/H_2O}$ and 
$\mathrm{CO_2/H_2O}$ abundance ratios could be powerful indicators of the birth distance of comets: the closer to the protosun, the earlier could the clearing have taken 
place and the stronger was the processing, resulting in significantly lowered present coma abundance ratios because of the withdrawn fronts. If coma abundance ratios 
truly reflected the nucleus bulk abundance, one might have expected a rather bimodal distribution: extremely low abundances in objects formed within the snow line 
of the species in question, and the full available abundance in objects formed beyond the snow line. The model of the chemical gradients of the Solar Nebula by 
\citet{dodsonrobinsonetal09} shows ice abundances (e.~g., of CO) raising from zero to full freeze--out over just $\stackrel{<}{_{\sim}}\,3\,\mathrm{AU}$. Because the 
radial distributions of $\mathrm{CO}$ and $\mathrm{CO_2}$ essentially are step--functions, it is difficult to comprehend why such narrow transition regions should 
be so well--represented in the observational record. The model by \citet{dodsonrobinsonetal09} suggests that all planetesimals formed beyond $\sim 8\,\mathrm{AU}$
 should have roughly the same $\mathrm{CO/H_2O}$ ratio. Similar bulk compositions among comet nuclei, combined with monotonically 
decreasing $\mathrm{CO_2}$ and amorphous water front depths with increasing heliocentric distance, could possibly be the best explanation for the wide range 
and lack of bimodality in the observed $\mathrm{CO/H_2O}$ and $\mathrm{CO_2/H_2O}$ distributions.

Comets that were exposed to solar radiation late and/or at large distance would have comparably shallow $\mathrm{CO_2}$ and amorphous water front depths 
(and, as may be the case with 67P, would contain $\mathrm{CO:CO_2}$ mixtures). If placed in the Scattered Disk, and later evolving dynamically into Centaurs, such objects would readily 
produce observable amounts of dust, $\mathrm{CO}$ and 
$\mathrm{CO_2}$. Other objects, formed closer to the Sun and/or exposed to its heat earlier, would have deeper fronts and perhaps be observed as inactive Centaurs. 
A Centaur that lacks observable activity may not necessarily have exhausted its near--surface supervolatiles during its time as a Centaur, but its $\mathrm{CO}$-- 
and $\mathrm{CO_2}$--poor crust could have developed already in the Primordial Disk. If such Centaurs are evolving dynamically into JFCs, one can imagine 
a time--period during which water--driven activity and erosion gradually brings the $\mathrm{CO_2}$ and amorphous $\mathrm{H_2O}$ fronts closer to 
the surface. The $\mathrm{CO/H_2O}$ and $\mathrm{CO_2/H_2O}$ ratios of such bodies would increase with time. \citet{ahearnetal12} pointed out that 
periodic ``comets with the fewest perihelion passages are depleted in CO relative to the others, exactly contrary to what one would expect from 
simple evolutionary models.'' If the apparent over--abundance of objects with low $\mathrm{CO/H_2O}$ ratios among recently injected JFCs is statistically significant, 
that is certainly inconsistent with the notion that physically ageing comets gradually ``run out'' of supervolatiles. However, a comparably large fraction of 
low--$\mathrm{CO/H_2O}$ objects among dynamically young JFCs would be consistent with the idea that some comets may need to erode significantly before 
their $\mathrm{CO_2}$ and amorphous $\mathrm{H_2O}$ fronts are sufficiently shallow to inject large amounts of $\mathrm{CO}$ and $\mathrm{CO_2}$ into the coma.

If variability in $\mathrm{CO_2}$ and amorphous $\mathrm{H_2O}$ front depths exist in dynamically new comets, Centaurs, and recently injected JFCs, and if 
that variability indeed was created because of Primordial Disk processing, it has an important corollary. The time period during which the variability needs to 
be established, during the first $\sim 0.2\,\mathrm{Myr}$ at the centre of the disk, is short compared to the Primordial Disk lifetime. If this near--surface 
stratification is destroyed at a later time during Primordial Disk evolution, it cannot be re--established. This includes a consideration of heating of Oort cloud comets 
by passing O-- and B--stars, as well as by supernovae. \citet{sternshull88} demonstrated that the former have heated $\sim 10\%$ of the Oort cloud comets to $34\,\mathrm{K}$, 
while there is a 50\% probability that the latter have heated the comets to $60\,\mathrm{K}$ \citep[also see][]{stern03}. These temperatures are not sufficient to remove $\mathrm{CO_2}$ 
ice or to crystallise amorphous water ice near the surface (according to Table~\ref{tab_results}, a $t_{\rm c}=1\,\mathrm{Myr}$ object is heated to $70\,\mathrm{K}$ without losing $\mathrm{CO_2}$ 
and amorphous water ice). Therefore, if one could demonstrate the existence of such primordial stratification, it means that comet nuclei did not experience significant catastrophic 
collisional processing in the Primordial Disk. Arguments for the perceived lack of heavy collisional processing in 67P and other JFCs, as well as a comet formation and early evolution 
scenario in which collisional cascades are avoided, were presented by \citet{davidssonetal16}.

\section{Conclusions} \label{sec_conclusions}

The thermophysical code \texttt{NIMBUS} (Numerical Icy Minor Body evolUtion Simulator) is described at length for the first time, and it 
is used to investigate the thermophysical evolution of small icy bodies in the Primordial Disk (here, $r_{\rm h}=23\,\mathrm{AU}$) during the earliest epoch of Solar System history. 
Specifically, porous bodies with diameters $D=\{4,\, 74,\, 203\}\,\mathrm{km}$ have been considered, consisting of amorphous water ice and 
$\mathrm{CO_2}$ ice (that both trap $\mathrm{CO}$ in different proportions), separate ``condensed'' $\mathrm{CO}$ ice, and refractories that 
contain long--lived radionuclides. Time--dependent protosolar illumination consistent with solar--mass Hayahsi-- and Henyey--tracks was applied, 
but coupled with a simple parameterisation of Solar Nebula opacity that exposed the icy bodies to the protosun at different clearing times, 
$5\,\mathrm{kyr}$, $1\,\mathrm{Myr}$, or $3\,\mathrm{Myr}$ into Solar System history. The main conclusions are the following:

\begin{trivlist}
\item $\bullet$ Small comets ($D=4\,\mathrm{km}$) lose all $\mathrm{CO}$ ice that is not trapped inside a less volatile medium 
in 70--$170\,\mathrm{kyr}$ depending on the Solar Nebula clearing time. Comets that were placed in the Oort Cloud, Kuiper Belt, or Scattered Disk 
at the time of the Primordial Disk dispersal ($\stackrel{>}{_{\sim}}15\,\mathrm{Myr}$ after Solar System formation) cannot possibly contain large deposits of CO ice.
\item $\bullet$ Large comets ($D=70$--$200\,\mathrm{km}$) lose all $\mathrm{CO}$ ice that is not trapped inside a less volatile medium 
in 10--$13\,\mathrm{Myr}$ if they are dusty (here, a dust--to--water ice mass ratio of 4). Dust--free $200\,\mathrm{km}$--class objects lose 
all free CO ice in $\sim 200\,\mathrm{Myr}$. 
\item $\bullet$ Small comets ($D=4\,\mathrm{km}$) do not preserve any $\mathrm{CO:CO_2}$ mixtures if they are exposed to the protosun too early. 
Exposure $\sim 1\,\mathrm{Myr}$ after CAI or later could imply almost complete $\mathrm{CO:CO_2}$ survival.
\item $\bullet$ Large comets ($D=70$--$200\,\mathrm{km}$) may have a thin near--surface layer of $\mathrm{CO:CO_2}$ mixtures at 
$\stackrel{<}{_{\sim}}1$--$2\,\mathrm{km}$ that survives radiogenic heating (and if applicable, early exposure to protosolar radiation).
\item $\bullet$ Primordial Disk bodies, of any size, that get exposed to protosolar radiation within the first $\sim 0.2\,\mathrm{Myr}$ of the 
Solar Nebula, lose $\mathrm{CO_2}$ ice in a surface layer being up to $\sim 30\,\mathrm{m}$ thick, and experience partial crystallisation 
in the upper $\sim 200\,\mathrm{m}$ that reaches completeness at the very surface.  
\end{trivlist}

Based one these conclusions and associated discussions, I propose that:

\begin{enumerate}
\item Dynamically new comets are not necessarily pristine -- the majority have experienced global loss of free CO ice, and some may have 
crystallised and lost $\mathrm{CO_2}$ ice in the top few tens of meters, because of thermal processing in the Primordial Disk prior to their 
emplacement in the Oort Cloud. 
\item The large ranges in $\mathrm{CO/H_2O}$ and $\mathrm{CO_2/H2O}$ observed in dynamically new comets is not necessarily due to 
intrinsic bulk abundance variability, but because the previously described diversity of $\mathrm{CO_2}$ and amorphous $\mathrm{H_2O}$ front depths.
\item The apparent increase of $\mathrm{CO/H_2O}$ and $\mathrm{CO_2/H_2O}$ in short--period comets with number of perihelion passages is 
because their $\mathrm{CO_2}$ and amorphous $\mathrm{H_2O}$ fronts are gradually brought closer to the comet surfaces by erosion driven by sublimation of 
crystalline water ice. 
\item If propositions i--iii are correct, the Primordial Disk cannot have experienced a collisional cascade, because that would destroy the near--surface 
stratification that only forms in the first $\sim 0.2\,\mathrm{Myr}$.
\item CO gas observed in the debris disks of young stars could originate from radiogenic heating of sufficiently large planetesimals, in addition 
to, or instead of, collisional cascades and UV desorption. 
\item Comet C/2017~K2~(PANSTARRS) probably did not form in the Primordial Disk, but could be a rare example of a body originating 
in the giant planet region and ejected very early, or was captured from a foreign system in the solar birth cluster. 
\item The $11\pm 1\,\mathrm{AU}$ activity switch--on observed in long--period comets and Centaurs is due to $\mathrm{CO_2}$ ice 
sublimation and/or $\mathrm{CO:CO_2}$ mixture segregation. Presence of $\mathrm{CO_2}$ ice prevents crystallisation from being an 
important source of released hypervolatiles beyond $\sim 8\,\mathrm{AU}$.
\item The lack of activity in large Centaurs beyond $12\,\mathrm{AU}$ implies that they are sufficiently dusty, and/or the Primordial Disk was sufficiently 
long--lived, to deprive them of all free CO ice. The observed CO is likely released from $\mathrm{CO_2}$ in the 8--$12\,\mathrm{AU}$ region, and 
from a combination of $\mathrm{CO_2}$ and amorphous water ice within $8\,\mathrm{AU}$. 
\end{enumerate}

\section*{Acknowledgements} 

Parts of the research was carried out at the Jet Propulsion Laboratory, California Institute of Technology, under a 
contract with the National Aeronautics and Space Administration. The author is indebted to Dr. Pedro Guti\'{e}rrez from Instituto de Astrof\'{i}sica 
de Andaluc\'{i}a, Granada, Spain, for invaluable discussions, suggestions, and support during the writing of this paper. The author 
appreciates the detailed, constructive, and important recommendations of the reviewer, Dina Prialnik, that 
substantially improved the paper.\\

\noindent
\emph{COPYRIGHT}.  \textcopyright\,2021. California Institute of Technology. Government sponsorship acknowledged.

\section*{Data Availability}

The data underlying this article will be shared on reasonable request to the corresponding author.

\bibliography{MN-21-0533-MJ.R1.bbl}

\begin{thebibliography}{}
\makeatletter
\relax
\def\mn@urlcharsother{\let\do\@makeother \do\$\do\&\do\#\do\^\do\_\do\%\do\~}
\def\mn@doi{\begingroup\mn@urlcharsother \@ifnextchar [ {\mn@doi@}
  {\mn@doi@[]}}
\def\mn@doi@[#1]#2{\def\@tempa{#1}\ifx\@tempa\@empty \href
  {http://dx.doi.org/#2} {doi:#2}\else \href {http://dx.doi.org/#2} {#1}\fi
  \endgroup}
\def\mn@eprint#1#2{\mn@eprint@#1:#2::\@nil}
\def\mn@eprint@arXiv#1{\href {http://arxiv.org/abs/#1} {{\tt arXiv:#1}}}
\def\mn@eprint@dblp#1{\href {http://dblp.uni-trier.de/rec/bibtex/#1.xml}
  {dblp:#1}}
\def\mn@eprint@#1:#2:#3:#4\@nil{\def\@tempa {#1}\def\@tempb {#2}\def\@tempc
  {#3}\ifx \@tempc \@empty \let \@tempc \@tempb \let \@tempb \@tempa \fi \ifx
  \@tempb \@empty \def\@tempb {arXiv}\fi \@ifundefined
  {mn@eprint@\@tempb}{\@tempb:\@tempc}{\expandafter \expandafter \csname
  mn@eprint@\@tempb\endcsname \expandafter{\@tempc}}}

\bibitem[\protect\citeauthoryear{{A'Hearn} et~al.,}{{A'Hearn}
  et~al.}{2012}]{ahearnetal12}
{A'Hearn} M.~F.,  et~al., 2012, Astrophys. J., 758, 29

\bibitem[\protect\citeauthoryear{Altwegg et~al.,}{Altwegg
  et~al.}{2015}]{altweggetal15}
Altwegg K.,  et~al., 2015, Science, 347, 1261952

\bibitem[\protect\citeauthoryear{Andr\'{e} \& Montmerle}{Andr\'{e} \&
  Montmerle}{1994}]{andremontmerle94}
Andr\'{e} P.,  Montmerle T.,  1994, Astrophys. J., 420, 837

\bibitem[\protect\citeauthoryear{Azreg-A\"inou}{Azreg-A\"inou}{2005}]{azregainou05}
Azreg-A\"inou M.,  2005, Monatshefte f\"{u}r Chemie, 136, 2017

\bibitem[\protect\citeauthoryear{Bailey \& Malhotra}{Bailey \&
  Malhotra}{2009}]{baileymalhotra09}
Bailey B.~L.,  Malhotra R.,  2009, Icarus, 203, 155

\bibitem[\protect\citeauthoryear{Bar-Nun, Herman, Laufer  \& Rappaport}{Bar-Nun
  et~al.}{1985}]{barnunetal85}
Bar-Nun A.,  Herman G.,  Laufer D.,   Rappaport M.~L.,  1985, Icarus, 63, 317

\bibitem[\protect\citeauthoryear{Bar-Nun, Dror, Kochavi  \& Laufer}{Bar-Nun
  et~al.}{1987}]{barnunetal87}
Bar-Nun A.,  Dror J.,  Kochavi E.,   Laufer D.,  1987, Phys. Rev. B, 35, 2427

\bibitem[\protect\citeauthoryear{Bardyn et~al.,}{Bardyn
  et~al.}{2017}]{bardynetal17}
Bardyn A.,  et~al., 2017, Mon. Not. R. Ast. Soc., 469, S712

\bibitem[\protect\citeauthoryear{Benkhoff \& Boice}{Benkhoff \&
  Boice}{1996}]{benkhoffboice96}
Benkhoff J.,  Boice D.~C.,  1996, Planet. Space Sci., 44, 665

\bibitem[\protect\citeauthoryear{Benkhoff \& Huebner}{Benkhoff \&
  Huebner}{1995}]{benkhoffandhuebner95}
Benkhoff J.,  Huebner W.~F.,  1995, Icarus, 114, 348

\bibitem[\protect\citeauthoryear{Benkhoff \& Spohn}{Benkhoff \&
  Spohn}{1991}]{benkhoffandspohn91}
Benkhoff J.,  Spohn T.,  1991, in K\"{o}mle N.~I.,  Bauer S.~J.,   Spohn T.,
  eds, , Theoretical modelling of comet simulation experiments.
Verlag der \"{O}sterreichischen Akademie der Wissenschaften, Wien, pp 31--47

\bibitem[\protect\citeauthoryear{Benkhoff, Seidensticker, Seiferlin  \&
  Spohn}{Benkhoff et~al.}{1995}]{benkhoffetal95}
Benkhoff J.,  Seidensticker K.~J.,  Seiferlin K.,   Spohn T.,  1995, Planet.
  Space Sci., 43, 353

\bibitem[\protect\citeauthoryear{Bentley et~al.,}{Bentley
  et~al.}{2016}]{bentleyetal16}
Bentley M.~S.,  et~al., 2016, Nature, 537, 73

\bibitem[\protect\citeauthoryear{Biver et~al.,}{Biver
  et~al.}{2019}]{biveretal19}
Biver N.,  et~al., 2019, Astron. Astrophys., 630, A19

\bibitem[\protect\citeauthoryear{Blum et~al.,}{Blum et~al.}{2017}]{blumetal17}
Blum J.,  et~al., 2017, Mon. Not. R. Astron. Soc., 469, S755

\bibitem[\protect\citeauthoryear{Bockel\'{e}e-Morvan, Lellouch, Biver, Paubert,
  Bauer, Colom  \& Lis}{Bockel\'{e}e-Morvan
  et~al.}{2001}]{bockeleemorvanetal01}
Bockel\'{e}e-Morvan D.,  Lellouch E.,  Biver N.,  Paubert G.,  Bauer J.,  Colom
  P.,   Lis D.~C.,  2001, Astron. Astrophys., 377, 343

\bibitem[\protect\citeauthoryear{Bockel\'{e}e-Morvan, Crovisier, Mumma  \&
  Weaver}{Bockel\'{e}e-Morvan et~al.}{2004}]{bockeleemorvanetal04}
Bockel\'{e}e-Morvan D.,  Crovisier J.,  Mumma M.~J.,   Weaver H.~A.,  2004, in
  Festou M.~C.,  Keller H.~U.,   Weaver H.~A.,  eds, , Comets II.
Univ. of Arizona Press, Tucson, pp 391--423

\bibitem[\protect\citeauthoryear{Boss \& Vanhala}{Boss \&
  Vanhala}{2000}]{bossvanhala00}
Boss A.,  Vanhala H. A. .~T.,  2000, Space Sci. Rev., 92, 13

\bibitem[\protect\citeauthoryear{Boss, Ipatov, Keiser, Myhill  \& Vanhala}{Boss
  et~al.}{2008}]{bossetal08}
Boss A.~P.,  Ipatov S.~I.,  Keiser S.~A.,  Myhill E.~A.,   Vanhala H. A.~T.,
  2008, Astrophys. J., 686, L119

\bibitem[\protect\citeauthoryear{Bottke, Vokrouhlick\'{y}, Minton,
  Nesvorn\'{y}, Morbidelli, Brasser, Simonson  \& Levison}{Bottke
  et~al.}{2012}]{bottkeetal12}
Bottke W.~F.,  Vokrouhlick\'{y} D.,  Minton D.,  Nesvorn\'{y} D.,  Morbidelli
  A.,  Brasser R.,  Simonson B.,   Levison H.~F.,  2012, Nature, 485, 78

\bibitem[\protect\citeauthoryear{Brasser \& Morbidelli}{Brasser \&
  Morbidelli}{2013}]{brassermorbidelli13}
Brasser R.,  Morbidelli A.,  2013, Icarus, 225, 40

\bibitem[\protect\citeauthoryear{Brasser, Duncan  \& Levison}{Brasser
  et~al.}{2007}]{brasseretal07}
Brasser R.,  Duncan M.~J.,   Levison H.~F.,  2007, Icarus, 191, 413

\bibitem[\protect\citeauthoryear{Brownlee et~al.,}{Brownlee
  et~al.}{2006}]{brownleeetal06}
Brownlee D.,  et~al., 2006, Science, 314, 1711

\bibitem[\protect\citeauthoryear{Cabral et~al.,}{Cabral
  et~al.}{2019}]{cabraletal19}
Cabral N.,  et~al., 2019, Astron. Astrophys., 621, A102

\bibitem[\protect\citeauthoryear{Calmonte et~al.,}{Calmonte
  et~al.}{2016}]{calmonteetal16}
Calmonte U.,  et~al., 2016, Mon. Not. R. Astron. Soc., 462, S253

\bibitem[\protect\citeauthoryear{Capria, Capaccioni, Coradini, {De Sanctis},
  Espinasse, Federico, Orosei  \& Salomone}{Capria et~al.}{1996}]{capriaetal96}
Capria M.~T.,  Capaccioni F.,  Coradini A.,  {De Sanctis} M.~C.,  Espinasse S.,
   Federico C.,  Orosei R.,   Salomone M.,  1996, Planet. Space Sci., 44, 987

\bibitem[\protect\citeauthoryear{Capria, Coradini, {De Sanctis}  \&
  Orosei}{Capria et~al.}{2000}]{capriaetal00b}
Capria M.~T.,  Coradini A.,  {De Sanctis} M.~C.,   Orosei R.,  2000, Astron.
  J., 119, 3112

\bibitem[\protect\citeauthoryear{Capria, Coradini, {De Sanctis}  \&
  Blecka}{Capria et~al.}{2001}]{capriaetal01}
Capria M.~T.,  Coradini A.,  {De Sanctis} M.~C.,   Blecka M.~I.,  2001, Planet.
  Space Sci., 49, 907

\bibitem[\protect\citeauthoryear{Capria, Coradini  \& {De Sanctis}}{Capria
  et~al.}{2002}]{capriaetal02}
Capria M.~T.,  Coradini A.,   {De Sanctis} M.~C.,  2002, Earth, Moon, Planets,
  90, 217

\bibitem[\protect\citeauthoryear{Capria, Coradini, {De Sanctis}, {Mazzotta
  Epifani}  \& Palumbo}{Capria et~al.}{2009}]{capriaetal09}
Capria M.~T.,  Coradini A.,  {De Sanctis} M.~C.,  {Mazzotta Epifani} E.,
  Palumbo P.,  2009, Astron. Astrophys., 504, 249

\bibitem[\protect\citeauthoryear{Castillo-Rogez, Johnson, Lee, Turner  \&
  Matson}{Castillo-Rogez et~al.}{2009}]{castillorogezetal09}
Castillo-Rogez J.~C.,  Johnson T.~V.,  Lee M.~H.,  Turner N.~J.,   Matson
  D.~L.,  2009, Icarus, 204, 658

\bibitem[\protect\citeauthoryear{Castillo-Rogez, Johnson, Thomas, Choukroun,
  Matson  \& Lunine}{Castillo-Rogez et~al.}{2012}]{castillorogezetal12}
Castillo-Rogez J.~C.,  Johnson T.~V.,  Thomas P.~C.,  Choukroun M.,  Matson
  D.~L.,   Lunine J.~I.,  2012, Icarus, 219, 86

\bibitem[\protect\citeauthoryear{Cercignani}{Cercignani}{2000}]{cercignani00}
Cercignani C.,  2000, Rarefied gas dynamics. From basic concepts to actual
  calculations.
Cambridge University Press, Cambridge

\bibitem[\protect\citeauthoryear{Choi, Cohen, Merk  \& Prialnik}{Choi
  et~al.}{2002}]{choietal02}
Choi Y.-J.,  Cohen M.,  Merk R.,   Prialnik D.,  2002, Icarus, 160, 300

\bibitem[\protect\citeauthoryear{Choukroun et~al.,}{Choukroun
  et~al.}{2020}]{choukrounetal20}
Choukroun M.,  et~al., 2020, Space Sci. Rev., 216, 44

\bibitem[\protect\citeauthoryear{Combi et~al.,}{Combi
  et~al.}{2020}]{combietal20}
Combi M.,  et~al., 2020, Icarus, 335, 113421

\bibitem[\protect\citeauthoryear{Cooke, \"{O}berg, Fayolle, Peeler  \&
  Bergner}{Cooke et~al.}{2018}]{cookeetal18}
Cooke I.~R.,  \"{O}berg K.~I.,  Fayolle E.~C.,  Peeler Z.,   Bergner J.~B.,
  2018, Astrophys. J., 852, 75

\bibitem[\protect\citeauthoryear{Davidsson}{Davidsson}{2008}]{davidsson08}
Davidsson B. J.~R.,  2008, Space Sci. Rev., 138, 207

\bibitem[\protect\citeauthoryear{Davidsson \& Guti\'{e}rrez}{Davidsson \&
  Guti\'{e}rrez}{2004}]{davidssongutierrez04}
Davidsson B. J.~R.,  Guti\'{e}rrez P.~J.,  2004, Icarus, 168, 392

\bibitem[\protect\citeauthoryear{Davidsson \& Guti\'{e}rrez}{Davidsson \&
  Guti\'{e}rrez}{2005}]{davidssongutierrez05}
Davidsson B. J.~R.,  Guti\'{e}rrez P.~J.,  2005, Icarus, 176, 453

\bibitem[\protect\citeauthoryear{Davidsson \& Guti\'{e}rrez}{Davidsson \&
  Guti\'{e}rrez}{2006}]{davidssongutierrez06}
Davidsson B. J.~R.,  Guti\'{e}rrez P.~J.,  2006, Icarus, 180, 224

\bibitem[\protect\citeauthoryear{Davidsson \& Skorov}{Davidsson \&
  Skorov}{2002}]{davidssonandskorov02b}
Davidsson B. J.~R.,  Skorov Y.~V.,  2002, Icarus, 159, 239

\bibitem[\protect\citeauthoryear{Davidsson \& Skorov}{Davidsson \&
  Skorov}{2004}]{davidssonandskorov04}
Davidsson B. J.~R.,  Skorov Y.~V.,  2004, Icarus, 168, 163

\bibitem[\protect\citeauthoryear{Davidsson et~al.,}{Davidsson
  et~al.}{2016}]{davidssonetal16}
Davidsson B. J.~R.,  et~al., 2016, Astron. Astrophys., 592, A63

\bibitem[\protect\citeauthoryear{Davidsson, Samarasinha, Farnocchia  \&
  Guti\'{e}rrez}{Davidsson et~al.}{2021a}]{davidssonetal21b}
Davidsson B. J.~R.,  Samarasinha N.,  Farnocchia D.,   Guti\'{e}rrez P.,
  2021a, Mon. Not. R. Astron. Soc. (In preparation)

\bibitem[\protect\citeauthoryear{Davidsson et~al.,}{Davidsson
  et~al.}{2021b}]{davidssonetal21}
Davidsson B. J.~R.,  et~al., 2021b, Icarus, 354, 114004

\bibitem[\protect\citeauthoryear{{De Sanctis}, Capaccioni, Capria, Coradini,
  Federico, Orosei  \& Salomone}{{De Sanctis} et~al.}{1999}]{desanctisetal99}
{De Sanctis} M.~C.,  Capaccioni F.,  Capria M.~T.,  Coradini A.,  Federico C.,
  Orosei R.,   Salomone M.,  1999, Planet. Space Sci., 47, 855

\bibitem[\protect\citeauthoryear{{De Sanctis}, Capria, Coradini  \& Orosei}{{De
  Sanctis} et~al.}{2000}]{desanctisetal00}
{De Sanctis} M.~C.,  Capria M.~T.,  Coradini A.,   Orosei R.,  2000, Astron.
  J., 120, 1571

\bibitem[\protect\citeauthoryear{{De Sanctis}, Capria  \& Coradini}{{De
  Sanctis} et~al.}{2001}]{desanctisetal01}
{De Sanctis} M.~C.,  Capria M.~T.,   Coradini A.,  2001, Astron. J., 121, 2792

\bibitem[\protect\citeauthoryear{{De Sanctis}, Capria  \& Coradini}{{De
  Sanctis} et~al.}{2007}]{desanctisetal07}
{De Sanctis} M.~C.,  Capria M.~T.,   Coradini A.,  2007, Mem. S. A. It. Suppl,
  11, 135

\bibitem[\protect\citeauthoryear{{De Sanctis} et~al.,}{{De Sanctis}
  et~al.}{2015}]{desanctisetal15}
{De Sanctis} M.~C.,  et~al., 2015, Nature, 525, 500

\bibitem[\protect\citeauthoryear{{Dello Russo}, Mumma, {DiSanti}, Magee-Sauer,
  Novak  \& Rettig}{{Dello Russo} et~al.}{2000}]{dellorussoetal00}
{Dello Russo} N.,  Mumma M.~J.,  {DiSanti} M.~A.,  Magee-Sauer K.,  Novak R.,
  Rettig T.~W.,  2000, Icarus, 143, 324

\bibitem[\protect\citeauthoryear{{Di Sisto} \& Brunini}{{Di Sisto} \&
  Brunini}{2007}]{disistobrunini07}
{Di Sisto} R.~P.,  Brunini A.,  2007, Icarus, 190, 224

\bibitem[\protect\citeauthoryear{Dodson-Robinson, Willacy, Bodenheimer, Turner
  \& Beichman}{Dodson-Robinson et~al.}{2009}]{dodsonrobinsonetal09}
Dodson-Robinson S.~E.,  Willacy K.,  Bodenheimer P.,  Turner N.~J.,   Beichman
  C.~A.,  2009, Icarus, 200, 672

\bibitem[\protect\citeauthoryear{Duncan \& Levison}{Duncan \&
  Levison}{1997}]{duncanlevison97}
Duncan M.~J.,  Levison H.~F.,  1997, Science, 276, 1670

\bibitem[\protect\citeauthoryear{Enzian, Cabot  \& Klinger}{Enzian
  et~al.}{1997}]{enzianetal97}
Enzian A.,  Cabot H.,   Klinger J.,  1997, Astron. Astrophys., 319, 995

\bibitem[\protect\citeauthoryear{Enzian, Klinger, Schwehm  \& Weissman}{Enzian
  et~al.}{1999}]{enzianetal99}
Enzian A.,  Klinger J.,  Schwehm G.,   Weissman P.~R.,  1999, Icarus, 138, 74

\bibitem[\protect\citeauthoryear{Espinasse, Klinger, Ritz  \&
  Schmitt}{Espinasse et~al.}{1991}]{espinasseetal91}
Espinasse S.,  Klinger J.,  Ritz C.,   Schmitt B.,  1991, Icarus, 92, 350

\bibitem[\protect\citeauthoryear{Espinasse, Coradini, Capria, Capaccioni,
  Orosei, Salomone  \& Federico}{Espinasse et~al.}{1993}]{espinasseetal93}
Espinasse S.,  Coradini A.,  Capria M.~T.,  Capaccioni F.,  Orosei R.,
  Salomone M.,   Federico C.,  1993, Planet. Space Sci., 41, 409

\bibitem[\protect\citeauthoryear{Fanale \& Salvail}{Fanale \&
  Salvail}{1984}]{fanaleandsalvail84}
Fanale F.~P.,  Salvail J.~R.,  1984, Icarus, 60, 476

\bibitem[\protect\citeauthoryear{Fern\'{a}ndez}{Fern\'{a}ndez}{1980}]{fernandez80}
Fern\'{a}ndez J.~A.,  1980, Mon. Not. R. Astr. Soc., 192, 481

\bibitem[\protect\citeauthoryear{Fern\'{a}ndez \& Ip}{Fern\'{a}ndez \&
  Ip}{1984}]{fernandezandip84}
Fern\'{a}ndez J.~A.,  Ip W.-H.,  1984, Icarus, 58, 109

\bibitem[\protect\citeauthoryear{Fornasier et~al.,}{Fornasier
  et~al.}{2013}]{fornasieretal13}
Fornasier S.,  et~al., 2013, Astron. Astrophys., 555, A15

\bibitem[\protect\citeauthoryear{Fulle et~al.,}{Fulle
  et~al.}{2017}]{fulleetal17}
Fulle M.,  et~al., 2017, Mon. Not. R. Astron. Soc., 469, S45

\bibitem[\protect\citeauthoryear{Garenne, Beck, Montes-Hernandez, Chiriac,
  Toche, Quirico, Bonal  \& Schmitt}{Garenne et~al.}{2014}]{garenneetal14}
Garenne A.,  Beck P.,  Montes-Hernandez G.,  Chiriac R.,  Toche F.,  Quirico
  E.,  Bonal L.,   Schmitt B.,  2014, Geochim. Cosmochim. Acta, 137, 93

\bibitem[\protect\citeauthoryear{Gasc et~al.,}{Gasc et~al.}{2017}]{gascetal17}
Gasc S.,  et~al., 2017, Mon. Not. R. Astron. Soc., 469, S108

\bibitem[\protect\citeauthoryear{Gerakines et~al.,}{Gerakines
  et~al.}{1999}]{gerakinesetal99}
Gerakines P.~A.,  et~al., 1999, Astrophys. J., 522, 357

\bibitem[\protect\citeauthoryear{Ghormley}{Ghormley}{1968}]{ghormley68}
Ghormley J.~A.,  1968, J. Chem. Phys., 48, 503

\bibitem[\protect\citeauthoryear{Giauque \& Egan}{Giauque \&
  Egan}{1937}]{giauqueandegan37}
Giauque W.~F.,  Egan C.~J.,  1937, J. Chem. Phys., 5, 45

\bibitem[\protect\citeauthoryear{Giauque \& Stout}{Giauque \&
  Stout}{1936}]{giauqueandstout36}
Giauque W.~F.,  Stout J.~W.,  1936, J. Am. Chem. Soc., 58, 1144

\bibitem[\protect\citeauthoryear{Goldin, Knight, Macklin  \& Macklin}{Goldin
  et~al.}{1949}]{goldinetal49}
Goldin A.~S.,  Knight G.~B.,  Macklin P.~A.,   Macklin R.~L.,  1949, Phys.
  Rev., 76, 336

\bibitem[\protect\citeauthoryear{Gomes, Morbidelli  \& Levison}{Gomes
  et~al.}{2004}]{gomes04}
Gomes R.~S.,  Morbidelli A.,   Levison H.~F.,  2004, Icarus, 170, 492

\bibitem[\protect\citeauthoryear{Gomes, Levison, Tsiganis  \& Morbidelli}{Gomes
  et~al.}{2005}]{gomesetal05b}
Gomes R.~S.,  Levison H.~F.,  Tsiganis K.,   Morbidelli A.,  2005, Nature, 435,
  466

\bibitem[\protect\citeauthoryear{Gonz\'{a}lez, Guti\'{e}rrez, Lara  \&
  Rodrigo}{Gonz\'{a}lez et~al.}{2008}]{gonzalezetal08}
Gonz\'{a}lez M.,  Guti\'{e}rrez P.~J.,  Lara L.~M.,   Rodrigo R.,  2008,
  Astron. Astrophys., 486, 331

\bibitem[\protect\citeauthoryear{Gonz\'{a}lez, Guti\'{e}rrez  \&
  Lara}{Gonz\'{a}lez et~al.}{2014}]{gonzalezetal14}
Gonz\'{a}lez M.,  Guti\'{e}rrez P.~J.,   Lara L.~M.,  2014, Astron. Astrophys.,
  563, A98

\bibitem[\protect\citeauthoryear{Gortsas, K\"{u}hrt, Motschmann  \&
  Keller}{Gortsas et~al.}{2011}]{gortsasetal11}
Gortsas N.,  K\"{u}hrt E.,  Motschmann U.,   Keller H.~U.,  2011, Icarus, 212,
  858

\bibitem[\protect\citeauthoryear{Gr\"{u}n et~al.,}{Gr\"{u}n
  et~al.}{1991}]{grunetal91}
Gr\"{u}n E.,  et~al., 1991, in R.~L.~Newburn J.,  Neugebauer M.,   Rahe J.,
  eds, , Vol.~1, Comets in the {Post--Halley} era.
Kluwer Academic Publishers, pp 277--297

\bibitem[\protect\citeauthoryear{Gr\"{u}n et~al.,}{Gr\"{u}n
  et~al.}{1993}]{grunetal93}
Gr\"{u}n E.,  et~al., 1993, J. Geophys. Res., 98, 15091

\bibitem[\protect\citeauthoryear{Grundy et~al.,}{Grundy
  et~al.}{2007}]{grundyetal07}
Grundy W.~M.,  et~al., 2007, Icarus, 191, 286

\bibitem[\protect\citeauthoryear{Guilbert-Lepoutre}{Guilbert-Lepoutre}{2011}]{guilbertlepoutre11}
Guilbert-Lepoutre A.,  2011, Astron. J., 141, 103

\bibitem[\protect\citeauthoryear{Guilbert-Lepoutre}{Guilbert-Lepoutre}{2012}]{guilbertlepoutre12}
Guilbert-Lepoutre A.,  2012, Astron. J., 144, 97

\bibitem[\protect\citeauthoryear{Guilbert-Lepoutre, Lasue, Federico, Coradini,
  Orosei  \& Rosenberg}{Guilbert-Lepoutre
  et~al.}{2011}]{guilbertlepoutreetal11}
Guilbert-Lepoutre A.,  Lasue J.,  Federico C.,  Coradini A.,  Orosei R.,
  Rosenberg E.~D.,  2011, Astron. Astrophys., 529, A71

\bibitem[\protect\citeauthoryear{Gundlach, Blum, Keller  \& Skorov}{Gundlach
  et~al.}{2015}]{gundlachetal15}
Gundlach B.,  Blum J.,  Keller H.~U.,   Skorov Y.~V.,  2015, Astron.
  Astrophys., 583, A12

\bibitem[\protect\citeauthoryear{Guti\'{e}rrez, Ortiz, Rodrigo  \&
  L\'{o}pez-Moreno}{Guti\'{e}rrez et~al.}{2000}]{gutierrezetal00}
Guti\'{e}rrez P.~J.,  Ortiz J.~L.,  Rodrigo R.,   L\'{o}pez-Moreno J.~J.,
  2000, Astron. Astrophys., 355, 809

\bibitem[\protect\citeauthoryear{Guti\'{e}rrez, Ortiz, Rodrigo  \&
  L\'{o}pez-Moreno}{Guti\'{e}rrez et~al.}{2001}]{gutierrezetal01}
Guti\'{e}rrez P.~J.,  Ortiz J.~L.,  Rodrigo R.,   L\'{o}pez-Moreno J.~J.,
  2001, Astron. Astrophys., 374, 326

\bibitem[\protect\citeauthoryear{G\"{u}ttler, Krause, Geretshauser, Speith  \&
  Blum}{G\"{u}ttler et~al.}{2009}]{guettlereta09}
G\"{u}ttler C.,  Krause M.,  Geretshauser R.~J.,  Speith R.,   Blum J.,  2009,
  Astron. J., 701, 130

\bibitem[\protect\citeauthoryear{Haisch, Lada  \& Lada}{Haisch
  et~al.}{2001}]{haischetal01}
Haisch K.~E.,  Lada E.~A.,   Lada C.~J.,  2001, Astrophys. J., 553, L153

\bibitem[\protect\citeauthoryear{Hales, Gorti, Carpenter, Hughes  \&
  Flaherty}{Hales et~al.}{2019}]{halesetal19}
Hales A.~S.,  Gorti U.,  Carpenter J.~M.,  Hughes M.,   Flaherty K.,  2019,
  Astrophys. J., 878, 113

\bibitem[\protect\citeauthoryear{Henke, Gail, Trieloff, Schwarz  \&
  Kleine}{Henke et~al.}{2012}]{henkeetal12}
Henke S.,  Gail H.-P.,  Trieloff M.,  Schwarz W.~H.,   Kleine T.,  2012,
  Astron. Astrophys., 537, A45

\bibitem[\protect\citeauthoryear{Hevey \& Sanders}{Hevey \&
  Sanders}{2006}]{heveysanders06}
Hevey P.~J.,  Sanders I.~S.,  2006, Meteo. Planet. Sci., 41, 95

\bibitem[\protect\citeauthoryear{Hoang et~al.,}{Hoang
  et~al.}{2020}]{hoangetal20}
Hoang M.,  et~al., 2020, Astron. Astrophys., 638, A106

\bibitem[\protect\citeauthoryear{H\"{o}fner et~al.,}{H\"{o}fner
  et~al.}{2017}]{hofneretal17}
H\"{o}fner S.,  et~al., 2017, Astron. Astrophys., 608, A121

\bibitem[\protect\citeauthoryear{Horai}{Horai}{1971}]{horai71}
Horai K.-I.,  1971, J. Geophys. Res., 76, 1278

\bibitem[\protect\citeauthoryear{Horner, Evans  \& Bailey}{Horner
  et~al.}{2004}]{horneretal04}
Horner J.,  Evans N.~W.,   Bailey M.~E.,  2004, Mon. Not. R. Astron. Soc., 354,
  798

\bibitem[\protect\citeauthoryear{H\"{o}rst \& Tolbert}{H\"{o}rst \&
  Tolbert}{2013}]{horsttolbert13}
H\"{o}rst S.~M.,  Tolbert M.~A.,  2013, Astrophys. J. Lett., 770, L10

\bibitem[\protect\citeauthoryear{Hu et~al.,}{Hu et~al.}{2017}]{huetal17}
Hu X.,  et~al., 2017, Astron. Astrophys., 604, A114

\bibitem[\protect\citeauthoryear{Huebner, Benkhoff, Capria, Coradini,
  {De~Sanctis}, Enzian, Orosei  \& Prialnik}{Huebner
  et~al.}{1999}]{huebneretal99}
Huebner W.~F.,  Benkhoff J.,  Capria M.~T.,  Coradini A.,  {De~Sanctis} M.~C.,
  Enzian A.,  Orosei R.,   Prialnik D.,  1999, Adv. Space Res., 23, 1283

\bibitem[\protect\citeauthoryear{Huebner, Benkhoff, Capria, Coradini, {De
  Sanctis}, Orosei  \& Prialnik}{Huebner et~al.}{2006}]{huebneretal06}
Huebner W.~F.,  Benkhoff J.,  Capria M.-T.,  Coradini A.,  {De Sanctis} C.,
  Orosei R.,   Prialnik D.,  2006, Heat and gas diffusion in comet nuclei.
ESA Publications Division, Noordwijk, The Netherlands

\bibitem[\protect\citeauthoryear{Hui, Jewitt  \& Clark}{Hui
  et~al.}{2018}]{huietal18}
Hui M.-T.,  Jewitt D.,   Clark D.,  2018, Astron. J., 155, 25

\bibitem[\protect\citeauthoryear{Hui, Farnocchia  \& Micheli}{Hui
  et~al.}{2019}]{huietal19}
Hui M.-T.,  Farnocchia D.,   Micheli M.,  2019, Astron. J., 157, 162

\bibitem[\protect\citeauthoryear{Jewitt}{Jewitt}{2009}]{jewitt09}
Jewitt D.,  2009, Astron. J., 137, 4296

\bibitem[\protect\citeauthoryear{Jewitt \& Luu}{Jewitt \&
  Luu}{1993}]{jewittluu93}
Jewitt D.,  Luu J.,  1993, Nature, 362, 730

\bibitem[\protect\citeauthoryear{Jewitt \& Matthews}{Jewitt \&
  Matthews}{1999}]{jewittmatthews99}
Jewitt D.,  Matthews H.,  1999, Astron. J., 117, 1056

\bibitem[\protect\citeauthoryear{Jewitt, Garland  \& Aussel}{Jewitt
  et~al.}{2008}]{jewittetal08}
Jewitt D.,  Garland C.~A.,   Aussel H.,  2008, Astron. J., 135, 400

\bibitem[\protect\citeauthoryear{Jewitt, Hui, Mutchler, Weaver, Li  \&
  Agarwal}{Jewitt et~al.}{2017}]{jewittetal17}
Jewitt D.,  Hui M.-T.,  Mutchler M.,  Weaver H.,  Li J.,   Agarwal J.,  2017,
  Astron. J. Lett., 847, L19

\bibitem[\protect\citeauthoryear{Jewitt, Agarwal, Hui, Li, Mutchler  \&
  Weaver}{Jewitt et~al.}{2019}]{jewittetal19}
Jewitt D.,  Agarwal J.,  Hui M.-T.,  Li J.,  Mutchler M.,   Weaver H.,  2019,
  Astron. J., 157, 65

\bibitem[\protect\citeauthoryear{Jewitt, Luu  \& Marsden}{Jewitt
  et~al.}{1992}]{jewittetal92}
Jewitt D.,  Luu J.,   Marsden B.~G.,  21992, IAU Circ., 5611, 1

\bibitem[\protect\citeauthoryear{Johansen \& {Mac Low}}{Johansen \& {Mac
  Low}}{2009}]{johansenetal09}
Johansen A.,  {Mac Low} A. N. Y. M.-M.,  2009, Astrophys. J., 704, L75

\bibitem[\protect\citeauthoryear{Johansen, Oishi, Low, Klahr, Henning  \&
  Youdin}{Johansen et~al.}{2007}]{johansenetal07}
Johansen A.,  Oishi J.~S.,  Low M.-M.~M.,  Klahr H.,  Henning T.,   Youdin
  A.~N.,  2007, Nature, 448, 1022

\bibitem[\protect\citeauthoryear{Johnson \& Lunine}{Johnson \&
  Lunine}{2005}]{johnsonlunine05}
Johnson T.~V.,  Lunine J.~I.,  2005, Nature, 435, 69

\bibitem[\protect\citeauthoryear{Keller et~al.,}{Keller
  et~al.}{2015}]{kelleretal15}
Keller H.~U.,  et~al., 2015, Astron. Astrophys., 583, A34

\bibitem[\protect\citeauthoryear{Keller et~al.,}{Keller
  et~al.}{2017}]{kelleretal17}
Keller H.~U.,  et~al., 2017, Mon. Not. R. Astron. Soc., 469, S357

\bibitem[\protect\citeauthoryear{Klinger}{Klinger}{1980}]{klinger80}
Klinger J.,  1980, Science, 209, 271

\bibitem[\protect\citeauthoryear{Klinger}{Klinger}{1981}]{klinger81}
Klinger J.,  1981, Icarus, 47, 320

\bibitem[\protect\citeauthoryear{K\"{o}mle \& Dettleff}{K\"{o}mle \&
  Dettleff}{1991}]{komleanddettleff91}
K\"{o}mle N.~I.,  Dettleff G.,  1991, Icarus, 89, 73

\bibitem[\protect\citeauthoryear{K\"{o}mle \& Steiner}{K\"{o}mle \&
  Steiner}{1992}]{komleandsteiner92}
K\"{o}mle N.~I.,  Steiner G.,  1992, Icarus, 96, 204

\bibitem[\protect\citeauthoryear{Kossacki, K\"{o}mle, Kargl  \&
  Steiner}{Kossacki et~al.}{1994}]{kossackietal94}
Kossacki K.~J.,  K\"{o}mle N.~I.,  Kargl G.,   Steiner G.,  1994, Planet. Space
  Sci., 42, 383

\bibitem[\protect\citeauthoryear{Kossacki, K\"{o}mle, Leliwa-Kopysty\'{n}ski
  \& Kargl}{Kossacki et~al.}{1997}]{kossackietal97}
Kossacki K.~J.,  K\"{o}mle N.~I.,  Leliwa-Kopysty\'{n}ski J.,   Kargl G.,
  1997, Icarus, 128, 127

\bibitem[\protect\citeauthoryear{Kossacki, Spohn, Hagermann, Kaufmann  \&
  K\"{u}hrt}{Kossacki et~al.}{2015}]{kossackietal15}
Kossacki K.~J.,  Spohn T.,  Hagermann A.,  Kaufmann E.,   K\"{u}hrt E.,  2015,
  Icarus, 260, 464

\bibitem[\protect\citeauthoryear{Kral, Matr\'{a}, Kennedy, Marino  \&
  Wyatt}{Kral et~al.}{2020}]{kraletal20}
Kral Q.,  Matr\'{a} L.,  Kennedy G.~M.,  Marino S.,   Wyatt M.~C.,  2020, Mon.
  Not. R. Astron. Soc., 497, 2811

\bibitem[\protect\citeauthoryear{K\"{u}hrt}{K\"{u}hrt}{1984}]{kuhrt84}
K\"{u}hrt E.,  1984, Icarus, 60, 512

\bibitem[\protect\citeauthoryear{K\"{u}hrt \& Keller}{K\"{u}hrt \&
  Keller}{1994}]{kuehrtandkeller94}
K\"{u}hrt E.,  Keller H.~U.,  1994, Icarus, 109, 121

\bibitem[\protect\citeauthoryear{Kulyk, Rousselot, Korsun, Afanasiev, Sergeev
  \& Velichko}{Kulyk et~al.}{2018}]{kulyketal18}
Kulyk I.,  Rousselot P.,  Korsun P.~P.,  Afanasiev V.~L.,  Sergeev A.~V.,
  Velichko S.~F.,  2018, Astron. Astrophys., 611, A32

\bibitem[\protect\citeauthoryear{Larson}{Larson}{2003}]{larson03}
Larson R.~B.,  2003, Rep. Prog. Phys., 66, 1651

\bibitem[\protect\citeauthoryear{Levison \& Duncan}{Levison \&
  Duncan}{1997}]{levisonduncan97}
Levison H.~F.,  Duncan M.~J.,  1997, Icarus, 127, 13

\bibitem[\protect\citeauthoryear{Levison, Morbidelli, van Laerhoven, Gomes  \&
  Tsiganis}{Levison et~al.}{2008}]{levisonetal08}
Levison H.~F.,  Morbidelli A.,  van Laerhoven C.,  Gomes R.,   Tsiganis K.,
  2008, Icarus, 196, 258

\bibitem[\protect\citeauthoryear{Levison, Duncan, Brasser  \& Kaufmann}{Levison
  et~al.}{2010}]{levisonetal10}
Levison H.~F.,  Duncan M.~J.,  Brasser R.,   Kaufmann D.~E.,  2010, Science,
  329, 187

\bibitem[\protect\citeauthoryear{Levison, Morbidelli, Tsiganis, Nesvorn\'{y}
  \& Gomes}{Levison et~al.}{2011}]{levisonetal11}
Levison H.~F.,  Morbidelli A.,  Tsiganis K.,  Nesvorn\'{y} D.,   Gomes R.,
  2011, Astron. J., 142, 152

\bibitem[\protect\citeauthoryear{Li, Jewitt, Mutchler, Agarwal  \& Weaver}{Li
  et~al.}{2020}]{lietal20}
Li J.,  Jewitt D.,  Mutchler M.,  Agarwal J.,   Weaver H.,  2020, Astron. J.,
  159, 209

\bibitem[\protect\citeauthoryear{Lieman-Sifry, Hughes, Carpenter, Gorti, Hales
  \& Flaherty}{Lieman-Sifry et~al.}{2016}]{liemansifryetal16}
Lieman-Sifry J.,  Hughes A.~M.,  Carpenter J.~M.,  Gorti U.,  Hales A.,
  Flaherty K.~M.,  2016, Astrophys. J., 828, 25

\bibitem[\protect\citeauthoryear{Lodders}{Lodders}{2003}]{lodders03}
Lodders K.,  2003, Astrophys. J., 591, 1220

\bibitem[\protect\citeauthoryear{Lorek, Gundlach, Lacerda  \& Blum}{Lorek
  et~al.}{2016}]{loreketal16}
Lorek S.,  Gundlach B.,  Lacerda P.,   Blum J.,  2016, Astron. Astrophys., 587,
  A128

\bibitem[\protect\citeauthoryear{Lovell et~al.,}{Lovell
  et~al.}{2020}]{lovelletal20}
Lovell J.~B.,  et~al., 2020, Mon. Not. R. Astron. Soc., TBD, TBD

\bibitem[\protect\citeauthoryear{Luna, Mill\'{a}n, Domingo  \& Satorre}{Luna
  et~al.}{2008}]{lunaetal08}
Luna R.,  Mill\'{a}n C.,  Domingo M.,   Satorre M.~{\'{A}}.,  2008, Astrophys.
  Space Sci., 314, 113

\bibitem[\protect\citeauthoryear{Luspay-Kuti et~al.,}{Luspay-Kuti
  et~al.}{2015}]{luspaykutietal15}
Luspay-Kuti A.,  et~al., 2015, Astron. Astrophys., 583, A4

\bibitem[\protect\citeauthoryear{Luu \& Jewitt}{Luu \&
  Jewitt}{1990}]{luujewitt90b}
Luu J.,  Jewitt D.,  1990, Astron. J., 100, 913

\bibitem[\protect\citeauthoryear{MacPherson, Davis  \& Zinner}{MacPherson
  et~al.}{1995}]{macphersonetal95}
MacPherson G.~J.,  Davis A.~N.,   Zinner E.~K.,  1995, Meteoritics, 30, 365

\bibitem[\protect\citeauthoryear{Malhotra}{Malhotra}{1993}]{malhotra93}
Malhotra R.,  1993, Nature, 365, 819

\bibitem[\protect\citeauthoryear{Marboeuf, Mousis, Petit  \& Schmitt}{Marboeuf
  et~al.}{2010}]{marboeufetal10}
Marboeuf U.,  Mousis O.,  Petit J.-M.,   Schmitt B.,  2010, Astrophys. J., 708,
  812

\bibitem[\protect\citeauthoryear{Marboeuf, Mousis, Petit, Schmitt, Cochran  \&
  Weaver}{Marboeuf et~al.}{2011}]{marboeufetal11}
Marboeuf U.,  Mousis O.,  Petit J.-M.,  Schmitt B.,  Cochran A.~L.,   Weaver
  H.~A.,  2011, Astron. Astrophys., 525, A144

\bibitem[\protect\citeauthoryear{Marchi et~al.,}{Marchi
  et~al.}{2013}]{marchietal13b}
Marchi S.,  et~al., 2013, Nature Geosci., 6, 303

\bibitem[\protect\citeauthoryear{Marino et~al.,}{Marino
  et~al.}{2016}]{marinoetal16}
Marino S.,  et~al., 2016, Mon. Not. R. Astron. Soc., 460, 2933

\bibitem[\protect\citeauthoryear{Markiewicz, Skorov, Keller  \&
  K\"{o}mle}{Markiewicz et~al.}{1998}]{markiewiczetal98}
Markiewicz W.~J.,  Skorov Y.~V.,  Keller H.~U.,   K\"{o}mle N.~I.,  1998,
  Planet. Space. Sci., 46, 357

\bibitem[\protect\citeauthoryear{Marty et~al.,}{Marty
  et~al.}{2017}]{martyetal17}
Marty B.,  et~al., 2017, Science, 356, 1069

\bibitem[\protect\citeauthoryear{Masiero, Davidsson, Liu, Moore  \&
  Tuite}{Masiero et~al.}{2021}]{masieroetal21}
Masiero J.~R.,  Davidsson B. J.~R.,  Liu Y.,  Moore K.,   Tuite M.,  2021,
  Planet. Sci. J.

\bibitem[\protect\citeauthoryear{Masset \& Snellgrove}{Masset \&
  Snellgrove}{2001}]{massetsnellgrove01}
Masset F.,  Snellgrove M.,  2001, Mon. Not. R. Astron. Soc., 320, L55

\bibitem[\protect\citeauthoryear{Matr\'{a} et~al.,}{Matr\'{a}
  et~al.}{2017a}]{matraetal17b}
Matr\'{a} L.,  et~al., 2017a, Mon. Not. R. Astron. Soc., 464, 1415

\bibitem[\protect\citeauthoryear{Matr\'{a} et~al.,}{Matr\'{a}
  et~al.}{2017b}]{matraetal17}
Matr\'{a} L.,  et~al., 2017b, Astrophys. J., 842, 9

\bibitem[\protect\citeauthoryear{Matr\'{a}, \"{O}berg, Wilner, Olofsson  \&
  Bayo}{Matr\'{a} et~al.}{2019}]{matraetal19}
Matr\'{a} L.,  \"{O}berg K.~I.,  Wilner D.~J.,  Olofsson J.,   Bayo A.,  2019,
  Astron. J., 157, 117

\bibitem[\protect\citeauthoryear{Matson, Castillo-Rogez, Schubert, Sotin  \&
  McKinnon}{Matson et~al.}{2009}]{matsonetal09}
Matson D.~L.,  Castillo-Rogez J.~C.,  Schubert G.,  Sotin C.,   McKinnon W.~B.,
   2009, in Dougherty M.~K.,  Esposito L.~W.,   Krimigis S.~M.,  eds, , Saturn
  from Cassini--Huygens.
Springer Science, pp 577--612

\bibitem[\protect\citeauthoryear{{McKinnon} et~al.,}{{McKinnon}
  et~al.}{2017}]{mckinnonetal17}
{McKinnon} W.~B.,  et~al., 2017, Icarus, 287, 2

\bibitem[\protect\citeauthoryear{Meech, Buie, Samarasinha, Mueller  \&
  Belton}{Meech et~al.}{1997}]{meechetal97}
Meech K.~J.,  Buie M.~W.,  Samarasinha N.~H.,  Mueller B. E.~A.,   Belton M.
  J.~S.,  1997, Astron. J., 113, 844

\bibitem[\protect\citeauthoryear{Meech et~al.,}{Meech
  et~al.}{2017a}]{meechetal17b}
Meech K.~J.,  et~al., 2017a, Astrophys. J., 153, 206

\bibitem[\protect\citeauthoryear{Meech et~al.,}{Meech
  et~al.}{2017b}]{meechetal17}
Meech K.~J.,  et~al., 2017b, Astrophys. J. Lett., 849, L8

\bibitem[\protect\citeauthoryear{Mekler, Prialnik  \& Podolak}{Mekler
  et~al.}{1990}]{mekleretal90}
Mekler Y.,  Prialnik D.,   Podolak M.,  1990, Astrophys. J., 356, 682

\bibitem[\protect\citeauthoryear{Merk \& Prialnik}{Merk \&
  Prialnik}{2003}]{merkprialnik03}
Merk R.,  Prialnik D.,  2003, Earth Moon Planets, 92, 359

\bibitem[\protect\citeauthoryear{Merk \& Prialnik}{Merk \&
  Prialnik}{2006}]{merkprialnik06}
Merk R.,  Prialnik D.,  2006, Icarus, 183, 283

\bibitem[\protect\citeauthoryear{Mo\'{o}r et~al.,}{Mo\'{o}r
  et~al.}{2015}]{mooretal15}
Mo\'{o}r A.,  et~al., 2015, Astrophys. J., 814, 42

\bibitem[\protect\citeauthoryear{Morbidelli \& Crida}{Morbidelli \&
  Crida}{2007}]{morbidellicrida07}
Morbidelli A.,  Crida A.,  2007, Icarus, 191, 158

\bibitem[\protect\citeauthoryear{Morbidelli, Levison, Tsiganis  \&
  Gomes}{Morbidelli et~al.}{2005}]{morbidellietal05}
Morbidelli A.,  Levison H.~F.,  Tsiganis K.,   Gomes R.,  2005, Nature, 435,
  462

\bibitem[\protect\citeauthoryear{Morbidelli, Tsiganis, Crida, Levison  \&
  Gomes}{Morbidelli et~al.}{2007}]{morbidellietal07}
Morbidelli A.,  Tsiganis K.,  Crida A.,  Levison H.~F.,   Gomes R.,  2007,
  Astron. J., 134, 1790

\bibitem[\protect\citeauthoryear{Morbidelli, Marchi, Bottke  \&
  Kring}{Morbidelli et~al.}{2012}]{morbidellietal12}
Morbidelli A.,  Marchi S.,  Bottke W.~F.,   Kring D.,  2012, Earth Planet. Sci.
  Lett., 355, 144

\bibitem[\protect\citeauthoryear{Mousis et~al.,}{Mousis
  et~al.}{2017}]{mousisetal17}
Mousis O.,  et~al., 2017, Astrophys. J. Lett., 839, L4

\bibitem[\protect\citeauthoryear{Muntean, Lacerda, Field, Fitzsimmons, Fraser,
  Hunniford  \& {McCullough}}{Muntean et~al.}{2016}]{munteanetal16}
Muntean E.~A.,  Lacerda P.,  Field T.~A.,  Fitzsimmons A.,  Fraser W.~C.,
  Hunniford A.~C.,   {McCullough} R.~W.,  2016, Mon. Not. R. Astron. Soc., 462,
  3361

\bibitem[\protect\citeauthoryear{Nesvorn\'{y} \& Morbidellli}{Nesvorn\'{y} \&
  Morbidellli}{2012}]{nesvornyandmorbidelli12}
Nesvorn\'{y} D.,  Morbidellli A.,  2012, Astron. J., 144, 117

\bibitem[\protect\citeauthoryear{Nesvorn\'{y}, Youdin  \&
  Richardson}{Nesvorn\'{y} et~al.}{2010}]{nesvornyetal10}
Nesvorn\'{y} D.,  Youdin A.~N.,   Richardson D.~C.,  2010, Astron. J., 140, 785

\bibitem[\protect\citeauthoryear{Norris, Gancarz, Rokop  \& Thomas}{Norris
  et~al.}{1983}]{norrisetal83}
Norris T.~L.,  Gancarz A.~J.,  Rokop D.~J.,   Thomas K.~W.,  1983, J. Geophys.
  Res., 88, B331

\bibitem[\protect\citeauthoryear{{O'Rourke} et~al.,}{{O'Rourke}
  et~al.}{2020}]{orourkeetal20}
{O'Rourke} L.,  et~al., 2020, Nature, 586, 697

\bibitem[\protect\citeauthoryear{\"{O}berg, Fayolle, Cuppen, {van Dishoeck}  \&
  Linnartz}{\"{O}berg et~al.}{2009}]{obergetal09}
\"{O}berg K.~I.,  Fayolle E.~C.,  Cuppen H.~M.,  {van Dishoeck} E.~F.,
  Linnartz H.,  2009, Astron. Astrophys., 505, 183

\bibitem[\protect\citeauthoryear{Orosei, Capaccioni, Capria, Coradini,
  Espinasse, Federico, Salomone  \& Schwehm}{Orosei
  et~al.}{1995}]{oroseietal95}
Orosei R.,  Capaccioni F.,  Capria M.~T.,  Coradini A.,  Espinasse S.,
  Federico C.,  Salomone M.,   Schwehm G.~H.,  1995, Astron. Astrophys., 301,
  613

\bibitem[\protect\citeauthoryear{Orosei, Capaccioni, Capria, Coradini, {De
  Sanctis}, Federico, Salomone  \& Huot}{Orosei et~al.}{1999}]{oroseietal99}
Orosei R.,  Capaccioni F.,  Capria M.~T.,  Coradini A.,  {De Sanctis} M.~C.,
  Federico C.,  Salomone M.,   Huot J.-P.,  1999, Planet. Space Sci., 47, 839

\bibitem[\protect\citeauthoryear{Palla \& Stahler}{Palla \&
  Stahler}{1993}]{pallastahler93}
Palla F.,  Stahler S.~W.,  1993, Astrophys. J., 418, 414

\bibitem[\protect\citeauthoryear{Parker \& Kavelaars}{Parker \&
  Kavelaars}{2010}]{parkerkavelaars10}
Parker A.~H.,  Kavelaars J.~J.,  2010, Astrophys. J. Lett., 722, L204

\bibitem[\protect\citeauthoryear{Parker, Kavelaars, Petit, Jones, Gladman  \&
  Parker}{Parker et~al.}{2011}]{parkeretal11}
Parker A.~H.,  Kavelaars J.~J.,  Petit J.-M.,  Jones L.,  Gladman B.,   Parker
  J.,  2011, Astrophys. J., 743, 1

\bibitem[\protect\citeauthoryear{P\"{a}tzold et~al.,}{P\"{a}tzold
  et~al.}{2019}]{patzoldetal19}
P\"{a}tzold M.,  et~al., 2019, Mon. Not. R. Astron. Soc., 483, 2337

\bibitem[\protect\citeauthoryear{Podolak \& Prialnik}{Podolak \&
  Prialnik}{1996}]{podolakandprialnik96}
Podolak M.,  Prialnik D.,  1996, Planet. Space Sci., 44, 655

\bibitem[\protect\citeauthoryear{Podolak \& Prialnik}{Podolak \&
  Prialnik}{2006}]{podolakprialnik06}
Podolak M.,  Prialnik D.,  2006, in Thomas P.~J.,  Chyba C.,   McKay C.,  eds,
  , {Comets and the origin and evolution of life}.
Springer--Verlag, Berlin

\bibitem[\protect\citeauthoryear{Pontoppidan et~al.,}{Pontoppidan
  et~al.}{2008}]{pontoppidanetal08}
Pontoppidan K.~M.,  et~al., 2008, Astrophys. J., 678, 1005

\bibitem[\protect\citeauthoryear{Preusker et~al.,}{Preusker
  et~al.}{2015}]{preuskeretal15}
Preusker F.,  et~al., 2015, Astron. Astrophys., 583, A33

\bibitem[\protect\citeauthoryear{Prialnik}{Prialnik}{1992}]{prialnik92}
Prialnik D.,  1992, Astrophys. J., 388, 196

\bibitem[\protect\citeauthoryear{Prialnik \& Bar-Nun}{Prialnik \&
  Bar-Nun}{1987}]{prialnikandbarnun87}
Prialnik D.,  Bar-Nun A.,  1987, Astrophys. J., 313, 893

\bibitem[\protect\citeauthoryear{Prialnik \& Bar-Nun}{Prialnik \&
  Bar-Nun}{1988}]{prialnikandbarnun88}
Prialnik D.,  Bar-Nun A.,  1988, Icarus, 74, 272

\bibitem[\protect\citeauthoryear{Prialnik \& Bar-Nun}{Prialnik \&
  Bar-Nun}{1990}]{prialnikandbarnun90}
Prialnik D.,  Bar-Nun A.,  1990, Astrophys. J., 363, 274

\bibitem[\protect\citeauthoryear{Prialnik \& Podolak}{Prialnik \&
  Podolak}{1995}]{prialnikpodolak95}
Prialnik D.,  Podolak M.,  1995, Icarus, 117, 420

\bibitem[\protect\citeauthoryear{Prialnik \& Podolak}{Prialnik \&
  Podolak}{1999}]{prialnikpodolak99}
Prialnik D.,  Podolak M.,  1999, Space Sci. Rev., 90, 169

\bibitem[\protect\citeauthoryear{Prialnik, Bar-Nun  \& Podolak}{Prialnik
  et~al.}{1987}]{prialniketal87}
Prialnik D.,  Bar-Nun A.,   Podolak M.,  1987, Astrophys. J., 319, 993

\bibitem[\protect\citeauthoryear{Prialnik, Benkhoff  \& Podolak}{Prialnik
  et~al.}{2004}]{prialniketal04}
Prialnik D.,  Benkhoff J.,   Podolak M.,  2004, in Festou M.~C.,  Keller H.~U.,
    Weaver H.~A.,  eds, , Comets II.
Univ. of Arizona Press, Tucson, pp 359--387

\bibitem[\protect\citeauthoryear{Prialnik, Sarid, Rosenberg  \& Merk}{Prialnik
  et~al.}{2008}]{prialniketal08}
Prialnik D.,  Sarid G.,  Rosenberg E.~D.,   Merk R.,  2008, Space Sci. Rev.,
  138, 147

\bibitem[\protect\citeauthoryear{Rickman \& Fern\'{a}ndez}{Rickman \&
  Fern\'{a}ndez}{1986}]{rickmanandfernandez86}
Rickman H.,  Fern\'{a}ndez J.~A.,  1986, in , The {Comet Nucleus Sample Return
  Mission}.
ESA SP--249, pp 185--194

\bibitem[\protect\citeauthoryear{Rickman, Fern\'{a}ndez  \& Gustafson}{Rickman
  et~al.}{1990}]{rickmanetal90}
Rickman H.,  Fern\'{a}ndez J.~A.,   Gustafson B. {\AA}.~S.,  1990, Astron.
  Astrophys., 237, 524

\bibitem[\protect\citeauthoryear{Robie, Hemingway  \& Takei}{Robie
  et~al.}{1982}]{robieetal82}
Robie R.~A.,  Hemingway B.~S.,   Takei H.,  1982, American Mineralogist, 67,
  470

\bibitem[\protect\citeauthoryear{Rosenberg \& Prialnik}{Rosenberg \&
  Prialnik}{2009}]{rosenbergandprialnik09}
Rosenberg E.~D.,  Prialnik D.,  2009, Icarus, 201, 740

\bibitem[\protect\citeauthoryear{Rosenberg \& Prialnik}{Rosenberg \&
  Prialnik}{2010}]{rosenbergandprialnik10}
Rosenberg E.~D.,  Prialnik D.,  2010, Icarus, 209, 753

\bibitem[\protect\citeauthoryear{Santos-Sanz et~al.,}{Santos-Sanz
  et~al.}{2012}]{santossanzetal12}
Santos-Sanz P.,  et~al., 2012, Astron. Astrophys., 541, A92

\bibitem[\protect\citeauthoryear{Sarid \& Prialnik}{Sarid \&
  Prialnik}{2009}]{saridandprialnik09}
Sarid G.,  Prialnik D.,  2009, Meteo. Planet. Sci, 44, 1905

\bibitem[\protect\citeauthoryear{S\'{a}rneczky et~al.,}{S\'{a}rneczky
  et~al.}{2016}]{sarneczkyetal16}
S\'{a}rneczky K.,  et~al., 2016, Astron. J., 152, 220

\bibitem[\protect\citeauthoryear{Satorre, Luna, Mill\'{a}n, Santonja  \&
  Cant\'{o}}{Satorre et~al.}{2009}]{satorreetal09}
Satorre M.~{\'{A}}.,  Luna R.,  Mill\'{a}n C.,  Santonja C.,   Cant\'{o} J.,
  2009, Planet. Space Sci., 57, 250

\bibitem[\protect\citeauthoryear{Schmitt, Espinasse, Grim, Greenberg  \&
  Klinger}{Schmitt et~al.}{1989}]{schmittetal89}
Schmitt B.,  Espinasse S.,  Grim R. J.~A.,  Greenberg J.~M.,   Klinger J.,
  1989, ESA~SP, 302, 65

\bibitem[\protect\citeauthoryear{Senay \& Jewitt}{Senay \&
  Jewitt}{1994}]{senayandjewitt94}
Senay M.~C.,  Jewitt D.,  1994, Nature, 371, 229

\bibitem[\protect\citeauthoryear{Shchuko, Shchuko, Kartashov  \&
  Orosei}{Shchuko et~al.}{2014}]{shchukoetal14}
Shchuko O.~B.,  Shchuko S.~D.,  Kartashov D.~V.,   Orosei R.,  2014, Planet.
  Space Sci., 104, 147

\bibitem[\protect\citeauthoryear{Shoshani, Heifetz, Prialnik  \&
  Podolak}{Shoshani et~al.}{1997}]{shoshanietal97}
Shoshani Y.,  Heifetz E.,  Prialnik D.,   Podolak M.,  1997, Icarus, 126, 342

\bibitem[\protect\citeauthoryear{Shoshany, Prialnik  \& Podolak}{Shoshany
  et~al.}{2002}]{shoshanyetal02}
Shoshany Y.,  Prialnik D.,   Podolak M.,  2002, Icarus, 157, 219

\bibitem[\protect\citeauthoryear{Sicilia-Aguilar et~al.,}{Sicilia-Aguilar
  et~al.}{2006}]{siciliaaguilaretal06}
Sicilia-Aguilar A.,  et~al., 2006, Astrophys. J., 638, 897

\bibitem[\protect\citeauthoryear{Simon, \"{O}berg, Rajappan  \&
  Maksiutenko}{Simon et~al.}{2019}]{simonetal19}
Simon A.,  \"{O}berg K.~I.,  Rajappan M.,   Maksiutenko P.,  2019, Astrophys.
  J., 883, 21

\bibitem[\protect\citeauthoryear{Skorov \& Rickman}{Skorov \&
  Rickman}{1995}]{skorovandrickman95}
Skorov Y.~V.,  Rickman H.,  1995, Planet. Space. Sci., 43, 1587

\bibitem[\protect\citeauthoryear{Skorov, K\"{o}mle, Markiewicz  \&
  Keller}{Skorov et~al.}{1999}]{skorovetal99}
Skorov Y.~V.,  K\"{o}mle N.~I.,  Markiewicz W.~J.,   Keller H.~U.,  1999,
  Icarus, 140, 173

\bibitem[\protect\citeauthoryear{Stern}{Stern}{2003}]{stern03}
Stern S.~A.,  2003, Nature, 424, 639

\bibitem[\protect\citeauthoryear{Stern \& Shull}{Stern \&
  Shull}{1988}]{sternshull88}
Stern S.~A.,  Shull J.~M.,  1988, Nature, 332, 407

\bibitem[\protect\citeauthoryear{Suhasaria, Thrower  \& Zacharias}{Suhasaria
  et~al.}{2017}]{suhasariaetal17}
Suhasaria T.,  Thrower J.~D.,   Zacharias H.,  2017, Mon. Not. R. Astron. Soc.,
  472, 389

\bibitem[\protect\citeauthoryear{Szab\'{o}, Kiss, P\'{a}l, Kiss, S\'{a}rneczky,
  Juh\'{a}sz  \& Hogerheijde}{Szab\'{o} et~al.}{2012}]{szaboetal12}
Szab\'{o} G.~M.,  Kiss L.~L.,  P\'{a}l A.,  Kiss C.,  S\'{a}rneczky K.,
  Juh\'{a}sz A.,   Hogerheijde M.~R.,  2012, Astrophys. J., 761, 8

\bibitem[\protect\citeauthoryear{Tancredi, Rickman  \& Greenberg}{Tancredi
  et~al.}{1994}]{tancredietal94}
Tancredi G.,  Rickman H.,   Greenberg J.~M.,  1994, Astron. Astrophys., 286,
  659

\bibitem[\protect\citeauthoryear{{Tesfaye~Firdu} \& Taskinen}{{Tesfaye~Firdu}
  \& Taskinen}{2010}]{tesfayefirdutaskinen10}
{Tesfaye~Firdu} F.,  Taskinen P.,  2010, {Densities of molten and solid alloys
  of (Fe, Cu, Ni, Co) -- S at elevated temperatures -- Literature review and
  analysis}.
Aalto University Publications in Materials Science and Engineering, Aalto
  University School of Science and Technology, Espoo

\bibitem[\protect\citeauthoryear{Thomas et~al.,}{Thomas
  et~al.}{2015a}]{thomasetal15a}
Thomas N.,  et~al., 2015a, Science, 347, aaa0440

\bibitem[\protect\citeauthoryear{Thomas et~al.,}{Thomas
  et~al.}{2015b}]{thomasetal15b}
Thomas N.,  et~al., 2015b, Astron. Astrophys., 583, A17

\bibitem[\protect\citeauthoryear{Tscharnuter, Sch\"{o}nke, Gail, Trieloff  \&
  L\"{u}ttjohan}{Tscharnuter et~al.}{2009}]{tscharnuteretal09}
Tscharnuter W.~M.,  Sch\"{o}nke J.,  Gail H.-P.,  Trieloff M.,   L\"{u}ttjohan
  E.,  2009, Astron. Astrophys., 504, 109

\bibitem[\protect\citeauthoryear{Tsiganis, Gomes, Morbidelli  \& f.
  Levison}{Tsiganis et~al.}{2005}]{tsiganisetal05}
Tsiganis K.,  Gomes R.,  Morbidelli A.,   f. Levison H.,  2005, Nature, 435,
  459

\bibitem[\protect\citeauthoryear{Walsh, Morbidelli, Raymond, {O'Brien}  \&
  Mandell}{Walsh et~al.}{2011}]{walshetal11}
Walsh K.~J.,  Morbidelli A.,  Raymond S.~N.,  {O'Brien} D.~P.,   Mandell A.~M.,
   2011, Nature, 475, 206

\bibitem[\protect\citeauthoryear{Weast}{Weast}{1974}]{weast74}
Weast R.~C.,  1974, Handbook of chemistry and physics, 55 edn.
CRC Press, Cleveland

\bibitem[\protect\citeauthoryear{Weidenschilling}{Weidenschilling}{1997}]{weidenschilling97}
Weidenschilling S.~J.,  1997, Icarus, 127, 290

\bibitem[\protect\citeauthoryear{Weissman \& Kieffer}{Weissman \&
  Kieffer}{1981}]{weissmanandkieffer81}
Weissman P.~R.,  Kieffer H.~H.,  1981, Icarus, 47, 302

\bibitem[\protect\citeauthoryear{Wierzchos, Womack  \& Sarid}{Wierzchos
  et~al.}{2017}]{wierzchosetal17}
Wierzchos K.,  Womack M.,   Sarid G.,  2017, Astron. J., 153, 230

\bibitem[\protect\citeauthoryear{Womack \& Stern}{Womack \&
  Stern}{1999}]{womackandstern99}
Womack M.,  Stern S.~A.,  1999, Sol. Sys. Res., 33, 187

\bibitem[\protect\citeauthoryear{Yasui \& Arakawa}{Yasui \&
  Arakawa}{2009}]{yasuiarakawa09}
Yasui M.,  Arakawa M.,  2009, J. Geophys. Res., 114, E09004

\bibitem[\protect\citeauthoryear{Yomogida \& Matsui}{Yomogida \&
  Matsui}{1983}]{yomogidamatsui83}
Yomogida K.,  Matsui T.,  1983, J. Geophys. Res., 88, 9513

\bibitem[\protect\citeauthoryear{Youdin \& Goodman}{Youdin \&
  Goodman}{2005}]{youdingoodman05}
Youdin A.~N.,  Goodman J.,  2005, Astrophys. J., 620, 459

\bibitem[\protect\citeauthoryear{Zuckerman, Foreille  \& Kastner}{Zuckerman
  et~al.}{1995}]{zuckermanetal95}
Zuckerman B.,  Foreille T.,   Kastner J.~H.,  1995, Nature, 373, 494

\makeatother
\end{thebibliography}

\bsp	
\label{lastpage}
\end{document}